\documentclass[12pt, preprint] {emulateapj}
\usepackage{natbib}

\def\grtsim{\mathrel{\hbox{\rlap{\hbox{\lower2pt\hbox{$\sim$}}}\raise2pt\hbox{$>$}}}} 
\def\lesssim{\mathrel{\hbox{\rlap{\hbox{\lower2pt\hbox{$\sim$}}}\raise2pt\hbox{$<$}}}}

\def\degree{\nobreak\ifmmode{^\circ}\else{$^\circ$}\fi}

\newcommand{\mum}{$\mu$m}

\newcommand{\whzsr}{W~Hz$^{-1}$~sr$^{-1}$}

\newcommand{\kms}{km~s$^{-1}$}

\newcommand{\av}{$A_{\rm V}$}

\newcommand{\lrada}{$L_{1.4 \rm GHz}$}

\newcommand{\lbol}{$L_{\rm bol}$}

\newcommand{\lir}{$L_{\rm IR}$}
\newcommand{\lfir}{$L_{\rm FIR}$}

\newcommand{\mdust}{$M_{\rm d}$}
\newcommand{\mgas}{$M_{\rm g}$}

\newcommand{\snu}{$S_{\nu}$}

\def\lsol{L$_{\odot}$}

\def\msol{M$_{\odot}$}
\def\msolyr{M$_{\odot}$~yr$^{-1}$}

\shorttitle{Mm observations of high-$z$ obscured quasars}
\shortauthors{A. Mart\'\i nez-Sansigre et~al.}

\begin{document}

\title{Millimetre observations of a sample of high-redshift obscured quasars$^{1}$} 

\author{Alejo Mart\'\i nez-Sansigre$^{2,3}$, Alexander Karim$^{2}$,  Eva
  Schinnerer$^{2}$, Alain Omont$^{4}$,
  Daniel J.B. Smith$^{5}$,  Jingwen Wu$^{6}$,  Gary J. Hill$^{7}$,  Hans-Rainer Kl\"
  ockner$^{3}$, Mark Lacy$^{8}$, Steve Rawlings$^{3}$,  Chris J. Willott$^{9}$  }

 \altaffiltext{1}{Based on observations carried out with the IRAM 30~m Telescope and the Plateau de Bure Interferometer. IRAM is supported by INSU/CNRS (France), MPG (Germany) and IGN (Spain).}
 \altaffiltext{2}{Max-Planck-Institut f\"ur Astronomie, K\"onigstuhl 17,
    D-69117 Heidelberg, Germany; ams@astro.ox.ac.uk, karim@mpia.de, schinner@mpia.de}
\altaffiltext{3}{Astrophysics, Department of Physics, University of
Oxford, Keble Road, Oxford OX1 3RH, UK; ams@astro.ox.ac.uk, hrk@astro.ox.ac.uk, sr@astro.ox.ac.uk }
\altaffiltext{4}{Institut d'Astrophysique de Paris, CNRS, and Universit\'e de Paris VI, 
  98bis boulevard Arago, F-75014 Paris, France; omont@iap.fr}
\altaffiltext{5}{The Centre for Astronomy \& Particle Theory, School of Physics
  and Astronomy, University of Nottingham, University Park, Nottingham NG7
  2RD, UK; daniel.j.b.smith@gmail.com}
\altaffiltext{6}{Harvard-Smithsonian Center for Astrophysics, 60 Garden St. MS78, Cambridge, MA, 02138, USA: jwu@cfa.harvard.edu}
\altaffiltext{7}{McDonald Observatory, University of Texas at Austin, 1
  University Station C1402, Austin, TX 78712-0259, USA; hill@astro.as.utexas.edu}
\altaffiltext{8}{Spitzer Science Center, California
  Institute of Technology, MS220-6, 1200 E. California Boulevard, Pasadena, CA
  91125, USA; mlacy@ipac.caltech.edu}
\altaffiltext{9}{Herzberg Institute of Astrophysics, National Research Council Canada, 5071 West Saanich Rd., Victoria, BC V9E 2E7, Canada; chris.willott@nrc-cnrc.gc.ca}

\begin{abstract}  
  We present observations at 1.2~mm with MAMBO-II of a sample of
  $z\grtsim2$ radio-intermediate obscured quasars, as well as CO
  observations of two sources with the Plateau de Bure Interferometer.
  The typical rms noise achieved by the MAMBO observations is 0.55~mJy
  beam$^{-1}$ and 5 out of 21 sources (24\%) are detected at a
  significance of $\geq3\sigma$. Stacking all sources leads to a
  statistical detection of $\langle S_{1.2~\rm mm} \rangle =
  0.96\pm0.11$~mJy and stacking only the non-detections also yields a
  statistical detection, with $\langle S_{1.2~\rm mm} \rangle =
  0.51\pm0.13$~mJy. At the typical redshift of the sample, $z=2$, 1~mJy
  corresponds to a far-infrared luminosity
  \lfir$\sim4\times10^{12}$~\lsol. If the far-infrared luminosity is
  powered entirely by star-formation, and not by AGN-heated dust, then
  the characteristic inferred star-formation rate is
  $\sim$700~\msolyr.  This far-infrared luminosity implies a dust mass
  of \mdust$\sim3\times10^{8}$~\msol, which is expected to be
  distributed on $\sim$kpc scales. We estimate that such large dust
  masses on kpc scales can plausibly cause the obscuration of the
  quasars.  Combining our observations at 1.2~mm with mid- and
  far-infrared data, and additional observations for two objects at
  350~\mum\, using SHARC-II, we present dust SEDs for our sample and
  derive a mean SED for our sample. This mean SED is not well fitted
  by clumpy torus models, unless additional extinction and
  far-infrared re-emission due to cool dust are included. This
  additional extinction can be consistently achieved by the mass of
  cool dust responsible for the far-infrared emission, provided the
  bulk of the dust is within a radius $\sim$2-3~kpc.  Comparison of
  our sample to other samples of $z\sim2$ quasars suggests that
  obscured quasars have, on average, higher far-infrared luminosities
  than unobscured quasars. There is a hint that the host galaxies of
  obscured quasars must have higher cool-dust masses and are therefore
  often found at an earlier evolutionary phase than those of
  unobscured quasars. For one source at $z=2.767$, we detect the
  CO(3-2) transition, with $S_{\rm CO}\Delta \nu=$630$\pm$50 mJy~\kms,
  corresponding to $L_{\rm CO (3-2)}=$3.2$\times10^{7}$~\lsol, or a
  brightness-temperature luminosity of $L'_{\rm CO
    (3-2)}=2.4\times10^{10}$ K~\kms~pc$^{2}$. For another source at
  $z=4.17$, the lack of detection of the CO(4-3) line suggests the
  line to have a brightness-temperature luminosity $L'_{\rm CO
    (4-3)}<1\times10^{10}$ K~\kms~pc$^{2}$. Under the assumption that
  in these objects the high-J transitions are thermalised, we can
  estimate the molecular gas contents to be $M_{\rm
    H_{2}}=1.9\times10^{10}$ \msol and $<8\times10^{9}$ \msol,
  respectively.  The estimated gas depletion timescales are $\tau_{\rm
    g}=4$~Myr and $<$16~Myr, and low gas-to-dust mass ratios of
  \mgas$/$\mdust$=19$ and $\leq8$ are inferred. These values are at
  the low end but consistent with those of other high-redshift
  galaxies.
 \end{abstract}

\keywords{galaxies:nuclei -- galaxies:active -- quasars:general -- infrared:galaxies -- galaxies:starbursts -- ISM:molecules}

\section{Introduction} 

Quasars are believed to be powered by supermassive black holes (SMBHs)
rapidly growing by accreting matter at a large fraction of their
Eddington rate\footnote{In this article we use the term quasar for the
  most powerful active galactic nuclei (AGNs). In the case of
  unobscured AGNs, the dividing line is taken at a B-band magnitude of
  $M_{\rm B}=-23.5$ which, assuming the typical quasar SED of
  \citet{1994ApJS...95....1E}, corresponds to $L_{\rm bol}\grtsim
  2\times10^{12}$~\lsol. We will therefore also refer to obscured AGNs
  as obscured quasars if they have $L_{\rm bol}\grtsim
  2\times10^{12}$~\lsol.}. The accreted matter forms a disk, is heated
to temperatures of $\sim10^{3}-10^{5}$~K, and emits thermal emission
at optical and ultraviolet wavelengths. Clouds of gas ionised by this
radiation will emit line radiation, and lines which are relatively
close to the black hole have rapid orbits which will show broad
velocity dispersions ($>$2000 \kms), while clouds further away will
show narrower lines.

Dust can survive when the equilibrium temperature with the ultraviolet
photons is $\lesssim$2000~K. This dust absorbs optical and ultraviolet
radiation and re-emits it at infrared wavelengths characteristic of
the dust temperature. If the line of sight of an observer to the
central region is blocked by dust, the optical and ultraviolet
continuum, as well as the broad lines, will not be observable. Such a
source is known as an obscured quasar.

While radio-loud obscured quasars have been known for a long time, in
the form of high-excitation narrow-line radio galaxies \citep[see ][
for a review]{1995PASP..107..803U}, the population of radio-quiet
obscured quasars had remained elusive, with only a few
individual objects known. This has changed recently, when large
numbers of these sources have been found in the spectroscopic database
of the Sloan Digital Sky Survey \citep{2003AJ....126.2125Z}. Samples
of spectroscopically-confirmed obscured quasars have now also been
identified in surveys using X-ray selection
\citep[e.g.][]{2004ApJS..155..271S,2005AJ....129..578B} as well as
mid-infrared, sometimes combined with radio, selection
\citep[e.g.][]{2005ApJ...622L.105H,2005Natur.436..666M,2007ApJ...669L..61L}.
It seems that the obscured quasars are at least as common as the the
unobscured quasars \citep[e.g.][]{2008AJ....136.2373R}, and probably
outnumber these by a $\sim$2:1 ratio
\citep[e.g.][]{2005Natur.436..666M,2008ApJ...674..676M,2007ApJ...669L..61L,2008ApJ...675..960P}.

There is a large body of evidence supporting obscuration of quasars as
an orientation effect, where obscured and unobscured quasars are
intrinsically identical sources, only dust around the accretion disk
covers certain lines of sight so that the broad emission lines and the
optical and ultraviolet continuum are not visible \citep[the dusty
torus of the unified schemes, e.g.][]{1995PASP..107..803U}. However,
dust distributed within the host galaxy can also cause obscuration,
particularly if the galaxy is at an early evolutionary phase and is
rich in gas and dust \citep[see
e.g.][]{1982ApJ...256..410L,1988ApJ...325...74S,1999MNRAS.308L..39F}.

The importance of galaxy-scale dust is evident in recent studies of
quasars where obscuration cannot be solely assigned to dust from a
torus around the accretion disk: in some objects even the narrow-lines
are obscured
\citep{2005Natur.436..666M,2006MNRAS.370.1479M,2006ApJ...645..115R}. The
jets emanating from some obscured quasars are face-on
\citep{2006MNRAS.373L..80M,2007ApJ...667L..17S,2009arXiv0905.1605K},
something that is not expected for quasars obscured by the torus, but
can be explained if the obscuring dust is on a larger ($\sim$kpc)
scale. The deep silicate absorption features and spectral energy
distributions (SEDs) suggest foreground extinction is sometimes
present \citep{2007ApJ...654L..45L,2008ApJ...675..960P}. This all
provides indirect evidence suggesting obscuring dust on $\sim$~kpc
scales, something which has been spectroscopically confirmed through
the measurement of the H$\alpha$ to H$\beta$ Balmer decrement \citep
{2007ApJ...663..204B}.

Dust present on kpc scales is expected to be relatively cool
($T\sim$50~K) and will emit thermally in the far-infrared ($\lambda
\grtsim$40~\mum), regardless of whether it is heated by young stars or
by the central active galactic nucleus \citep[AGN, see e.g.][ and
Section~\ref{sec:scale}]{1987ApJ...320..537B,1989ApJ...347...29S,2004A&A...421..129S}.
Observing this dust around the peak of its emission (around rest-frame
$\sim$65~\mum) is not optimal, since warmer dust can still contribute
significantly there. The Rayleigh-Jeans tail of the emission, however,
will have a much smaller contribution from warm dust, and has the
additional advantage of being observable from the ground, for example
in the atmospheric windows at 850~\mum\, or 1.2~mm.
 
In this paper we present continuum observations of a sample of 21
high-redshift ($z\grtsim$2) radio-intermediate obscured quasars at
1.2~mm, combined with other infrared and submillimetre data, as well
as a search for CO in two of the sources.  The sample was selected to
approximately match the break in the unobscured luminosity function at
$z\sim2$ \citep[$M_{\rm B}\sim$-25.7, ][]{2004MNRAS.349.1397C}, so in
terms of radiation these quasars represent the energetically-dominant
population around the peak of the quasar activity. The sources are,
however, radio-intermediate in that their radio luminosities are
slightly higher than those expected from radio-quiet quasars
\citep[with \lrada\,
$\sim10^{24}$~\whzsr][]{2006MNRAS.373L..80M}. Nevertheless, observations
with very large baseline interferometry suggest the relative
importance of the compact cores makes them more similar to the
genuinely radio-quiet population than to the radio galaxies and
radio-loud quasars \citep[see][]{2009arXiv0905.1605K}.
Section~\ref{sec:obs} summarises the observations and data reduction.
In Section~\ref{sec:res} we show the resulting fluxes, inferred
luminosities, star-formation rates, dust masses, characteristic scale
of the cool dust and discuss the possible obscuration by kpc-scale
dust. Section~\ref{sec:broad} shows the results of broad-band SED
fitting and Section~\ref{sec:mean} presents the mean SED for our
sample. In Section~\ref{sec:co} we discuss the molecular gas
observations. We compare our sample to other millimetre or
submillimetre observations of $z\sim2$ quasars in
Section~\ref{sec:comp}. The discussion and summary are found in
Section~\ref{sec:disc}.

\section{observations and data reduction}\label{sec:obs}

\subsection{MAMBO continuum observations}\label{sec:mambo}

The sample of obscured quasars was observed at a wavelength of 1.2~mm
(250~GHz) using the 117-element Max-Planck Millimetre Bolometer Array
\citep[MAMBO-II][ hereafter simply MAMBO]{1998SPIE.3357..319K}, at the
IRAM Pico de Veleta 30~m telescope. The observations were carried out
in ``ON-OFF'' mode during the MAMBO pool at the IRAM 30~m telescope
throughout various semesters. AMS16 was observed in winter 2005-2006,
while the remaining objects were observed between summer 2007 (Sum07)
and winter 2007-2008 (Win07).  Due to a warning issued by IRAM (see
below), the objects detected during the Win07 pool were re-observed in
summer 2008 (Sum08) and winter 2008-2009 (Win08).

To minimise the effect of atmospheric absorption, the science targets
were observed whenever possible at elevations $\geq$45~deg, and often
at around 60~deg.  Observing at elevations higher than 75~deg is,
however, not possible with the 30~m telescope.

For the first-order removal of the sky, the observations make use of a
secondary mirror to chop (or ``wobble''), with a typical azimuthal throw of
32 to 45 arcsec and a chopping frequency of 2~Hz. The ON-OFF observations are
done by nodding, so that the sky-only position sampled by the chopping is
alternatively on the right or on the left of the sky-plus-source position.
Each nod position (ON or OFF) is observed for one minute, and to minimise
overheads an ON-OFF-OFF-ON sequence is used. Hence integration blocks
(``scans'') are multiples of 4 minutes.

The ON-OFF observations were carried out centred on the most sensitive
pixel, number 20, with the atmospheric opacity $\tau_{\rm 230~GHz}
\lesssim 0.3$ and medium-to-low sky noise ($\lesssim$200~mJy
beam$^{-1}$). Each source was typically observed in several blocks of
20 minute integrations.

The ``sky-only'' position can sometimes fall on a real astronomical
source.  Given that high-redshift AGNs are often found to have
(sub)millimetre-bright galaxies nearby \citep{2003Natur.425..264S},
there is a real risk of ruining the sky subtraction by chopping onto a
bright source.  To minimise the risk of this happening (and the impact
on the final data), the chop angle (``wobbler throw'') was
alternatively varied between 32, 35 and 42 arcsec, and observations
were carried out on different dates and at different times of the day.
This means the chop in azimuth corresponds to different physical
offsets in RA and Dec, so the chance of repeatedly falling on the same
astronomical source is minimised.

Throughout the observing runs, gain calibration was performed by observing
Neptune, Uranus or Mars, and the flux calibration was always found to
be accurate to $\leq$20\%. The gain calibration was also monitored
regularly by observing other sources with known bright millimetre
fluxes ($\geq$5~Jy).  Total power measurements at different elevations
were made to infer the atmospheric opacity (``skydip'').  These were
performed typically every 2 hours under stable weather conditions, and
at shorter intervals for less stable conditions. The skydips were
generally done with the telescope azimuthal angle matched to that of
the science targets. The focus of the telescope was also regularly
monitored on bright sources. Before the first science target and every
40 minutes the accuracy of the telescope pointing was checked using
the nearby source J1638$+$573 (0.6~Jy). Observations of J1638$+$573
and other bright sources were also used to check for anomalous
refraction, in which case no science observations were made.

During the observations, the sensitivity of MAMBO in a 1 second
integration was typically between 35 and 45 mJy beam$^{-1}$. Including
typically 60\% overheads, the observations of this sample of 21
obscured quasars totalled approximately 65~hours.

The data were reduced using the MOPSIC pipeline (developed by
R. Zylka).  MOPSIC removes the atmospheric emission using the two
chopper positions\footnote{see
  http://iram.fr/IRAMFR/ARN/dec05/node9.html for more details. }. The
sky noise is measured by taking the weighted mean of the correlated
signal from adjacent pixels, and subsequently removed. The MAMBO
pixels are separated in the array by two beams and weak point sources
($<$150~mJy) will therefore not be detected in adjacent pixels. Using
the measurements of the atmospheric opacity from the skydips, and the
elevation at which objects were observed, the extinction due to
atmospheric water vapour is then corrected. Gain calibration converts
counts to flux density, and finally, for each science target the
weighted mean of all the scans is taken. These are the flux densities
quoted in Table~\ref{tab:obs}: a relatively uniform root mean square
(rms) noise $\sim$0.55~mJy beam$^{-1}$ was achieved for most of the
sources, with only a few sources having an rms $\sim$0.7-0.9 mJy
beam$^{-1}$.

In September~2008 IRAM issued a warning to MAMBO observers: several
technical problems had ocurred during the Win07 semester, which
included thermal and electric malfunctions in the preamplifier,
heightened sensitivity of the bolometer to microphotonics and spikes
in the detected signal. These problems were expected to affect a
fraction of the scans, and the warning suggested that any $<$10~mJy
detections obtained during this semester should be checked. Four of
sources from this sample were detected during Win07: AMS07, AMS08
AMS12 and AMS17. All four were observed again during the Win08 period,
together with extra scans of AMS10, AMS11 and AMS20. The data for
Win07 and Win08 were reduced independently, with the results shown in
Table~\ref{tab:07_08}. 

The Win07 flux densities are systematically higher, so only sources
detected in the Sum08/Win08 period are considered safe
detections. However, the source AMS17 was independently detected by
\citet[][their source 22558]{2008ApJ...683..659S}, so their flux
density is used in Table~\ref{tab:obs}. The source AMS07 is not considered
a safe detection.  The potential worry was detecting sources
spuriously so for the sources that are clearly not detected the
combined data can be used to yield deeper
limits. Table~\ref{tab:07_08} marks in bold the flux densities that
are used in Table~\ref{tab:obs}.

\subsection{SHARC observations}\label{sec:sharc}

Two objects were observed in continuum at 350~\mum, AMS13 and AMS16,
on June 16th and 17th 2005, using the Submillimeter High Angular
Resolution Camera II (SHARC-II) at the 10.4 m telescope of the Caltech
Submillimeter Observatory (CSO) at Mauna Kea, Hawaii
\citep{dowell03}. SHARC-II is a background-limited camera utilizing a
``CCD-style'' bolometer array with 12 $\times$ 32 pixels. At 350 \mum,
the beam size is 8.5 arcsec, with a 2.59 $\times$ 0.97 arcmin$^{2}$
field of view. The observations were taken during very dry weather:
AMS16 was observed during 2 hours with an opacity at 225 GHz of
$\tau_{225~\rm GHz}=$0.05, and AMS13 was observed for 1 hour with
$\tau_{225~\rm GHz}=$0.06. The Dish Surface Optimization System
\citep[DSOS, ][]{leong05} was used during the observation to correct
the dish surface figure for imperfections and gravitational
deformations as the dish moves in elevation.

The raw data were reduced with the Comprehensive Reduction Utility for
SHARC-II, CRUSH - version 1.40a9-2 (Kov{\'{a}}cs 2006).  We used the
sweep mode of SHARC-II to observe the sources: in this mode the
telescope moves in a Lissajous pattern that keeps the central regions
of the maps fully sampled, but causes the edges to be much noisier
than the central regions. Pointing was checked with Uranus and
secondary object CRL\_16293-2422 several times during the observations.
The pointing accuracy is better than 3 arcsec. Uranus was also used as
the flux calibrator. We used Starlink's ``stats'' package to
measure the flux of the calibrator and estimate the noise of the
image. The resulting root-mean-square noise values for AMS13 and AMS16
are 7.0 mJy beam$^{-1}$ and 5.5 mJy beam$^{-1}$, respectively, and neither
source is detected at the 3$\sigma$ level.

\subsection{PdBI observations}\label{sec:pdbi}

Two sources, AMS12 and AMS16 were observed using the Plateau de Bure
Interferometer (PdBI) to search for molecular gas in the host
galaxies, by using carbon monoxide (CO) as a tracer of the molecular
gas.

AMS16 was the first object observed with MAMBO, in 2005.  At the
redshift of the source, $z=4.169$, the CO (4-3) rotational transition
falls in the 3~mm atmospheric window [where (4-3) stands for the
transition from the $J=4$ to the $J=3$ angular quantum numbers].

The observations were carried out on 28 August and 01 September 2007,
with 5 antennae in the most compact configuration (D). The source
3C~345 was used as amplitude and bandpass calibrator, MWC~349 as
amplitude calibrator and the phase calibrators were 1823$+$568 (28
Aug.) and 1637$+$574 (01 Sep.). The correlator was configured with
both polarizations observing at a central frequency of 89.2002~GHz,
with a bandwidth of 1~GHz. For the CO~(4-3) transition at $z=4.169$,
this corresponds to $\pm1680$~\kms, and should suffice to avoid
missing the line given the accuracy of the redshift.

The night of 28 August had moderately good conditions, and 25\% of the
data were flagged automatically\footnote{ Data can be flagged as having
  bad quality for many reasons. Examples for reasons to automatically
  flag the data during the observations are time errors, tracking
  problems or antenna shadowing. During the calibration and reduction
  of the data, visibilities with high phase noise or high pointing
  correction, tracking errors as well as amplitude loss can also be
  flagged.}. The night of 01 September had excellent conditions (with
only 1\% of the data flagged) although for one antenna, 1 spectral
unit out of the 8 (L07 in horizontal polarization for antenna 6) had
to be flagged for the entire night. The total usable time from both
nights amounted to 12.9 hours on-source.

The data were reduced using the GILDAS
software\footnote{http://www.iram.fr/IRAMFR/GILDAS}. After amplitude,
phase and bandpass calibration, data cubes of the phase calibrators
and of AMS16 were created. The phase calibrators came out perfectly
centred at the phase center, suggesting excellent phase calibration,
but AMS16 was not detected either in continuum or in line. A spectrum
was extracted at the position of the source, with the shape of the
beam. The root-mean square noise achieved in the spectrum was 0.65~mJy
beam$^{-1}$ per 10~MHz bin. Assuming a box-car CO line shape with full
width to zero intensity of 400~\kms, this corresponds to 74~mJy
beam$^{-1}$ \kms. The continuum sensitivity achieved over the entire
bandwidth was 0.065~mJy beam$^{-1}$ rms.

AMS12 was also observed using the PdBI, to look for the CO (3-2)
transition.  The observations, carried out during the 30 April and 13
May 2009, were centred on 91.796~GHz using a 1~GHz
bandwidth. Depending on the day, the calibrators 3C~345, MWC~349,
2145$+$067 and 0923$+$392 were used as flux or bandpass calibrators,
while the source 1637$+$574 was used as a phase calibrator on both
dates. On the 30 April, only 5 antennae were available and in addition
approximately 50\% of the data were flagged.  For both dates the water
vapour radiometer for antenna 2 did not work, so no atmospheric phase
corrections were applied to the data from this antenna. In total, 7.3
hours of usable data were obtained and an rms noise of 0.70~mJy
beam$^{-1}$ per 10~MHz channel was reached.

The CO (3-2) transition was clearly detected in AMS12, with an
integrated flux of 630~mJy beam$^{-1}$ \kms.  The continuum rms is
0.1~mJy beam$^{-1}$, and no continuum is detected at 3~mm, so that we
obtain a 3$\sigma$ limit of $S_{3 \rm~mm}<$0.3~mJy.  In the case of
AMS16, the lack of continuum sets an 3$\sigma$ upper limit on the
89~GHz (3~mm) flux density of $S_{3 \rm~mm}<$0.2~mJy.

\section{Results from continuum observations}\label{sec:res}

The millimetre continuum observations provide us with a wealth of
information about the far-infrared luminosities and cool-dust masses
of our sources. They can also be used to test whether there is enough
dust along the host galaxy to cause the obscuration in the quasars. In
the following section, we derive these physical properties from our
observations.

The data uniformly available for this sample includes the following
wavelengths: 3.6, 4.5, 5.8, 8.0, 24, 70 and 160~\mum, as well as
1.2~mm. For a small number of sources data is also available at 350 or
850~\mum. With such excellent wavelength coverage, one might expect
the far-infrared luminosities and cool-dust masses to be very
accurately constrained. However, the wavelengths corresponding to the
rest-frame mid-infrared are dominated by hot dust from the warm torus
around the central engine, and cannot be used to constrain the cool-dust
component as we explain below \citep [this was originally by
selection, but has additionally been confirmed by mid-infrared
spectroscopy, see ][] {2006MNRAS.370.1479M,2008ApJ...674..676M}.

The far-infrared luminosity is dominated by cool dust, expected to be
characterised by temperatures $\sim$35-50~K. This means that emission
is expected to peak around 65-90~\mum\, and drop exponentially at
shorter wavelengths. The emission from the warm  torus is caused by dust
at a range of temperatures, typically $\sim$100-2000~K, whose emission
will peak at shorter wavelengths than that of the cool dust. If the
AGN is powerful enough, as is the case in the sources making up this
sample, the dust emission from the torus will dominate at mid-infrared
wavelengths. Hence, the mid-infrared emission is not useful in
constraining the cool dust mass and luminosity.

\begin{figure} 
\plotone{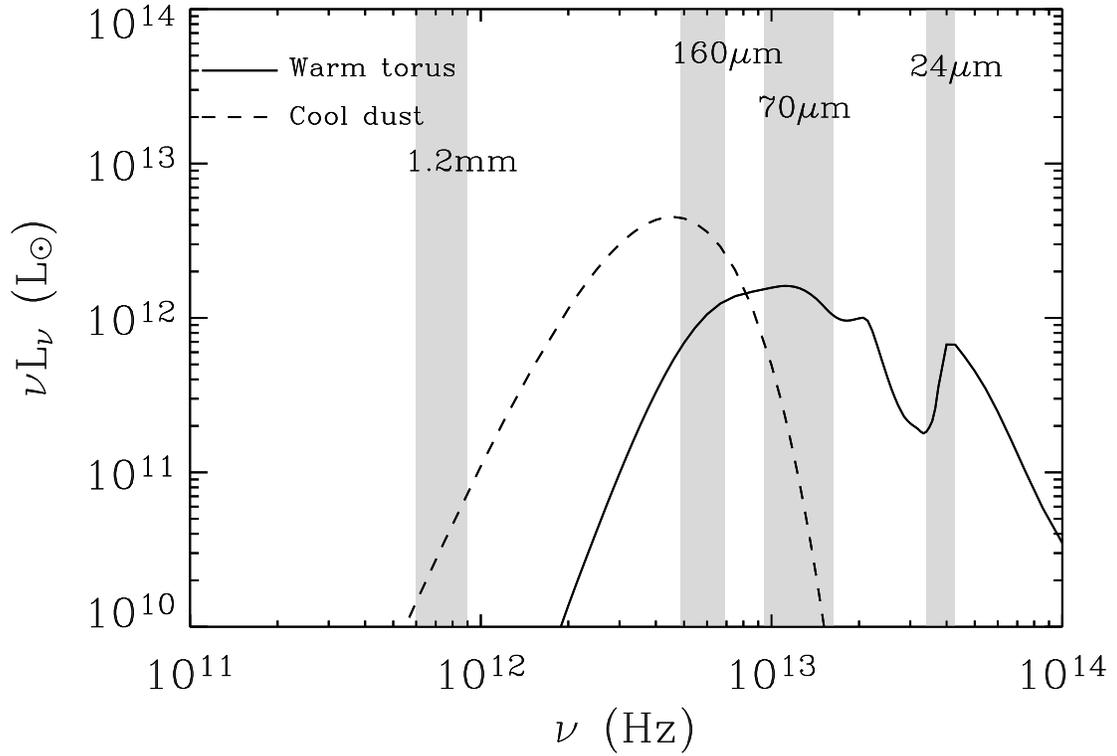}
\caption{ Illustration of the physical components that dominate at
  each observed band. The figure shows the SED of a model torus
  \citep[solid line, from ][]{2008ApJ...685..147N,2008ApJ...685..160N}
  and a cool dust component (dashed line) modelled as a gray body with
  temperature $T=$40~K and emissivity index $\beta=1.5$ (see
  Equation~\ref{eq:graybody}). The observed bands at 24, 70 and
  160~\mum, as well as 1.2~mm, are available for all 21 sources and
  have been marked in gray. This model source is assumed to be at
  $z=2$ so the bands have been accordingly shifted to the rest-frame frequencies.  The relative luminosities of the two components are
   arbitrary but representative of the sources in this
  sample. For completeness, the exact torus parameters used are:
  single-cloud optical depths $\tau_{\rm V}=60$, gaussian angular
  distribution of clumps around the equator, $\sigma=60$, the number
  of clouds along a ray in the equatorial plane $N=10$, the ratio of
  inner to outer torus radius $Y=30$, and the power of the radial
  density distribution, $q=0$.}
\label{fig:illu} 
\end{figure}

To illustrate this, Figure~\ref{fig:illu} shows a model rest-frame SED
consisiting of cool dust modelled by a gray body with temperature
$T=40$~K and emissivity $\beta=1.5$ together with a model AGN torus
from the radiative transfer models of
\citet{2008ApJ...685..147N,2008ApJ...685..160N}. The Spitzer MIPS
bands at 24, 70 and 160~\mum, as well as the MAMBO band at 1.2~mm have
been marked in gray, assuming this source to be at $z=2$. The emission
seen at 1.2~mm will be completely dominated by the cool dust, while
the emission at 24~\mum\, will be completely dominated by the warm
dust of the torus. The 70~\mum\, band is expected to be dominated by
the torus, with some contribution from the cooler dust, while the
situation for the emission observed at 160~\mum\, is expected to be
reverse: the cool dust dominates with some contribution from the
torus.

This is only an illustration, and the relative importance of each
component will vary depending on the actual temperature of the cool
dust, the actual SED of the torus and their relative
luminosities. However, Figure~\ref{fig:illu} does show that when a
powerful AGN is present, the data at 24 and 70~\mum\, and the
mid-infrared spectra will not provide very powerful constraints on the
cool-dust emission. It is the data at 1.2~mm that will provide the
best measurement of this emission. The data at 160~\mum\, should
provide useful constraints but for most sources in our sample it is
too shallow to provide detections. Data at 350 or 850~\mum\, is
extremely useful but is only available for 3 sources.

Hence, we will determine the far-infrared luminosities based on the
flux densities at 1.2~mm alone for most of the sources in this
sample. For sources where the data at 160, 350 or 850~\mum\, provide a
useful constrain, we will make use of this data too. However, the
24~\mum\, data will not be used to constrain the cool dust emission
(and in most cases, neither will the 70~\mum\, data).

\subsection{Flux densities at 1.2~mm}\label{sec:fluxes}

\begin{figure} 
\plotone{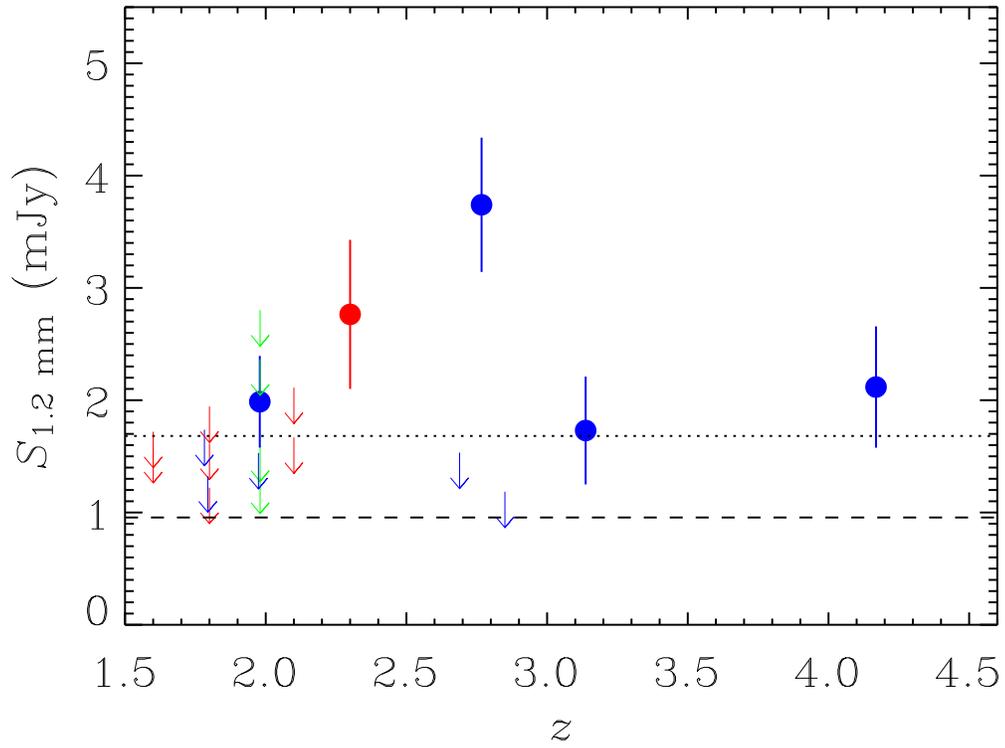}
\caption{Flux density, $S_{1.2~\rm mm}$,  versus redshift $z$ for the sample of
  obscured quasars. Sources marked in blue have redshifts from optical
  spectroscopy, those marked in red have redshifts only from their
  mid-infrared spectra. All limits are 3$\sigma$.  The four green
  sources have no spectroscopic redshifts and have been placed at
  $z=1.98$, the median value for sources with a spectroscopic
  redshift. At $z=2$ there are 4 green sources (one detection and
  three upper limits) and 2 blue sources (one detection and one upper
  limit).  The dotted line represents the approximate flux density
  limit of the MAMBO observations: $\sim$1.65~mJy.  The dashed line
  shows the value of $\langle S_{1.2~\rm mm} \rangle$, the mean flux density
  for all 21 sources.}
\label{fig:s_z} 
\end{figure}

The MAMBO observations yield detections at the $\geq$3$\sigma$ level
for 5 out of 21 sources observed (24\%).  The results are summarised
in Table~\ref{tab:obs} and Figure~\ref{fig:s_z} (where sources with no
spectroscopic redshift were placed at $z=1.98$, the median for sources
with a spectroscopic redshift). The mean noise achieved is 0.56~mJy
beam$^{-1}$, the median is 0.54~mJy beam$^{-1}$. Hence, for the non
detections, typical 3$\sigma$ limits $\sim$1.65~mJy are obtained.

\begin{figure} 
\plotone{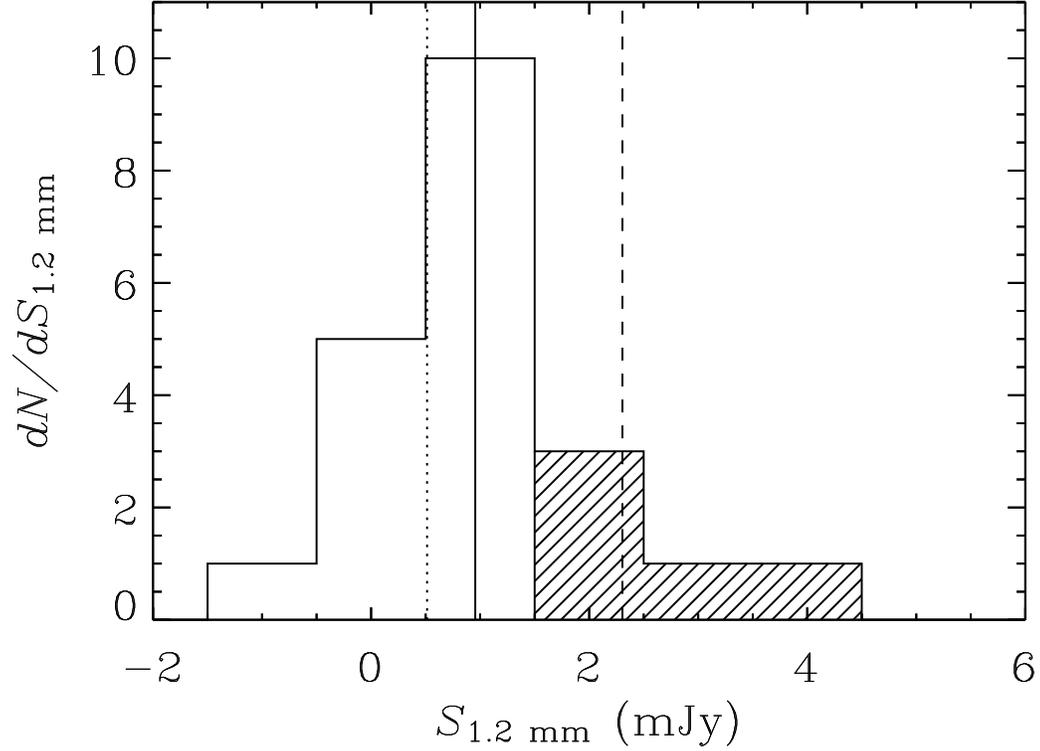}
\caption{Distribution of observed flux densities. The empty histogram
  represents the measured flux densities for all sources, the shaded
  histogram for detections.  The distribution is clearly centred on a
  positive value, and skewed towards the high-flux density end. The solid vertical line shows the stacked mean of all sources, the dashed line the stacked mean for detections, and the dotted line the stacked mean for non-detections.}
\label{fig:dbn_s} 
\end{figure}

With 76\% of the sources undetected, the median flux density does not
provide a useful characteristic number for the sample. A more
meaningful number is the mean flux density at 1.2~mm for the entire
sample.  The flux densities of all sources were therefore combined
using a weighted mean:

\begin{equation}
\langle S_{1.2~\rm mm} \rangle = { \left({\displaystyle\sum_{i} w_{i} S_{1.2~{\rm mm}~i}}\right)  \over
   \left( {{\displaystyle\sum_{i} w_{i}}}\right) } \pm \left( {{\displaystyle\sum_{i} w_{i}}}\right)^{-1/2}
\end{equation}

\noindent where $w_{i} = 1/\sigma_{i}^{2}$. This yielded a statistical
detection, with $\langle S_{1.2~\rm mm} \rangle = 0.96\pm0.11$ mJy. A
potential worry is that a few bright sources might be causing the
statistical detection, while the rest of the sources are intrinsically
faint. Stacking only the non-detections still yields a detection at
the 3.9$\sigma$ level: $\langle S_{1.2~\rm mm} \rangle = 0.51\pm0.13$
mJy. Stacking all the detections leads to a mean flux density of
$\langle S_{1.2~\rm mm} \rangle = 2.30\pm0.23$ mJy.
Figure~\ref{fig:dbn_s} shows the distribution of flux densities (from
Table~\ref{tab:obs}), which is centred on a positive value and skewed
towards the higher flux densities. This figure supports the statement
that the typical flux density of an object from this sample is
$\sim$0.5-1.0~mJy.  If the actual flux densities of the sources were
so low that the distribution were dominated by the noise,
Figure~\ref{fig:dbn_s} would be centred on 0, with a symmetric
characteristic width $\sim$0.55~mJy, and only a slight skew towards
the high fluxes (the detections with \snu$\grtsim$1.65~mJy).

We now consider whether the detection at 1.2~mm is related to the
presence or lack of narrow lines in the optical spectra.  Of the 5
detections at 1.2~mm, 4 show narrow lines (80\%) and only 1 (20\%) has
a blank optical spectrum (AMS19). Alternatively, of the 9 objects with
narrow lines, 4 have detections at 1.2~mm (44\%), while of the 12 with
no narrow lines, only 1 is detected (8\%).  Due to the small numbers,
no trend is significant, but at first sight the data hints that
sources with narrow lines are more likely to be detected at
1.2~mm. 

Looking at the stacked flux densities for narrow-line and for blank
objects (no narrow lines) suggests the dust content to be similar: the
weighted mean for narrow-line sources is 1.09$\pm$0.16~mJy, while for
blank sources it is 0.81$\pm$0.16~mJy, so both are statistically
detected.  Although at first sight the difference appears significant,
the narrow-line objects include AMS12 with a flux of 3.7~mJy which is
affecting the mean: the mean of the other 8 narrow line objects
excluding AMS12 is 0.76~mJy, very similar to that of the objects with
no narrow lines. Hence, the mean mm flux densities for both narrow
line and blank objects appear similar.

\subsection{Far-infrared luminosities}\label{sec:firlum}

Given a flux density, the infrared luminosity of a gray body is
given by:

\begin{eqnarray}\label{eq:graybody}
L_{\rm FIR} = \nonumber\\
{{\rm 15} \over \pi^{5}}{4\pi d_{\rm L}^{2} \over (1+z)}\left({k_{\rm B}\over h_{\rm
      P} \nu_{\rm rf} }\right)^{\beta}
{ \Gamma({\rm 4}+\beta)\zeta({\rm 4}+\beta) \sigma_{\rm SB} T^{{\rm 4}+\beta}
  S_{1.2~\rm mm} \over  B_{\nu_{\rm rf}}(T, \nu_{\rm rf}) }  
\end{eqnarray}

\noindent where $h_{\rm P}$ is Planck's constant, $k_{\rm B}$ is
Boltzmann's constant, $\sigma_{\rm SB}$ is the Stefan-Boltzmann
constant, and $d_{\rm L}$ is the luminosity distance at redshift
$z$. The terms $\nu$ and $\nu_{\rm rf}$ refer to observed and
rest-frame frequency respectively, $\nu_{\rm rf}=(1+z)\nu$. The term
$B_{\nu_{\rm rf}}(T, \nu_{\rm rf})$ is the Planck radiation function:

\begin{equation}
B_{\nu_{\rm rf}}(T, \nu_{\rm rf}) \equiv {2 h_{P} \nu_{\rm rf}^{3} \over c^{2} }{{\rm
    1} \over \left(e^{h_{\rm P} \nu_{\rm rf} \over k_{\rm B} T}-1\right)}
\end{equation}

\begin{figure*} 
\plottwo{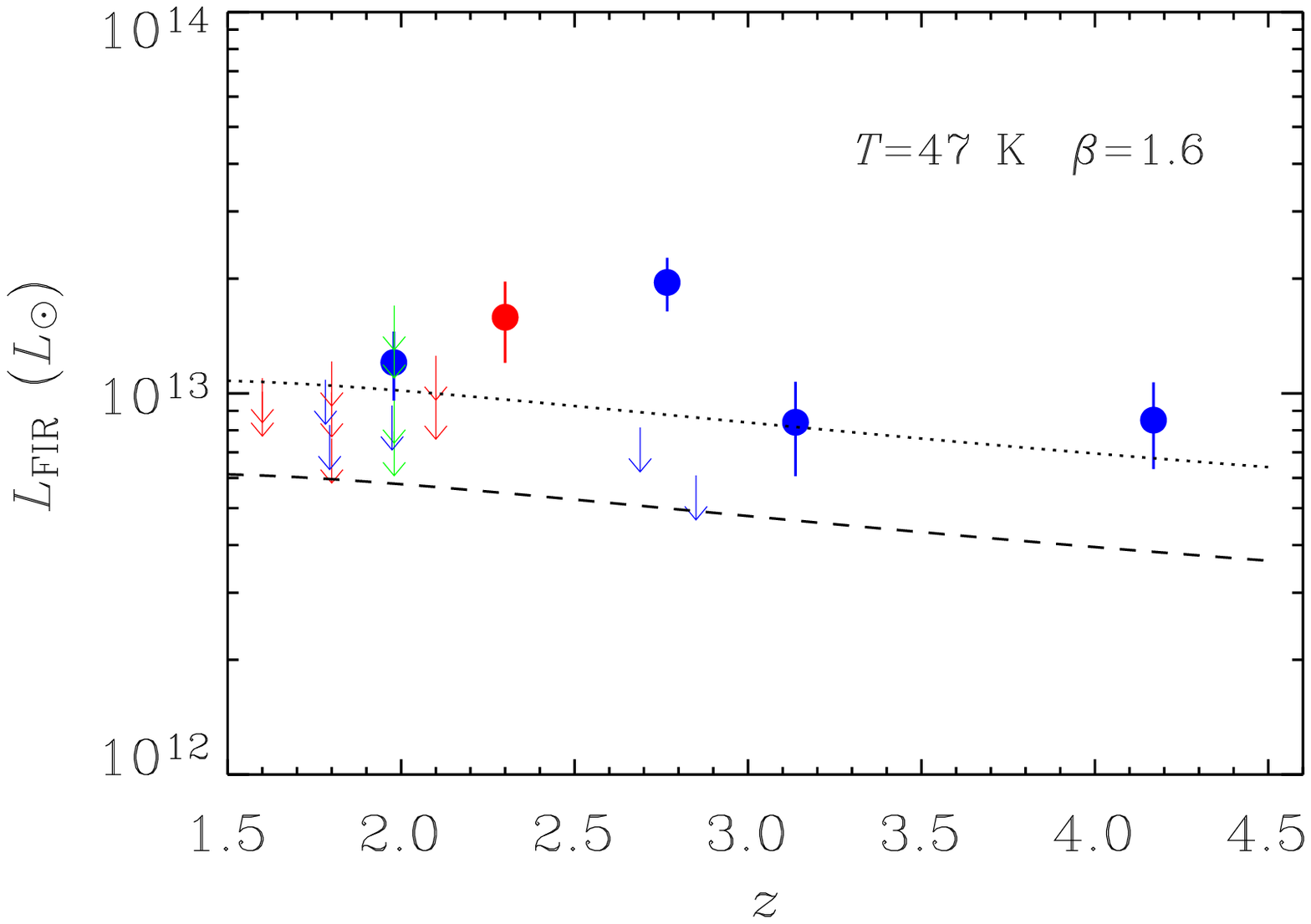}{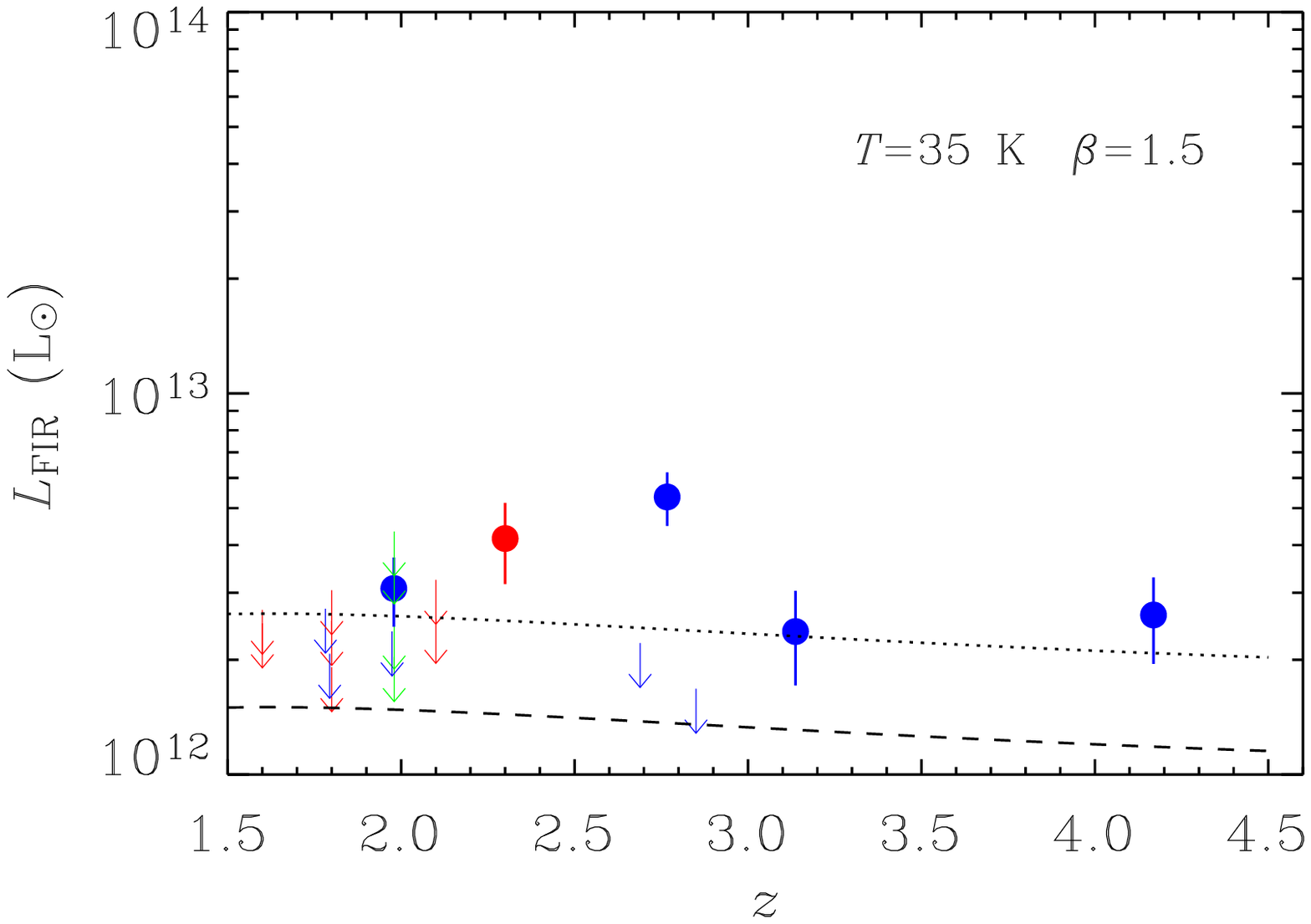}
\caption{ Far-infrared luminosity, \lfir, versus redshift $z$, for the two
  fiducial gray bodies. Symbols are the same as in Figure~\ref{fig:s_z}. The dotted
  line represents the value of \lfir\, corresponding to the approximate flux
  limit, while the horizontal dashed line represents the far-infrared luminosity corresponding to $\langle S_{1.2~\rm mm} \rangle$ at each redshift. }
\label{fig:lir}
\end{figure*}

Figure~\ref{fig:lir} and Table~\ref{tab:infer} show the inferred
values of \lfir\, assuming two different gray bodies, one with
$T=47$~K and $\beta=1.6$ \citep[representative of unobscured quasars
at high redshift, see][]{2006ApJ...642..694B}, the other with $T=35$~K
and $\beta=1.5$ \citep[representative of submillimetre-selected
galaxies, SMGs, see][]{2006ApJ...650..592K}.  The value of $\langle
S_{1.2~\rm mm} \rangle$, also assuming $z=1.98$, corresponds to
$\langle L_{\rm FIR}\rangle =$6.3$\pm$0.7$\times10^{12}$~\lsol\,
($T=47$~K and $\beta=1.6$) or 1.6$\pm$0.2$\times10^{12}$~\lsol\, (for
$T=35$~K and $\beta=1.5$). Hence, the obscured quasars are typically
ultra-luminous in the far-infrared, with a characteristic luminosity
of $\sim4\times10^{12}$~\lsol\, (the mean of both values).

Despite the range in spectroscopic redshifts (1.6 $\leq z \leq $4.2),
the relatively flat selection function at 1.2~mm means that the
detection of sources is primarily determined by their far-infrared
luminosities, and is only weakly dependent on redshift (see
Figure~\ref{fig:lir}). In fact, for a given flux density, the
corresponding far-infrared luminosity is actually smaller at high
redshift. Hence intrinsically less luminous objects are easier to
detect at higher redshift. This probably explains why in our sample,
the higher redshift sources ($z\grtsim$2) have a slightly higher
detection rate than the lower redshift sources ($z\lesssim$2).

The luminosity arising in the far-infrared is a considerable fraction
of the AGN bolometric luminosity (\lbol, between infrared and X-ray
energies, so between $10^{12}$ and $10^{19}$ Hz) which is typically
$2\times10^{13}$~\lsol\, for our sample (see Section~\ref{sec:broad}).
Hence, \lfir\, is typically $\sim$0.1-0.3\lbol\, depending on the
assumed gray body. This is slightly higher than what is found in
low-redshift unobscured quasars, where \lfir$\lesssim$0.1\lbol\,
\citep[e.g.][ and see also
Section~\ref{sec:comp}]{1989ApJ...347...29S,1994ApJS...95....1E,2006ApJ...649...79S}. For
SMGs, \lfir$\approx$\lbol\, since the far-infrared luminosity
consitutes the bulk of the bolometric emission
\citep[e.g.][]{2004ApJS..154..130E}.

In addition to the data at 1.2~mm, we have access to data at 70 and
160~\mum\, for all 21 sources \citep[from ][]{2006AJ....131..250F},
data at 350 for 2 sources (AMS13 and AMS16, see Section~\ref{sec:sharc}) and data at
850~\mum\, for one source \citep[AMS11, data from ][]{2004ApJS..154..137F}. In
most of the cases, the two fiducial gray bodies provide an acceptable
fit (see Figure~5). For the sources where neither the $T=47$~K and
$\beta=1.6$ or the $T=35$~K and $\beta=1.5$ gray bodies are
appropriate, we use the data to find better fits. We only fit sources
that have been detected at 1.2~mm.

For AMS08, Figure~5 suggests that the 70~\mum\, data are dominated by
the cool dust component and not by the warm torus, so we use the data
at 70 and 160~\mum, as well as 1.2~mm, and allow $\beta$, $T$ and
\lfir\, to vary. For AMS12 and AMS19, the 70~\mum\, data do not
provide tight constraints, so $\beta$ is kept fixed at a value of 1.5,
and only $T$ and \lfir\, are allowed to vary. The results are
summarised in Table~\ref{tab:gb_fit} and overplotted in Figure~5.

\subsection{Estimated star-formation rates}

Under the assumption that the far-infrared luminosity arises solely
from dust heated by young stars, and not by the AGN itself,
star-formation rates (SFRs) can be estimated from \lfir. Since the
dust is primarily heated by young stars, the conversion to total SFR
involves assuming an initial mass function (IMF). We use the
conversion from \citet{1998ARA&A..36..189K},

\begin{equation}
  \left( {{\rm SFR} \over {\rm M}_{\odot}~{\rm yr}^{-1} } \right) = 1.7\times10^{-10}
  \left( {L_{\rm IR} \over {\rm L}_{\odot} } \right)
\end{equation}

\noindent where \lir\, refers to the total infrared luminosity. This
conversion assumes a \citet{1955ApJ...121..161S} IMF. Given that at
mid-infrared wavelengths our sources are dominated by emission from
warm dust ($T>$100~K) heated by the AGN and not by stars
\citep[see][]{2008ApJ...674..676M}, the mid-infrared luminosities due
to star-formation cannot be easily derived. We therefore approximate
the total infrared luminosity to the far-infrared luminosity,
\lir$\approx$\lfir.  \lfir\, is determined by integrating analytically
the fiducial gray-bodies at all wavelengths, so there is no wavelength
at which the integration of the gray body is cut off. However, any
other components which would contribute to \lir\, and are not part of
the gray body (e.g. PAHs or mid-infrared continuum) are neglected.
Table~\ref{tab:infer} summarises the results.

For the sources detected at 1.2~mm, the inferred values of \lfir\, and
SFRs are in the range 300-3000 \msolyr, comparable to those of SMGs
\citep[e.g.][]{1998Natur.394..248B,1998Natur.394..241H,2004ApJ...611..732C,2005MNRAS.359.1165G}. The
SFR corresponding to $\langle L_{\rm FIR}
\rangle\sim4\times10^{12}$~\lsol\, suggests that the characteristic
SFR for the sample is $\sim$700~\msolyr.

\citet{2008ApJ...674..676M} detected polycyclic aromatic hydrocarbon
(PAH) emission in three of the sources in this sample (see
Table~\ref{tab:obs}), and found the luminosity of the detected
7.7~\mum\, PAHs to be comparable to the typical luminosities of PAHs
found in the sample of SMGs studied by
\citet{2007ApJ...660.1060V}. The far-infrared luminosities inferred
here are slightly lower but still comparable to those of SMGs.

If a significant fraction of \lfir\, is actually due to AGN-heated
dust, these values for the SFRs will be overestimated. This could
happen if dust at $\grtsim$kpc scales is being heated by the
\lbol$\sim10^{13}$~\lsol\, quasar \citep[see e.g.][ and
Section~\ref{sec:scale}]{1987ApJ...320..537B,1989ApJ...347...29S,2004A&A...421..129S}.

In addition to the infrared data, radio observations are also
available, which could in principle yield further constraints on the
SFRs. The multi-frequency and high-resolution radio data, however,
suggest the radio emission is completely AGN-dominated, and therefore
does not provide information on the SFRs \citep[see
][]{2006MNRAS.373L..80M,2009arXiv0905.1605K}.

It is not possible to discriminate whether the far-infrared luminosity
arises from AGN- or star-formation-heated dust: PAHs were only
detected in 3 objects in our sample yet the upper limits obtained for
the other sources correspond to very high PAH luminosities \citep[and
hence are not very stringent limits, see ][]{2008ApJ...674..676M}.
The SFRs inferred for our sample are extremely high but
physically-possible.

\subsection{Cool-dust masses}

The mass of cool dust responsible for the far-infrared emission can be
estimated as:

\begin{equation}
M_{\rm d} = { d_{\rm L}^{2} S_{1.2~\rm mm}  \over   (1+z) \kappa(\nu_{\rm
    rf})B_{\nu_{\rm rf}}(T, \nu_{\rm rf}) },
\end{equation}

\noindent where $\kappa$ is the mass absorption coefficent and the
main source of uncertainty, so we will consider two different fiducial
values to illustrate the estimated values of \mdust.

We follow \citet{2006ApJ...642..694B} and assume $\kappa=$0.04~m$^{2}$
kg$^{-1}$ at 1.2~mm \citep[within the values quoted by
][]{2004A&A...425..109A}. The values for individual objects are listed
in Table~\ref{tab:infer}.  Using $\langle S_{1.2~\rm mm} \rangle =
0.96\pm0.11$ and again assuming $z=1.98$ as the typical redshift, we
estimate the characteristic dust mass to be $\langle M_{\rm d}
\rangle$$=2.9\times10^{8}$~\msol\, and 5.1$\times10^{8}$~\msol\, (for $T=47$~K,
$\beta=1.6$ and $T=35$~K, $\beta=1.5$ respectively). If instead we had
followed \citet{2003MNRAS.341..589D} in assuming a value of
$\kappa=$2.64~m$^{2}$ kg$^{-1}$ at 125~\mum, the resulting estimates
of $\langle M_{\rm d} \rangle$ would have been $=1.6\times10^{8}$~\msol\, and
2.3$\times10^{8}$~\msol (for $T=47$~K, $\beta=1.6$ and for $T=35$~K,
$\beta=1.5$), which makes little difference to our arguments.  In any
case, the values of \mdust\, quoted should be taken as indicative
only, and we take $3\times10^{8}$~\msol\, as the fiducial cool-dust
mass for this sample (the mean of the four values quoted above).

\subsection{Characteristic scale of cool dust}\label{sec:scale}

We now estimate the characteristic distance of the dust under two assumptions:
that it is heated by young stars, and that it is heated by the AGN. 

We note that if the dust is heated by stars, then the inferred SFRs
are similar to those of SMGs. Interferometric observations of SMGs
suggest a characteristic scale of $\sim$2~kpc
\citep{2005MNRAS.359.1165G,2006ApJ...640..228T,2008ApJ...688...59Y},
so if the cool dust around our obscured quasars is heated by stars,
the scale of the dust is expected to be similar to this value.

To estimate the scale of AGN-heated dust, we equate the grain rate absorption
of energy from ultraviolet photons to the radiated energy \citep[see e.g.][
for a derivation]{1987ApJ...320..537B}:

\begin{equation}
{ L_{\rm uv}e^{-\tau_{\rm uv}} \over 4\pi r^{2}}  = 8\pi   \left({4 \kappa_{0} a
  \rho \over 3 \nu_{\rm 0}^{\beta}}\right) \left({h_{\rm P} \over
  { c^{2}}}\right) \left({ k_{\rm B} T \over h}\right)^{{\rm 4} + \beta} \Gamma({\rm 4}+\beta)\zeta({\rm 4}+\beta)
\end{equation}

\noindent where $L_{\rm uv}$ is the ultraviolet luminosity of the
central source, and $\tau_{\rm uv}$ is the opacity at ultraviolet
wavelengths.  The symbols $a$ and $\rho$ stand for the grain size and
material density, and $Q(a,\nu)={4 \kappa_{0} a \rho \over 3 }\left({
    \nu \over \nu_{0}} \right)^{\beta}$ is the emissivity. Following
\citet{2004A&A...425..109A} we assume $a=$0.1~\mum\, and
$\rho=$3000~kg m$^{-3}$ as fiducial values, which yields:

\begin{equation}
\left({ r \over {\rm kpc}}\right) =  16.3 \left({ T \over 40~{\rm
      K}}\right)^{-\left({ 4+\beta\over 2}\right)}\left({ L_{\rm
      uv}e^{-\tau_{\rm uv}} \over 5\times10^{12}~{\rm L}_{\odot} }\right)^{1\over2}
\label{eq:rad}
\end{equation}

The typical extinction-corrected bolometric luminosity of our sample
of quasars is $2\times10^{13}$~\lsol\, (see
Section~\ref{sec:broad}). Assuming $L_{\rm uv}\sim 0.25$~\lbol\,
\citep[e.g. ][]{1994ApJS...95....1E} the typical $L_{\rm uv}$ is about
5$\times10^{12}$~\lsol.  If we assume the ultraviolet photons travel
unhindered up to the characteristic radius of the dust, so that
$\tau_{\rm uv}=0$, and assume $T=47$~K and $\beta=1.6$, then the dust
responsible for the 1.2~mm emission is at a distance from the AGN of
$\sim$10~kpc (27~kpc for $T=35$~K and $\beta=1.5$).

Given that $\tau_{\rm uv}$ is likely to be $>0$ at kpc scales, since
the cool dust itself can absorb the uv photons from the AGN and shield
the dust on larger scales (see Section~\ref{sec:host}), these scales
should be taken as upper limits. Hence the cool dust is expected to be
distributed on scales $\lesssim$10~kpc.

\subsection{Host obscuration?}\label{sec:host}

The characteristic cool-dust mass derived from our observations can also be
used to estimate how much obscuration the host galaxy can cause.
We are considering the cool dust only ($T\lesssim$50~K), which emits in the
far-infrared. We are interested in the amount of extinction it can cause at
optical and mid-infrared wavelengths, where the emission from the cool dust is
negligible. Hence, a simple screen approximation neglecting the emission from
the obscuring dust is appropriate, so that a full radiative-transfer
calculation can be avoided.

 In order to do this, we  use the dust
extinction models of \citet{1992ApJ...395..130P} for the Milky Way
(MW) and Small Magellanic Cloud (SMC). These laws, however, correspond
to $\beta=2.0$ for $\lambda\grtsim$200~\mum, so extrapolating directly
from $\kappa$ at 1.2~mm to the visual band (0.55~\mum) using these
extinction models would not be consistent with the values of
$\beta=1.5-1.6$ we have assumed. 

We therefore estimate the opacity at 200~\mum, given a
typical dust mass of 3$\times10^{8}$~\msol\, and assuming
$\beta=1.6$. Assuming a spherically-symmetric smooth
distribution of dust with a characteristic radius $r$ and volume
$V={4\pi r^{3} \over 3}$, the column density and hence the oppacity
can be estimated:

\begin{equation}
\tau_{200~\mu \rm m}=  r {M_{\rm d}\over V} \kappa_{200~\mu \rm m} .
\label{eq:tau_m}
\end{equation}

\noindent Then, using an extinction law, the corresponding extinction
in the visual band, \av, can be derived (in units of magnitudes).
Assuming a characteristic radius of 10~kpc, $\kappa=$0.04~m$^{2}$
kg$^{-1}$ at 1.2~mm, we estimate $\tau_{200~\mu \rm
  m}=1.2\times10^{-3}$, which requires \av$=$4 for the MW model and
1.5 for SMC. If instead we assume $\kappa=$2.64~m$^{2}$ kg$^{-1}$ at
125~\mum\, and the same radius, the opacity increases to
$\tau_{200~\mu \rm m}=2\times10^{-3}$ and requires \av$=$7.5 (MW) or
2.5 (SMC).  

Assuming instead a radius of 2~kpc, characteristic of the gas and dust
in SMGs
\citep{2005MNRAS.359.1165G,2006ApJ...640..228T,2008ApJ...688...59Y},
the inferred value of $\tau_{200~\mu \rm m}$ is $0.03-0.05$,
corresponding to values of \av$=$35-65 (SMC dust and depending on
$\kappa$) or even $\geq100$ (MW). A radius of 20~kpc leads instead to
low values of \av\, $\sim$0.3-1.5 (using both MW and SMC extinction
curves).

Given that the condition for a quasar to be considered obscured is
\av$\geq$5 \citep[see ][]{1999MNRAS.306..828S}, and comparing to the
estimates of \av\, from Section~\ref{sec:broad} shown in
Table~\ref{tab:infer}, the main conclusion from this estimate is that
there is probably enough cool dust in the host galaxies of many of
these sources to obscure the quasars, regardless of the orientation of
the torus with our line of sight.

There are, however, examples of high-redshift unobscured quasars with
similar (sub)millimeter fluxes and hence similarly large dust masses
\citep[e.g.][]{2003A&A...398..857O,2006ApJ...642..694B}: the presence
of large dust masses does not guarantee heavy obscuration. This is
also the case for our sample (see Section~\ref{sec:fluxes}).  A likely
explanation for this is that the dust in the host galaxy has a very
irregular or clumpy distribution, so that only some lines of sight
will be obscured.  Rather than orientation-dependent obscuration, as
is the case for orientation caused by the torus, the obscuration (or
lack of) by dust in the host galaxy is likely to be dependent on luck,
although samples of obscured quasars are expected to be biased towards
``bad luck'' in the form of blocked lines of sight.

\section{Mid-infrared SED fitting}\label{sec:broad}

In this section, we make use of broad-band data to characterise the
SEDs of our sources. We use the IRAC and MIPS-24~\mum\, data to obtain
an estimate of $L_{\rm bol}$ and $A_{\rm V}$, from broad-band SED
fitting.  The approximation we use is that an obscured quasar SED can
be parametrised as an unobscured quasar SED, with bolometric
luminosity \lbol, with a foreground screen of dust causing an amount
\av\, of extinction.

The values for \lbol\, and \av\, obtained in this way are only a
parametrisation of the obscured SED. The \av\, is only the apparent
extinction, assuming the unobscured SED to be the underlying SED. This is an
accurate description at optical and ultraviolet wavelengths: the intrinsic
illuminating source is the accretion disk around the SMBH, which emits
thermally with a range of temperatures $\sim10^{3}-10^{5}$~K, and the dust
will simply absorb this and re-emit it at longer wavelenghts. 

The mid-infrared emission from unobscured quasars, however, is already
reprocessed light from warm dust ($T\sim$100-2000~K). The dust
responsible for this emission must be directly illuminated by the
accretion disk, and must be relatively close to the source of
ultraviolet photons: it is likely to originate in the inner region of
the torus. For example, following Section~\ref{sec:scale} and using
Equation~\ref{eq:rad}, $T\sim$300~K corresponds to $\sim$50~pc. For an
unobscured quasar, the mid-infrared emission will travel almost
unhindered until it reaches the observer, while in the case of an
obscured quasar, it still has to go through a considerable amount of
dust. Most of this dust will be further away (and thus colder) and
might not be directly illuminated by the central engine anyway. Thus,
the dust obscuring the mid-infrared emission is once again unlikely to
emit  very significantly at the same wavelengths. The
approximation of the obscured SED as an unobscured SED with a
foreground screen of dust is still defendable at mid-infrared
wavelengths, although it will be a poorer approximation than in the
optical regime. In any case, it is  a simplification of the situation.

This approximation breaks down completely in the far-infrared, so we
do not attempt to model the far-infrared SEDs in this way. Instead,
the far-infrared component of the SEDs was modelled as a gray body, as
described in Section~\ref{sec:firlum}.

To model the SEDs in this way, we use the 24-$\mu$m data from the
catalogue of \citet{2006AJ....131.2859F}, and the IRAC data from
\citet{2005ApJS..161...41L} as well as deeper observations around 4
objects from our program PID 20705 (PI M. Lacy, the objects are AMS05,
AMS12, AMS16 and AMS17). We use the \citet{1994ApJS...95....1E} median
quasar SED.  The quasar SED is then obscured by a screen of MW-type
dust, using the models of \citet{1992ApJ...395..130P}. It is this dust
law that will determine the depth of the silicate feature (around
9.7~\mum) in the models.  An elliptical galaxy SED \citep[from
][]{1980ApJS...43..393C} is also used, to represent the old stellar
population, and this stellar light is not subject to any obscuration.
The quasar bolometric luminosity, \av, and the luminosity of the
galaxy are all allowed to vary, and the best fit is kept\footnote{This is a essentially the same routine as used in
  \citet{2007MNRAS.379L...6M}, except the redshift is not varied and
  their ``blue'' component is not included: we use the spectroscopic
  redshifts, or assume $z=1.98$ when no redshift is available, while
  the blue component is irrelevant in the wavelength range discussed
  here (3.6-24~\mum). In addition, we are not selecting models here,
  only fitting parameters, so for each object we simply keep the
  values with the lowest $\chi^{2}$. Finally, due to the narrower
  wavelength range here, we allow the errors in photometry to be
  $<10$\% \citep[c.f.][]{2007MNRAS.379L...6M}.}.

For sources with many limits (namely AMS08, AMS10, AMS15, AMS18), the
marginalised posterior distribution function (PDF) was flat for a
range of values of \av\, (i.e. the same $\chi^{2}$ was obtained for
all values of \av\, above a certain value). Given that low-luminosity
objects are always more common than high-luminosity objects, and given
that the inferred value of \lbol\, will correlate with \av\footnote{In
  order to match the observed flux density of a given source at a given redshift,
  increasing the \av\, will cause the inferred value of \lbol\, to
  increase.}, we chose the lowest value of \av\, (to the closest
multiple of 5) and associated \lbol\, within the flat region of the
posterior PDF. The results are summarised in Table~\ref{tab:infer} and
all SEDs are presented in Figure~5. The best-fit SEDs are shown (blue
solid line), as well as the individual components (quasar and
elliptical galaxy, dashed lines). The broad-band data at 3.6, 4.5,
5.8, 8.0, 24, 70 and 160~\mum\, as well as the IRS spectra are
superimposed.  On the far-infrared end, the 1.2~mm point, and the two
gray bodies are also shown.  For the sources AMS13 and AMS16, the
350~\mum\, data from SHARC are also shown, and for AMS11 a data point
at 850~\mum\, from \citet{2004ApJS..154..137F} is also included. For
source AMS05, additional data at 1.2, 1.7 and 2.2~\mum\, are also
shown \citep[from][]{2009MNRAS.393..309S}.

The IRS spectra were not used in the fit, but used to provide instead an
independent quality check on the fits.  In most sources the best fit to the
broad-band SED is in reasonable agreement with the IRS spectrum. In a few
cases the IRS spectra show deeper silicate absorption features than those
resulting from the best-fit \av\, (e.g. AMS06, AMS19), in other cases the
agreement is very good (e.g. AMS05, AMS13). The overall agreement
between the IRS spectra and the best-fit broad-band SEDs is reasonably good.

Two sources have very low values of \av\, from the broad-band SED
fitting: AMS02 (\av$=$0.5) and AMS20 (\av$=$0.0). This would suggest
both objects are unobscured quasars, and should not make it into the
sample. However, both objects showed blank optical spectra, despite
long integrations \citep[see
][]{2005Natur.436..666M,2006MNRAS.370.1479M}. The IRS spectrum of
AMS02 shows the silicate feature in absorption while the spectrum of
AMS20 shows a hint of silicate absorption \citep[although in ][ it was
not considered secure due to the low signal-to-noise
ratio]{2008ApJ...674..676M}. Both sources have 3.6~\mum\, flux
densities at the edge of the selection criteria, and AMS02 also has a
24~\mum\, density at the edge of the criteria \citep[see Table~1 and
Figure~1 of ][]{2005Natur.436..666M}.  AMS02 is certainly a heavily
obscured source, the flat SED might be due to a relatively bright host
galaxy compared to the AGN emission. AMS20 is less clear, the flat SED
could also be due to the host galaxy contribution, and in any case the
blank optical spectrum shows it is certainly not an unobscured quasar.
It  might be a reddened quasar (0$<$\av$\lesssim$5), or an obscured quasar with a peculiar mid-infrared SED.

For our sample, we find no significant correlations between \lfir\,
and \lbol\, or \lfir\, and \av.

\section{A mean obscured quasar SED}\label{sec:mean}

We now derive a typical dust SED for our sample of obscured quasars
and compare it to clumpy models of AGN tori.

As mentioned earlier, due to the large number of upper limits the
median flux density at 1.2~mm is not a very useful quantity (since it
yields an upper limit), yet we find the weighted mean yields a
statistical detection. To be consistent, we will therefore consider
mean quantities for the mid-infrared too. We note that within our
sample there is no correlation between \lbol\, and \lfir, so we treat
the (rest-frame) near-infrared, mid-infrared and far-infrared
components independently. The mean SED of our
spectroscopically-confirmed objects is shown as a solid red line in
Figure~\ref{fig:mean_sed}.

We do warn about several caveats: The far-infrared,
mid-infrared and near-infrared components have been determined in
different ways. The far-infrared component is an analytic gray body to
the mean flux density at 1.2~mm, with an empirical quasar SED (from
the mean \lbol). It is still ill constrained, since only some objects
have been detected at 70 or 160~\mum. The mid-infrared component
results from stacking IRS spectra, and is probably the best determined
part, although it uses only sources with spectroscopic redshifts and
IRS spectra.  

For the entire SED, only sources with spectroscopic redshifts are
used.   For sources with no redshift from optical spectroscopy,
redshifts can only be determined if the silicate feature is deep
enough to be noticeable in noisy spectra. The mean mid-infrared SED is
therefore partially biased towards sources with deep absorption
features at 9.7~\mum. 

The near-infrared component uses the data from all sources, but
because these sources have a variety of redshifts, of values of \av,
and of host galaxy luminosities, it is really a mixture of
heavily-extinct AGN light and stellar light.

In addition, the sources in this sample probably include examples of
both torus- and host-obscured quasars, meaning that the mean SED will
represent a mixture of such objects.  Despite all these caveats, we
consider the mean SED to be a useful fiducial representation of
obscured quasars.

\subsection{Far-infrared component}

We parametrise the far-infrared component as being described by the
two fiducial gray bodies. The far-infrared luminosity is determined by
the weighted mean flux density, $\langle S_{1.2~\rm mm} \rangle =
0.96\pm0.11$ mJy at $z=1.98$.  Assuming the gray body with
$T=$47~K and $\beta=$1.6 results in $\langle L_{\rm FIR}\rangle =
6.3\times10^{12}$~\lsol, while assuming $T=$35~K and $\beta=$1.5, we
obtain $\langle L_{\rm FIR}\rangle = 1.6\times10^{12}$~\lsol.

\subsection{Mid-infrared component}

The mid-infrared component was obtained by shifting the IRS spectra to
rest-frame wavelengths, interpolating to a common wavelength grid and taking
the mean luminosity density of the 14 sources with spectroscopic redshifts and
mid-infrared spectra.  Different wavelength bins have a different total number
of sources contributing, and it was decided to cut below 4.5~\mum\, and above
12.5~\mum, where 3 or less sources were contributing.
Figure~\ref{fig:mean_sed} shows the mean IRS spectrum smoothed by a box-car
average of 0.7~\mum. 

The depth of the silicate absorption feature at 9.7~\mum, $\tau_{9.7}$
is usually measured by interpolating the continuum at 9.7~\mum\, using
the continuum bluewards and redwards. The difference in flux densities
between the interpolated continuum ($f_{\rm IC}$) and the observed
feature ($f_{\rm OF}$) then yields $\tau_{9.7}$:

\begin{equation}
\tau_{\rm 9.7} \equiv {\rm ln}\left({ f_{\rm IC} \over f_{\rm OF} }\right) 
\end{equation}

This is difficult to measure from the mean IRS spectrum, since there
are no data points redwards of the absorption feature, so no real
interpolation can be made. However, comparing the flux densities of the 9.7~\mum\, feature and the 8~\mum\, continuum in Figure~\ref{fig:mean_sed},
the depth of the feature is likely to be in the range $2.5 \lesssim
\tau_{9.7} \lesssim 3.0$. 

Figure~\ref{fig:mean_sed} also shows that this is deeper than what is
expected from the unobscured SED with the screen of dust (using the
mean \av: dashed line), which otherwise fits the region of
4.5-8.5~\mum\, quite well. We note that the mean \av\, from all
sources with spectroscopic redshifts (\av$=$32 is very similar to the
mean of the 14 sources used for the mid-infrared component, \av$=$34). This small difference leads to a practically
indistinguishable SED in the mid-infrared and cannot explain the
discrepancy in the depth of the silicate feature.

At this stage, the discrepancy with the silicate feature is only due
to the dust extinction model used: an extinction law with  deeper
silicate absorption (such as the small magellanic cloud dust model of Pei et
al. 1992) can plausibly reproduce the observed feature.

However, a typical silicate feature as deep as $\grtsim2.5$ suggests
some degree of geometrical obscuration caused by cold dust
\citep[see][for a discussion]{2007ApJ...654L..45L}. This lends further
support for some degree of host obscuration.

Finally, the mean spectrum from the IRS shows an absorption feature at
$\sim$7~\mum\, probably due to hydrogenated amorphous carbons \citep[see e.g.][]{2002A&A...385.1022S}.

\subsection{Near-infrared component}

For the near-infrared component, for each object the flux densities of the
IRAC bands were converted to luminosities at the rest-frame wavelength. They
were then binned in a grid ranging between 0.5 and 4.5~\mum. 
Looking at the SEDs in Figure~5 it seems that the upper limits
are generally close to the detections in adjacent bands, so we treat limits
like detections. In the bins with contributions from several sources, the mean
luminosity density was computed.  However, some of the bins have a
contribution from only one object and some from none (these last bins are not
shown in the plots). The solid red line in Figure~\ref{fig:mean_sed} shows the
mean photometry at rest-frame near-infrared, smoothed by a box-car average of
0.6~\mum.  The line is irregular, because of the vastly different
individual SEDs seen in Figure~5: in some objects, all IRAC bands
are likely to be dominated by the AGN (e.g.  AMS19), in others by the host
galaxy (e.g. AMS16). If instead of using limits as data, we do not use them at
all, the resulting SED is very similar.

This part of the SED should clearly be treated with caution. However, an
estimate of the luminosity of the typical host galaxy can be attempted.
Overplotted on Figure~\ref{fig:mean_sed} is a $z=0$ elliptical galaxy
from \citet{1980ApJS...43..393C}, normalised to be 12.6$\times$ brighter
than an $L^{\star}$ galaxy in the local K-band luminosity function
\citep{2001MNRAS.326..255C}. Assuming $\sim$2 magnitudes of passive evolution
between $z\sim2$ and $z=0$, it is the equivalent to a present day
2$L^{\star}$ galaxy \citep[this is partly by selection, as is
described in detail by ][]{2005Natur.436..666M,2006MNRAS.370.1479M}. We
can thus see that the luminosity of the near-infrared component of the typical
obscured quasar SED corresponds approximately to the progenitor of a
2$L^{\star}$ galaxy, although it might also have an important contribution
from AGN light (in which case the galaxy luminosity will be overestimated).

\subsection{Comparison to clumpy models of the torus}

It is of particular interest to compare the mean SED to clumpy models
of the AGN torus. This is shown in Figure~\ref{fig:mean_clumpy} (left
panel), where the mean SED is overplotted on the models of
\citet{2008ApJ...685..147N,2008ApJ...685..160N}.  The models cannot
reproduce simultaneously the relatively flat SED at wavelengths
shorter than 8~\mum, the sharp drop in the near-infrared and the deep
silicate absorption. The left panel of Figure~\ref{fig:mean_clumpy}
also shows that the models with deeper silicate features still struggle to
match the observed silicate depth ($\tau_{9.7}\grtsim2.5$) yet will
overpredict the far-infrared luminosity.

The right panel of Figure~\ref{fig:mean_clumpy} shows the two flattest
clumpy SEDs from the right panel, now with additional extinction of
\av$=$80 added due to foreground dust.

We see that the clumpy models, with the additional extinction and
re-emission from the cool dust can reproduce reasonably well the
observed SED. Since the dust is cold, it re-emits in the far-infrared,
meaning it does not ``fill-up'' the silicate absorption feature,
unlike dust at $\sim$300~K would \citep[again, see
][]{2007ApJ...654L..45L}.  To reproduce the deep silicate feature, a
value of \av$\sim$80 is required (see
Figure~\ref{fig:mean_clumpy}). 

The typical dust mass of $3\times10^{8}$~\msol\, can cause this
extinction of \av$=$80 provided that the dust is on scales
$<$2.3-3.0~kpc (from Equation~\ref{eq:tau_m} in Section~\ref{sec:host}
and assuming MW dust and the two values of $\kappa_{200~\mu\rm m}$).
This is in excellent agreement with the inferred scales of the cool
dust emission in SMGs \citep[typically
$\sim$2~kpc][]{2005MNRAS.359.1165G,2006ApJ...640..228T,2008ApJ...688...59Y}.

\citet{2008ApJ...675..960P} studied in detail the mid-infrared spectra
and mid- to near-infrared SEDs of 21 high-redshift heavily obscured
quasars, selected to be bright ($\geq$1~mJy) at 24~\mum, optically
faint (typically $R>24$) and with mid-infrared spectra characteristic
of heavily obscured AGN (see their Section~2 for more details). They
found that 9 out of 21 sources did not fit clumpy models well and
required an additional foreground screen of dust that did not re-emit
at mid- or near-infrared wavelengths.

Our findings for the mean SED are consistent with their results, and
in addition, we are able to estimate the \av\, in a consistent way
from \mdust, which was derived from our observations at 1.2~mm. The
resulting dust size required for \mdust\, to cause the \av\,
($\sim$2.3-3.0~kpc) is very close to the sizes estimated for the
far-infrared emission of SMGs.  Again, our results provide evidence
for obscuration due to dust on kpc scales, dust that has lower
temperatures and is distributed on larger scales than what is expected
for the dust of the torus.

\subsection{Comparison to local ULIRGs}\label{sec:ulirgs}

Figure~\ref{fig:mean_comp} shows the mid-infrared component of the
mean SED with the spectra of four local ultra-luminous infrared
galaxies overplotted (ULIRGs, with $L_{\rm IR}>10^{12}$
L$_{\odot}$). The spectra for these sources were obtained from
\citet{2007ApJ...656..148A} and \citet{2007ApJ...654L..49S}. These
ULIRGs represent one pure-AGN, IRAS~12514$+$1027 (hereafter
IRAS~12514), two AGN-starburst composites, Mrk~231 and Mrk~273, and
one starburst-dominated source, Arp~220. All spectra have been normalised to 
match our mean SED between 7.0 and 7.8~$\mu$m (around 4$\times10^{13}$~Hz). 

Compared to these sources, our mean SED is relatively flat, similarly
to Mrk~231 or IRAS~12514, yet the silicate absorption feature is
significantly deeper, more similar to that of Mrk~273 or even
Arp~220. Once again, we see the peculiar spectral shape of this mean
SED, where the silicate feature is suprisingly deep given the relative
flatness of the continuum. The most similar nearby ULIRG is
IRAS~12514, which is known to be an obscured quasar
\citep{2003MNRAS.338L..19W}, although this source has a shallower
feature at 9.7~$\mu$m.

\section{Results from  observations of CO}\label{sec:co}

\subsection{Line properties}

Two sources, AMS12 and AMS16, were observed with the IRAM PdBI to search for the
CO (3-2) or (4-3) rotational transitions,  tracers of molecular gas.

The CO (3-2) line in AMS12 is clearly detected, and is well described
by a gaussian with a peak flux of 2.14~mJy and a full-width
half-maximum (FWHM) of 275~\kms\, (see Figure~\ref{fig:coams12}). The
integrated line flux of this gaussian is $S_{\rm CO}\Delta \nu=$630$\pm$50 mJy~\kms,
where $\Delta \nu S_{\rm CO}$ is the velocity-integrated CO line
flux. This corresponds to a (frequency-integrated) luminosity of
$L_{\rm CO (3-2)}=$3.2$\times10^{7}$~\lsol, or a
brightness-temperature luminosity of $L'_{\rm CO
  (3-2)}=2.4\times10^{10}$ K~\kms~pc$^{2}$.

The central frequency of the line is at -16~\kms, corresponding to a
frequency shift of 4.9~MHz (so a central frequency of 91.801~GHz), so
the redshift of the CO line  is $z_{\rm CO}=2.7668$.  At our
signal to noise, the line shape is well described by a gaussian, there
is no double profile visible, which could suggest a disk. There is
also no hint of two separate lines which could suggest that the
molecular gas is undergoing a major merger.

In the case of AMS16, no line was detected (see
Figure~\ref{fig:coams16}). The full data cube was searched by looking
for line emission at various spectral resolutions, using a routine
provided by Roberto Neri, but no significant emission was detected at
the position of the source.

A signal at the 4.5~$\sigma$ level can be seen at 17~19~44 $+$58~47~00. If
real this would correspond to a line offset from the narrow-line redshift by
$\sim$600~\kms\, and with a full width at zero intensity of $\sim$850~\kms.
However, the position does not correspond to any source detectable in the
R-band, 3.6~$\mu$m or 1.4~GHz image. This position corresponds to a 17~arcsec
offset, or 116~kpc at the redshift of the source. We do not consider this
emission any further.

The lack of detection of the CO~(4-3) line can either be due to the line being
shifted outside of the observed bandwidth ($>\pm$1680 \kms\, shift), or the
line being within the bandwidth but too faint.  The redshift was determined
from three narrow lines, Ly~$\alpha$ (1216~\AA), N~IV (1240~\AA) and C~IV
(1549~\AA) with individual redshifts of $z=4.171$, 4.166 and 4.170
respectively (see Mart\'\i nez-Sansigre et al., 2006a). Thus the maximum
redshift offset of any individual line, with respect to the average $\langle z
\rangle=4.169$ is that of the N~IV line with $\Delta z=0.003$, which only
corresponds to a 175~\kms\, offset.

Given the bandwidth used, we consider the possibility that the
CO~(4-3) line falls outside of the observed range unlikely, so
detection of CO~(4-3) in AMS16 can be ruled out down to a 3$\sigma$
limit of $S_{\rm CO}\Delta \nu <$0.23 Jy \kms, assuming a
box-car line shape with full width to zero intensity of 400~\kms. From
these limits, we now estimate the limits on the CO (4-3) luminosities
and derived molecular gas properties. Figure~\ref{fig:coams16} shows
the spectrum extracted at the position of AMS16.

At $z=4.169$, $S_{\rm CO}\Delta \nu<$230 mJy \kms\, corresponds to a
frequency-integrated line luminosity of $L_{\rm CO
  (4-3)}<$3$\times10^{7}$~\lsol, or a brightness-temperature
luminosity of $L'_{\rm CO (4-3)}<1\times10^{10}$
K~\kms~pc$^{2}$. 

\subsection{Inferred physical properties}

If we assume the molecular gas responsible for the CO emission is
dense enough and warm enough for the (3-2) or (4-3) transitions to be
thermalised, then the brightness-temperature luminosities of these
high-J transitions are the same as that of the (1-0) transition:
$L'_{\rm CO (1-0)}=L'_{\rm CO (3-2)}$ and $L'_{\rm CO (1-0)}=L'_{\rm
  CO (4-3)}$.

Following \citet{1998ApJ...507..615D}, we can then
estimate an upper limit on the mass of molecular hydrogen, by using

\begin{equation}
\left({ M_{\rm H_{2}} \over {\rm M_{\odot}}}\right)=\alpha_{\rm CO}\left({ L'_{\rm CO (1-0)} \over {\rm L_{\odot}}}\right),
\end{equation}

\noindent where $\alpha_{\rm CO}=$0.8 (K \kms\, \msol)$^{-1}$ is the
conversion factor for local ULIRGs \citep{1998ApJ...507..615D}. Thus,
under the assumption that $L'_{\rm CO (1-0)}=L'_{\rm CO (3-2)}$, we
infer a total gas mass of $M_{\rm H_{2}}=1.9\times10^{10}$~\msol\, for
AMS12. Assuming that $L'_{\rm CO (1-0)}=L'_{\rm CO (4-3)}$, we can
derive a limit for the total H$_{2}$ mass in AMS16 of $M_{\rm
  H_{2}}<8\times10^{9}$~\msol.

The far-infared luminosity of AMS12 of $2.5\times10^{13}$~\lsol\, (see
Table~\ref{tab:gb_fit}), suggests a SFR of 4300~\msolyr. The ratio of
far to CO luminosity is log$_{10}[L_{\rm FIR}/L_{\rm CO}']=3.0$, the
gas-to-dust ratio, \mgas$/$\mdust, is 19, and the gas depletion
timescale is estimated to be $\tau_{\rm g}\sim$4~Myr.
 
For AMS16, we can only infer limits for these values: if we assume the
values from the $T=35$~K, $\beta=1.5$ gray body, we infer
log$_{10}[L_{\rm FIR}/L_{\rm CO}']\geq$2.4, \mgas$/$\mdust$\leq$8 and
$\tau_{\rm g}\leq$16~Myr. Assuming instead the $T=47$~K, $\beta=1.6$
gray body yields the following values: log$_{10}[L_{\rm
  FIR}/L_{\rm CO}']\geq$2.9, \mgas$/$\mdust$\leq$20 and $\tau_{\rm
  g}\leq$5~Myr. 

For both AMS12 and AMS16, the estimated values of log$_{10}[L_{\rm
  FIR}/L_{\rm CO}']$ are at the low end but consistent with the
observed values for other high-redshift sources with molecular gas
measurements \citep[e.g.][]{2005ARA&A..43..677S,2005MNRAS.359.1165G}.
The gas-to-dust ratios are much lower than the values $\sim$100-150
found for local spiral galaxies. The values of $\tau_{\rm g}$ are
ultimately derived from the quantity $L_{\rm FIR}/L_{\rm CO}'$ and are
thus also at the low end of the observed values. A possible
explanation for the high values of $L_{\rm FIR}/L_{\rm CO}'$ is that a
fraction of the far-infrared luminosity is due to AGN heated dust,
rather than by dust heated by young stars.

We warn that if the assumption that the (4-3) transition is
thermalised is inappropriate, $L'_{\rm CO (1-0)}>L'_{\rm CO (3-2)}$ or
$L'_{\rm CO (1-0)}>L'_{\rm CO (4-3)}$, we will have underestimated
\mgas, and hence $\tau_{\rm g}$ and the gas-to-dust ratio. For a
discussion about whether this assumption is appropriate, see
\citet{2001Natur.409...58P} and \citet{2006ApJ...650..604R}, although
for the high-redshift AGN studied in detail it seems to be a
reasonable assumption \citep[see
e.g.][]{2004A&A...423..441B,2005ApJ...621L...1K,2008A&A...491..173A}.

From the FWHM of the CO(3-2) line in AMS12, $\Delta\nu=$275~\kms,  and assuming a characteristic radius, we can estimate the dynamical mass given  \citep[see][]{2003ApJ...597L.113N}:

\begin{equation}
 \left({ M_{\rm dyn}~{\rm sin^{2}}i \over {\rm M}_{\odot}}\right) = 4\times10^{4}  \left({ \Delta\nu \over {\rm km~s^{-1}} }\right)^{2} \left({ r \over {\rm kpc} }\right)
\end{equation}

\noindent The CO is unresolved in the $\sim$5 arcsecond resolution
observations with the PdBI observations, yielding a size limit of
$r<$40~kpc. A better estimate for $r$ is 2~kpc
\citep{2006ApJ...640..228T}.  The inferred mass is then $M_{\rm dyn}
{\rm sin^{2}}i=6\times10^{9}$~\msol, and is obviously dependent on the
unknown inclination angle $i$.  The ratio of gas to dynamical mass is
then $M_{\rm H_{2}}/M_{\rm dyn}=3\times{\rm sin^{2}}i$. Thus the
estimated gas mass is larger than the estimated dynamical mass, unless
the inclination angle $i$ is $\leq35$. It seems likely that the gas
mass represents a high fraction of the total dynamical mass, as found
for other high-redshift galaxies observed in CO
\citep{2005MNRAS.359.1165G,2006ApJ...640..228T}.

\section{Comparison to other samples of $z\sim2$ quasars}\label{sec:comp}

Other samples of $z\sim2$ quasars have been observed at millimetre or
submillimetre wavelengths, allowing us to compare the far-infrared
luminosities inferred for the different samples. Far-infrared emission
being isotropic, the properties of obscured and unobscured quasars
should be similar if obscuration is only an orientation effect. To
test this, we consider here three other studies at approximately the
same redshift:

\begin{itemize}
\item The sample of \citet[][ hereafter O03]{2003A&A...398..857O},
  consisting of optically-selected unobscured quasars observed at
  1.2~mm. These sources are intrinsically very bright with $M_{\rm
    B}\lesssim-27$, corresponding to \lbol\, in the range
  $5\times10^{13}-5\times10^{14}$~\lsol, and at $ 1.94 \leq z \leq
  2.85$. They have a detection rate of 9 out of 26 sources (35\%).
\item The sample presented by \citet[][ P04]{2004ApJ...611L..85P},
  which consists of optically unobscured quasars known also to be
  X-ray unabsorbed.  These sources have values of
  \lbol$\sim10^{13}$~\lsol, similar to our sample, with $ 1.20 \leq z \leq
  2.35$. They were observed
  at 850~\mum, with a detection rate of 1 out of 20 (5\%). 
\item The optically unobscured but X-ray absorbed quasars from the
  sample of \citet[][ S05]{2005MNRAS.360..610S}, with similar values
  of \lbol\, to our sample, $ 1.01 \leq z \leq 2.80$. They were
  observed at 850~\mum\, with 6 out of 19 sources detected at the
  $\geq3\sigma$ level (32\%).
\end{itemize}

In all samples, the bolometric luminosities have been estimated
consistently using the \citet{1994ApJS...95....1E} SED, and correcting
for absorption at X-ray energies (S05) or dust-extinction at optical
wavelengths (this sample). The far-infrared luminosities have all been
computed assuming a gray body with $T=47$~K, $\beta=1.6$. Changing
these parameters will affect very similarly the far-infrared luminosities of
all samples.

Figure~\ref{fig:lfir_lbol1} shows \lfir\, vs \lbol\, for the four
samples of $z\sim2$ quasars. The coloured symbols represent the
individual sources, the black filled symbols the mean values for each
sample, and the black empty symbols the mean of the
non-detections. The dashed line shows a constant far-infrared fraction
of ${L_{\rm FIR} \over L_{\rm bol}}=0.05$, characteristic of the
low-redshift quasars studied by \citet{1994ApJS...95....1E}.

From the available data, the O03 and P04 samples as a whole have a
ratio of ${L_{\rm FIR} \over L_{\rm bol}}$ similar to the low-redshift
quasars.  However, our sample and the S05 samples both seem to have
larger far-infrared fractions than the O03 and P04 samples\footnote{It is difficult to quantify accurately the statistical
  significance of the differences between samples. As quoted in the
  ASURV manual: ``It is possible that all existing survival methods
  will be inaccurate for astronomical data sets containing many points
  very close to the detection limit.'' \citep{1990BAAS...22..917I},
  which is clearly the case in this comparison. To make things worse,
  the individual samples considered here are small in a statistical
  sense: they all consist of less than 30 objects.}.

If all obscuration were orientation dependent then the host galaxies of
obscured and unobscured quasars would be expected to have, on average, 
the same properties, such as star-formation rates
\citep[e.g.][]{2000A&A...357..839C} and cool dust content
\citep[e.g.][]{2004A&A...424..531H}. If, however, some obscured quasars are at
a different evolutionary stage, then their host galaxy can be reasonably
expected to have a larger gas and dust content, which might be causing the
obscuration, or at least contributing to it. Obscured quasars might therefore
be in a different evolutionary phase to unobscured quasars \citep[e.g.][]
{1988ApJ...325...74S,1999MNRAS.308L..39F}.

The extinction due to dust is negligible in the far-infrared (FIR), so that at
wavelengths $\grtsim50$~$\mu$m the emission becomes essentially isotropic.
Hence, if orientation is the only difference between obscured and unobscured
quasars, they should have virtually identical cool dust 
properties, irrespective of whether this emission is due to dust heated by the
quasar or by young stars in the host galaxy. 

Although it is difficult to quantify the significance of the
difference between samples, it seems that optically obscured quasars
(and optically-unobscured but X-ray absorbed quasars) have, on
average, higher far-infrared luminosities than unobscured quasars. In
turn, this implies that the cool-dust content of the host galaxies are
different so that, on average, obscured quasars are hosted by dustier
galaxies.

This is, of course, partly a result of selection. The samples selected
to contain obscured sources will of course be biased towards dustier
objects, but such a bias will only show in the far-infrared
observations if obscuring dust on kpc scales is important (as opposed
to dust in the torus only, which will emit mostly at mid-infrared
wavelengths). The S05 sample is particularly biased towards such host
obscuration, since it contains optically-bright quasars that are not
obscured by the torus, but which show absorption at X-ray energies.  

Hence, the difference in far-infrared luminosities between obscured
and unobscured quasars lends strong support to the scenario that some
of the obscuring dust is in the host galaxy, not only the torus, and
there is a hint that many obscured quasars might represent a different
evolutionary phase to unobscured quasars.

\section{Discussion and conclusions}\label{sec:disc}

We have observed a sample of $z\grtsim2$ radio-intermediate obscured
quasars at 1.2~mm (250~GHz) using MAMBO. The detection rate is 5 out of 21
(24\%). Sources with narrow lines seem to have a higher detection rate than sources with no narrow lines, but with such small numbers the difference is not significant. Larger samples will be required to study any differences. 

Stacking leads to a statistical detection of $\langle S_{1.2~\rm mm}
\rangle = 0.96\pm0.11$. Even if only the non-detections are stacked,
they still yield a statistical detection, with $\langle S_{1.2~\rm mm}
\rangle = 0.51\pm0.13$. Thus, the typical flux density of this sample
is $\sim$0.5-1.0~mJy, and this corresponds to a far-infrared
luminosity $\sim4\times10^{12}$~\lsol. 

If the far-infrared luminosity is powered entirely by star-formation, and not
by AGN-heated dust, then the typical inferred star-formation rate is
$\sim$700~\msolyr. This large star-formation rate is comparable to those
inferred in ULIRGs and SMGs, the most powerful starburst galaxies known. 

The observations at 1.2~mm also allow us to estimate the mass of the
cool dust, and we find a typical mass of $\sim3\times10^{8}$~\msol. If
heated by young stars, this dust is expected to be distributed on
$\sim$2~kpc scales (the scale found for SMGs). If it is heated by the
central AGN, with \lbol$\sim10^{13}$~\lsol, then it is expected to be
distributed on scales $\lesssim$10~kpc.

Indeed, we estimate that such a large mass of cool dust is capable of
causing alone the large values of \av\, in this sample, without help
of an obscuring torus.

Combining our observations at mid-infrared and millimetre wavelengths,
we present dust SEDs for our sample, and derive a typical SED for our
sample of high-redshift obscured quasars. The clumpy tori models
cannot reproduce the SED. However, clumpy tori models with an
additional screen of cold dust (and the corresponding far-infrared emission) do
reproduce this mean SED.  The amount of extinction from this
additional screen is derived from the cool dust mass, from the
observations at 1.2~mm. The depth of the silicate feature can be
consistently achieved by the inferred cool dust mass provided that the
dust is on scales $\lesssim$2~kpc, which is in excellent agreement
with the typical radius of the far-infrared emission in SMGs.  Again,
this lends support to kpc-scale dust along the host galaxy playing an
important role in the obscuration of these sources.

Obscuration by dust on kpc scales would also explain why in about half of the
sample the narrow emission lines from the central AGN are not detectable, as
well as why in some cases the jets seem to be pointing towards us, something
unexpected if they are only obscured by the torus of the unified schemes.

However, we remind the reader that unobscured quasars have also been
found to have similar cool-dust masses, and that in our sample there
is not a one-to-one correspondence between the presence (or lack of)
narrow lines and detection at 1.2~mm. The presence of a large mass of
dust does not guarantee obscuration.  A clumpy dusty interstellar
medium where different lines of sight have vastly different optical
depths is a likely explanation for why some quasars with large dust
masses are obscured while others are not.

When comparing to other samples of high-redshift quasars, we
find that the obscured quasars probably have a higher fraction of
their luminosity emerging at far-infrared wavelengths, compared to
unobscured quasars. The far-infrared emission is isotropic, so that
this difference cannot be ascribed to orientation-dependent
obscuration: obscured quasars are likely to have higher far-infrared
luminosities and cool-dust masses than unobscured quasars, suggesting
the host galaxies are in a dustier, presumably earlier,  phase.

Additionally, we have searched for molecular gas in two sources at
$z=2.767$ and 4.169, using the CO (3-2) and (4-3) transitions. In the
$z=2.767$ source we detect a line with $L_{\rm CO
  (3-2)}=$3.2$\times10^{7}$~\lsol\, (equivalent to a
brightness-temperature luminosity of $L'_{\rm CO
  (3-2)}=2.4\times10^{10}$ K~\kms~pc$^{2}$). In the other source, the
lack of detection suggests a line luminosity $L_{\rm CO
  (4-3)}<$3$\times10^{7}$~\lsol\, ($L'_{\rm CO (4-3)}<1\times10^{10}$
K~\kms~pc$^{2})$. Under the assumption that in these objects the (3-2)
and (4-3) transitions are thermalised, we can estimate the molecular
gas contents to be $M_{\rm H_{2}}=1.9\times10^{10}$ and
$<8\times10^{9}$ \msol, respectively.  The estimated gas depletion
timescales are $\tau_{\rm g}=4$ and $<$16~Myr, and low gas-to-dust
mass ratios of \mgas$/$\mdust$=19$ and $\leq20$ are inferred.
A dynamical mass of $M_{\rm dyn} {\rm
  sin^{2}}i=6\times10^{9}$~\msol\, is estimated from the CO(3-2) detection.

\acknowledgements 

We thank the IRAM staff for help with the observations and data
reduction for this program. We are particularly grateful to St\'ephane
Leon for extensive help with the observations at the 30m, Robert Zylka
for making the MOPSIC package publicly available, Jan Martin Winters
for help with the PdBI observations, Philippe Salom\'e for help
reducing the PdBI data on AMS16 and Roberto Neri for use of his
software. We also thank Javier Rod\'on, Veronica Roccatagliata and
Aurora Sicilia-Aguilar for help with software. We thank the support
and assistance provided by the CSO staff and SHARC-II team at Caltech
during the observations and data reduction. We are also grateful to
the CLUMPY group\footnote{http://www.pa.uky.edu/clumpy/index.html} for
making their models publicly available. This manuscript was improved
by the suggestions of an anonymous referee. The work was partially
supported by grants associated with Spitzer programs GO-20705 and
GO-30634, and is based on observations made with the Spitzer Space
Telescope, which is operated by the Jet Propulsion Laboratory,
California Institute of Technology under a contract with NASA.

\bibliographystyle{apj.bst}

\clearpage 

\begin{table}
  \caption{Observed properties of the  sample of obscured quasars.}
    \begin{center}
      \begin{tabular}{lccccccccc}
        \hline
\hline
Name & RA & Dec & $z_{\rm spec}^{a}$  & Optical$^{b}$ & Radio$^{c}$ & $f_{\rm
  core}^{d}$& Mid-infrared$^{e}$ & $S_{1.2~\rm mm}$$^{f}$  \\
 &     (J2000)  &      &                   &                &            &         &
 &  /mJy  \\\hline

AMS01	 & 17 13 11.17 & +59 55 51.5& - &B & SS & $<$0.3& C&  0.63$\pm$0.53	\\
AMS02	 &  17 13 15.88 & +60 02 34.2&  1.8 & B & SS & - & S& 0.87$\pm$0.54 \\
AMS03	 & 17 13 40.19 & +59 27 45.8  &  2.698 &NL & SS & 0.8& -& 0.76$\pm$0.51\\
AMS04	 &17 13 40.62 & +59 49 17.1  &    1.782 & NL & SS & -&S&0.55$\pm$0.58\\
AMS05	 & 17 13 42.77 & +59 39 20.2  &  2.850 &NL & GP & 1.1& S& -0.26$\pm$0.40\\
AMS06	 &  17 13 43.91 & +59 57 14.6  &  1.8 & B & SS & $<$0.3& S& 0.86$\pm$0.41\\
AMS07	 &  17 14 02.25 & +59 48 28.8 &  - &B  & FS & - & C&  1.05$\pm$0.44 \\
AMS08	 & 17 14 29.67 & +59 32 33.5 &  1.979 & NL& - & - & S& {\bf 1.99$\pm$0.34}  \\
AMS09	 &  17 14 34.87 & +58 56 46.4  & 2.1 & B & SS & 0.9& S& 0.51$\pm$0.56\\	
AMS10	 & 17 16 20.08 & +59 40 26.5  & - &  B & SS & - & C&  0.62$\pm$0.79   \\   
AMS11	 & 17 18 21.33 & +59 40 27.1  & 1.6 &  B & SS & - & P,S& 0.36$\pm$0.57  \\
AMS12	 &  17 18 22.65 & +59 01 54.3  & 2.767 & NL & SS &0.3 & -& {\bf 3.74$\pm$0.60}  \\
AMS13	 &  17 18 44.40 & +59 20 00.8  & 1.974 & NL & SS & -& S&  0.27$\pm$0.51$^{g}$\\
AMS14	 &  17 18 45.47 & +58 51 22.5  & 1.794 & NL & SS & - & S&   0.27$\pm$0.44\\
AMS15	 & 17 18 56.93 & +59 03 25.0   &  2.1 & B & FS & 0.5& S& 0.12$\pm$0.70\\
AMS16	 &   17 19 42.07 & +58 47 08.9   & 4.169 & NL & GP &$<$0.6 & -& {\bf 2.12$\pm$0.54}\\ 
AMS17	 &   17 20 45.17 & +58 52 21.3  & 3.137 & NL & SS& $<$0.3& P,S& {\bf 1.73$\pm$0.48}$^{h}$  \\
AMS18	 &  17 20 46.32 & +60 02 29.6   & 1.6 & B& FS & - & S&   1.11$\pm$0.53\\
AMS19	 &  17 20 48.00 & +59 43 20.7 & 2.3 & B & FS & 1.0 &  S& {\bf 2.76$\pm$0.66}\\
AMS20	 &   17 20 59.10 & +59 17 50.5 & - & B & GP &  - & C& -0.78$\pm$0.93  \\
AMS21	 & 17 21 20.09 & +59 03 48.6  & 1.8 & B& SS& 0.7& P,S&  0.52$\pm$0.65\\ 
\hline
\hline
        \end{tabular}  
\end{center}
  \tablenotetext{a}{Redshifts with 3 decimal places are from
    optical spectroscopy \citep{2006MNRAS.370.1479M}, while those with only 1 decimal place are from
    mid-infrared spectroscopy \citep{2008ApJ...674..676M}. The redshift for
    AMS05 is from the optical spectrum taken by \citet{2009MNRAS.393..309S}. }
  \tablenotetext{b}{ Summary of optical spectroscopy properties. B stands for blank spectrum, NL for narrow lines. }
  \tablenotetext{c}{ Summary of radio spectral properties from \citet {2006MNRAS.373L..80M}. SS stands for steep spectrum, FS for flat spectrum and GP for gigahertz-peaked source.}
  \tablenotetext{d}{ Approximate fraction of radio flux at 1.4~GHz recovered at
    VLBI resolution at 1.6~GHz \citep{2009arXiv0905.1605K}. }
  \tablenotetext{e}{ Spectral properties in the mid-infrared \citep[see ][]{2008ApJ...674..676M}. C stands for
    continuum only, S for silicate absorption, P for polycyclic aromatic
    hydrocarbons (PAHs). }
\tablenotetext{f}{Flux densities at 1.2~mm. Significant detections are marked
  in bold. }
 \tablenotetext{g}{ Flux density for AMS13  taken from
   \citet{2005ApJ...632L..13L}, their source 22204.}
\tablenotetext{h}{ Flux density for AMS17  taken from
   \citet{2008ApJ...683..659S}, their source 22558.}

 \label{tab:obs}
\end{table}

\begin{table}
  \caption{Comparison of resulting fluxes in Win07 and Sum08/Win08. Bold shows the values used in Table~\ref{tab:obs}}
    \begin{center}
      \begin{tabular}{lccc}
        \hline
Name & Win07 & Sum08/Win08 & Combined \\  
 & /mJy  & /mJy  &/mJy \\
\hline
\hline
AMS07	 &   2.43$\pm$0.83 & {\bf 1.05$\pm$0.44}  & 1.27$\pm$0.41 \\
AMS08	 &  2.04$\pm$0.59& {\bf 1.99$\pm$0.41}  &1.99$\pm$0.34 \\
AMS10	 &0.81$\pm$0.93 & 0.45$\pm$1.1  &{\bf 0.62$\pm$0.79}  \\   
AMS11	 &  0.83$\pm$0.64&-0.59$\pm$1.00 &{\bf 0.36$\pm$0.57}\\
AMS12	 &   4.03$\pm$1.03& {\bf 3.74$\pm$0.60}  &3.80$\pm$0.55\\
AMS17$^{a}$	 &  2.39$\pm$0.55& 1.45$\pm$0.62 &1.94$\pm$0.44 \\
AMS20	  & -0.16$\pm$1.08& -1.24$\pm$1.56 & {\bf -0.78$\pm$0.93} \\
\hline
\hline
        \end{tabular} 
  \end{center} 
\tablenotetext{a}{ The flux density for AMS17  used in Table~\ref{tab:obs}, 1.73$\pm$0.48,  is taken from   \citet{2008ApJ...683..659S}, their source 22558.}
 \label{tab:07_08}
\end{table}

\begin{table}
  \caption{Inferred properties of the  sample of obscured quasars.}
    \begin{center}
      \begin{tabular}{lccccccccc}
        \hline
\hline
Name &  log$_{10}$[\lbol/\lsol]$^{a}$ & \av$^{a}$ &  & $T=47$~K $\beta=1.6$$^{b}$   &  & &$T=35$~K $\beta=1.5$$^{c}$      \\
      &  &     &  log$_{10}$[\lfir/\lsol]   &   log$_{10}$[SFR/\msolyr]   &
      log$_{10}$[\mdust/\msol]          &     log$_{10}$[\lfir/\lsol]
      &     log$_{10}$[SFR/\msolyr]      & log$_{10}$[\mdust/\msol]   
    \\\hline
\hline 
AMS01  &  $\geq$13.2  &$\geq$ 40.5  &  $<$13.0&    $<$ 3.2&    $<$ 8.7&    $<$12.4&    $<$ 2.6&    $<$ 8.9\\
AMS02  &        12.4  &        0.5  &  $<$13.0&    $<$ 3.2&    $<$ 8.7&    $<$12.4&    $<$ 2.6&    $<$ 8.9\\
AMS03  &  $\geq$13.2  &$\geq$ 21.5  &  $<$12.9&    $<$ 3.2&    $<$ 8.6&    $<$12.3&    $<$ 2.6&    $<$ 8.9\\
AMS04  &        13.7  &       68.5  &  $<$13.0&    $<$ 3.3&    $<$ 8.7&    $<$12.4&    $<$ 2.7&    $<$ 9.0\\
AMS05  &        13.6  &        5.5$^{d}$  &  $<$12.8&    $<$ 3.0&    $<$ 8.5&    $<$12.2&    $<$ 2.5&    $<$ 8.8\\
AMS06  &        13.1  &       12.0  &  $<$12.9&    $<$ 3.1&    $<$ 8.6&    $<$12.3&    $<$ 2.5&    $<$ 8.8\\
AMS07  &  $\geq$12.9  &$\geq$ 17.0  &  $<$12.9&    $<$ 3.1&    $<$ 8.6&    $<$12.3&    $<$ 2.5&    $<$ 8.8\\
AMS08  &  $\geq$13.6  &$\geq$ 65.0  &     13.1&        3.3&        8.8&       12.5&        2.7&        9.0\\
AMS09  &        13.1  &       16.0  &  $<$13.0&    $<$ 3.2&    $<$ 8.7&    $<$12.4&    $<$ 2.6&    $<$ 8.9\\
AMS10  &  $\geq$13.1  &$\geq$ 55.0  &  $<$13.2&    $<$ 3.4&    $<$ 8.9&    $<$12.6&    $<$ 2.8&    $<$ 9.1\\
AMS11  &        12.7  &       13.0  &  $<$13.0&    $<$ 3.3&    $<$ 8.7&    $<$12.4&    $<$ 2.7&    $<$ 9.0\\
AMS12  &        13.3  &       21.0  &     13.3&        3.5&        9.0&       12.7&        3.0&        9.3\\
AMS13  &        13.9  &       22.5  &  $<$13.0&    $<$ 3.2&    $<$ 8.7&    $<$12.4&    $<$ 2.6&    $<$ 8.9\\
AMS14  &  $\geq$13.8  &$\geq$ 73.5  &  $<$12.9&    $<$ 3.2&    $<$ 8.6&    $<$12.3&    $<$ 2.6&    $<$ 8.9\\
AMS15  &  $\geq$13.0  &$\geq$ 40.0  &  $<$13.1&    $<$ 3.3&    $<$ 8.8&    $<$12.5&    $<$ 2.8&    $<$ 9.1\\
AMS16  &        13.9  &       21.0  &     12.9&        3.2&        8.6&       12.4&        2.7&        9.0\\
AMS17  &        13.8  &       33.5  &     12.9&        3.2&        8.6&       12.4&        2.6&        8.9\\
AMS18  &  $\geq$14.2  &$\geq$100.0  &  $<$13.0&    $<$ 3.2&    $<$ 8.7&    $<$12.4&    $<$ 2.6&    $<$ 8.9\\
AMS19  &        13.4  &        7.5  &     13.2&        3.4&        8.9&       12.6&        2.9&        9.2\\
AMS20  &        12.7  &        0.2  &  $<$13.2&    $<$ 3.5&    $<$ 8.9&    $<$12.6&    $<$ 2.9&    $<$ 9.2\\
AMS21  &        13.1  &       24.0  &  $<$13.1&    $<$ 3.3&    $<$ 8.8&    $<$12.5&    $<$ 2.7&    $<$ 9.0\\
\hline
\hline
       \end{tabular}
  \end{center} 
  \tablenotetext{a}{Bolometric luminosities and extinctions derived from the broad-band data between 3.6 and 24~\mum, as described in  Section~\ref{sec:broad}.}
  \tablenotetext{b}{Far-infrared luminosities, star-formation rates and cool-dust masses inferred from the 1.2~mm flux densities, assuming a gray body with $T=47$~K and $\beta=1.6$ (see Section~\ref{sec:res}).}
  \tablenotetext{c}{Far-infrared luminosities, star-formation rates and cool-dust masses inferred from the 1.2~mm flux densities, assuming a gray body with $T=35$~K and $\beta=1.5$ (see Section~\ref{sec:res}).}
  \tablenotetext{d}{The value of \av$=$5.5 was inferred by   \citet{2009MNRAS.393..309S} using additional data at 1.2, 1.7 and 2.2~\mum.}
 \label{tab:infer}
\end{table}

\begin{table}
\caption{Summary of gray body fits}
  \begin{center}
    \begin{tabular}{cccccccc}
      \hline
      \hline
      Source & $\beta$ & $T$ & log$_{10}$[\lfir  & log$_{10}$[\mdust \\
      & /K & /\lsol] & /\msol]   \\
      \hline  
      AMS08$^{a}$ & 1.1 & 57 & 13.0 & 8.9\\
      AMS12$^{b}$  & 1.5& 52 & 13.4  & 9.0\\
       AMS19$^{b}$ & 1.5 &  42 & 12.9 & 9.0\\
      \hline
      \hline
    \end{tabular}
  \end{center}
  \tablenotetext{a}{In the case of AMS08, three data points are used: 70 and 160~\mum\, as well as 1.2~mm, so that three parameters can be fitted: $\beta$, $T$ and \lfir. }
  \tablenotetext{b}{For these two sources, the 70~\mum\, data do not provide useful constraints, and only two parameters are fitted:   $T$ and \lfir, while $\beta$ is kept fixed at  a value of 1.5.  }
  \label{tab:gb_fit}
\end{table}

\clearpage 

\clearpage

\begin{figure*}
  \includegraphics{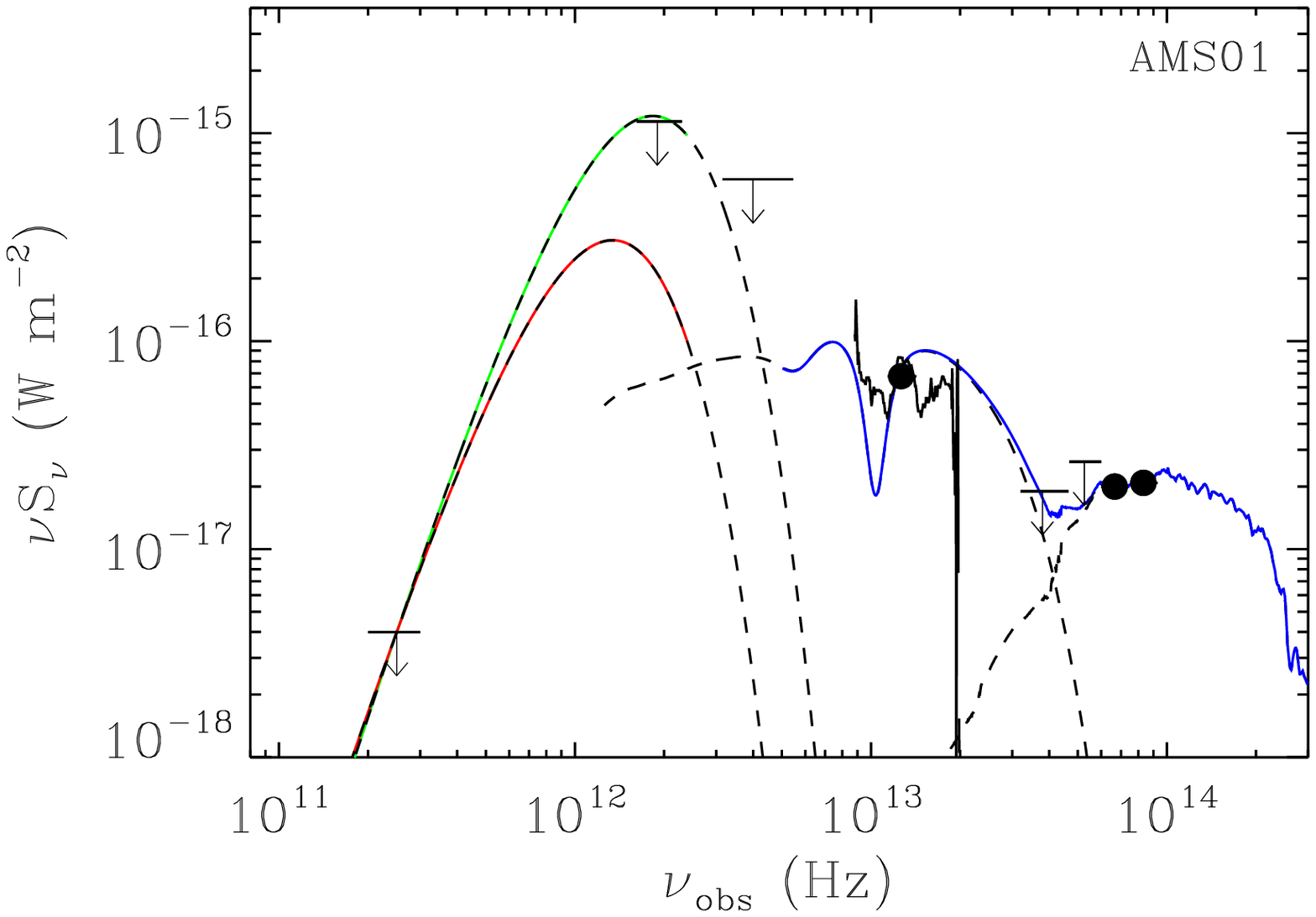} 
  
  \includegraphics{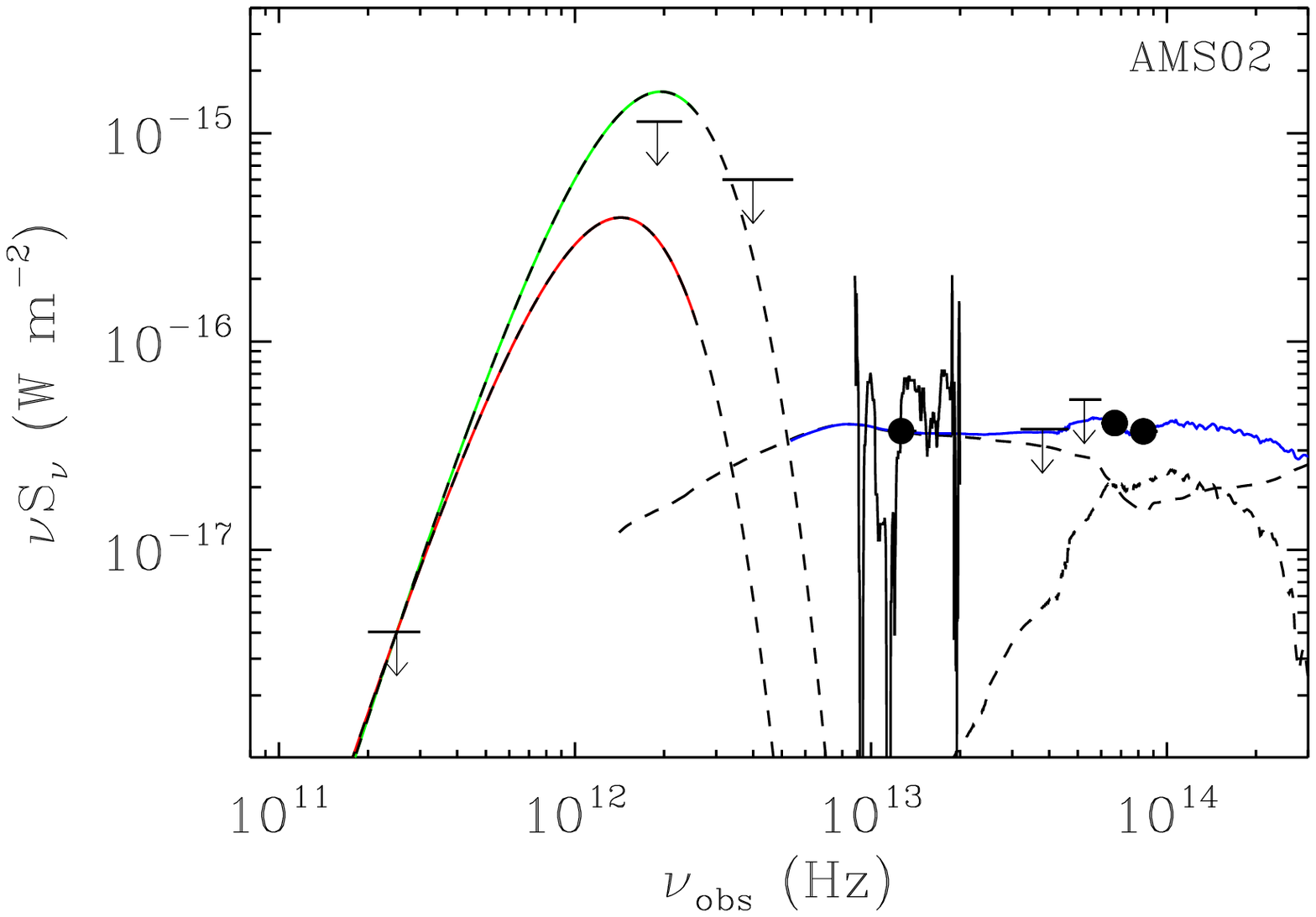} 
  
\includegraphics{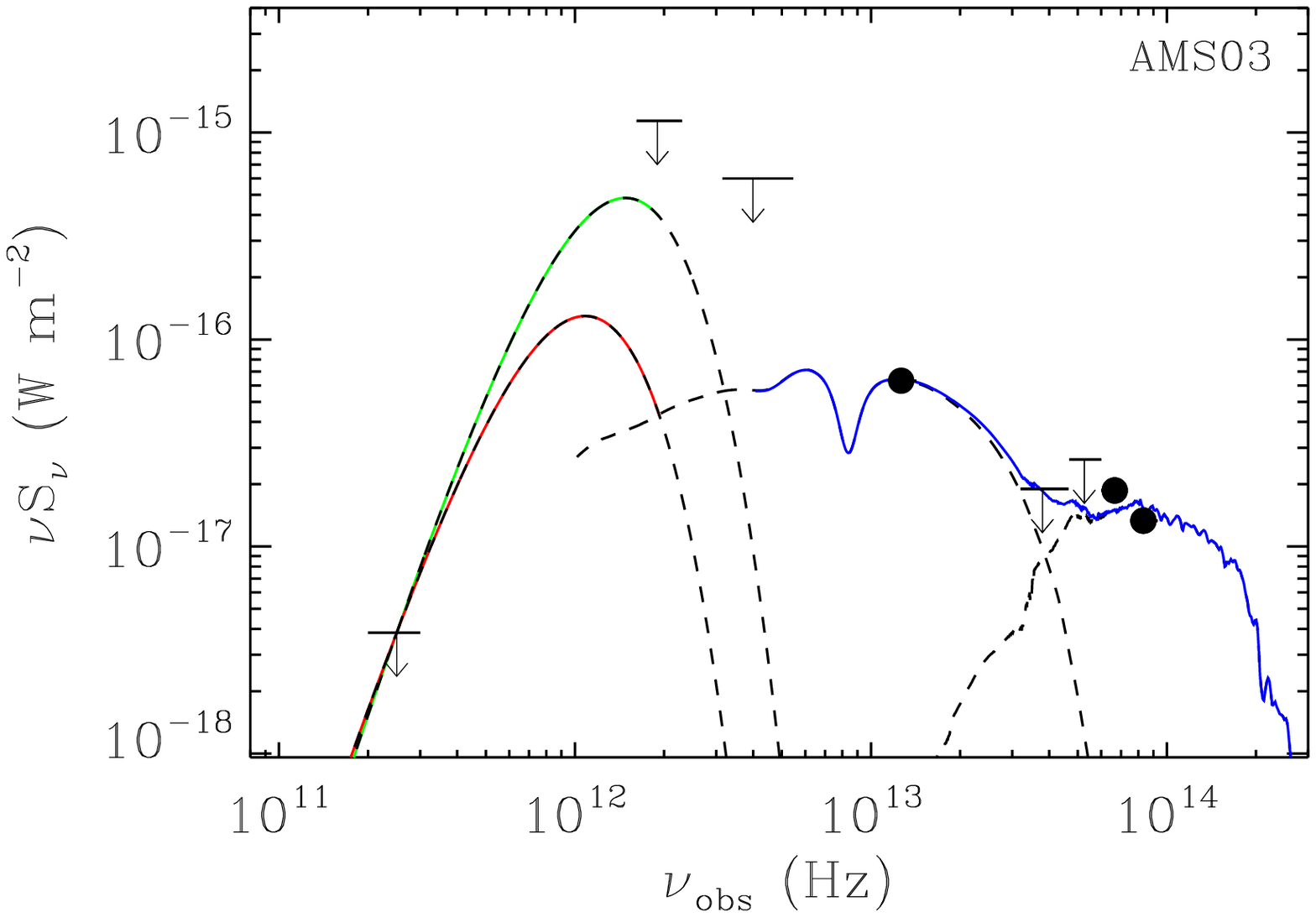} 

\includegraphics{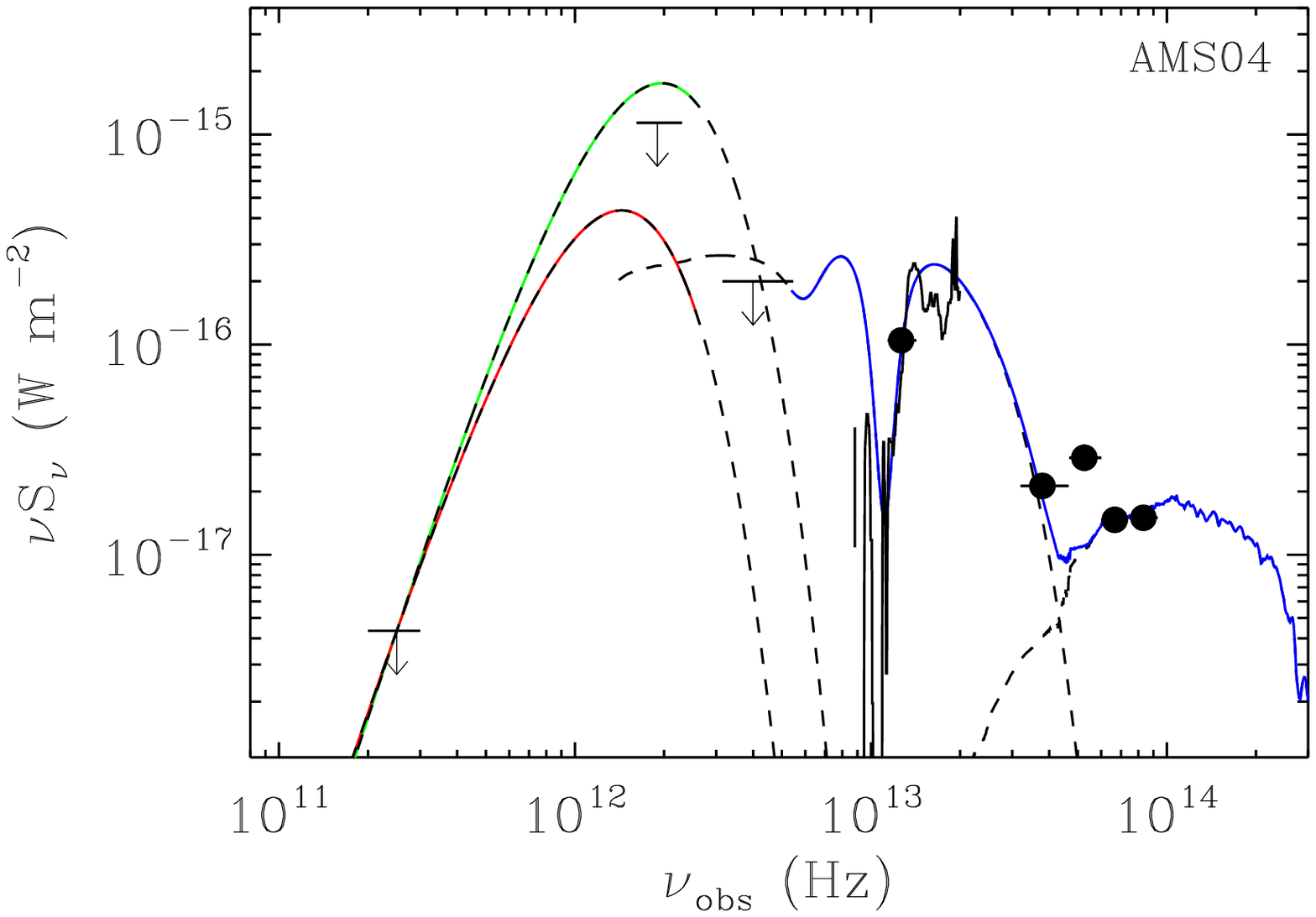} 

\includegraphics{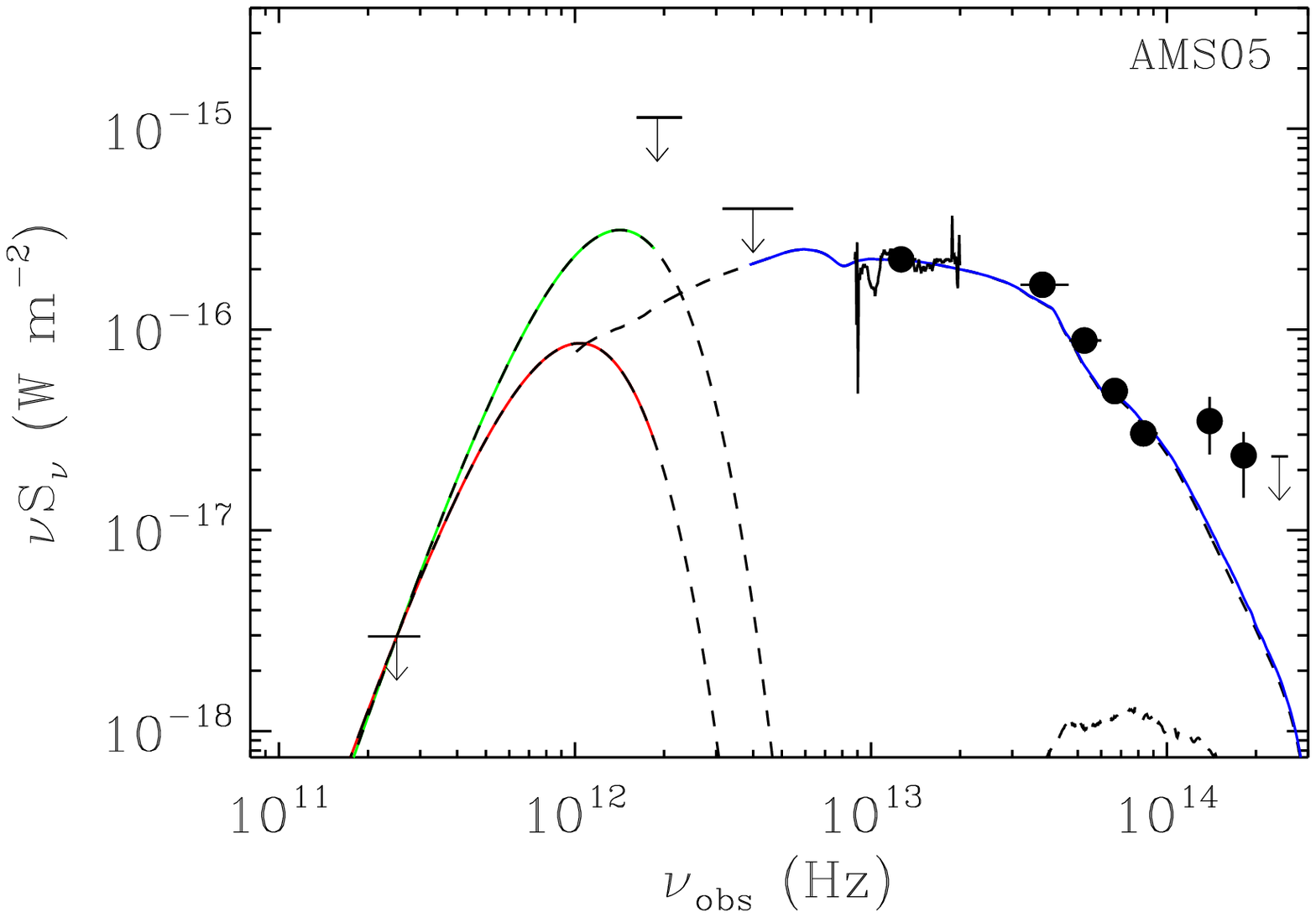} 

\includegraphics{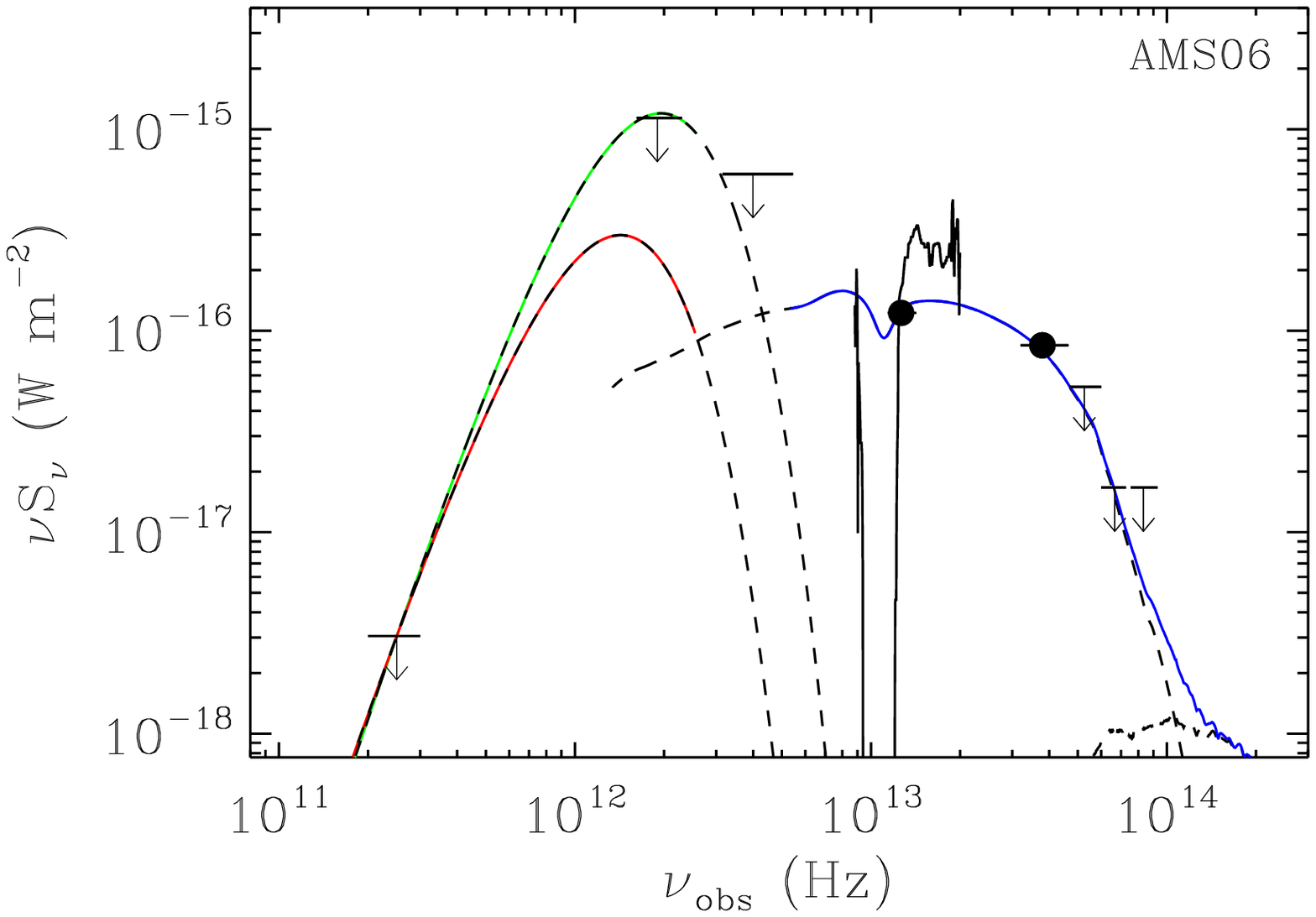} 
\label{fig:seds}
\vspace{18.5cm}
{ \caption{ Spectral energy distributions of the sample of
    high-redshift obscured quasars. Data points are marked as black
    points for detections or arrows for limits (left to right):
    1.2~mm, 160, 70, 24, 8, 5.8, 4.5, 3.6~\mum\, \citep[data from][
    and deeper unpublished IRAC
    data]{2005ApJS..161...41L,2006AJ....131.2859F,2006AJ....131..250F}. All
    limits are 3$\sigma$.  Overlayed in black are also the IRS
    spectra, when available. The far-infrared SED is modelled by a gray body, as described in Section~\ref{sec:firlum}. The mid-infrared component is modelled by an unobscured quasar SED with foreground extinction (see Section~\ref{sec:broad}). The dashed lines are the
    \citet{1994ApJS...95....1E} quasar SED with a screen of dust
    \citep[with the value of \av\, as quoted in Table~\ref{tab:infer}
    and using the MW dust model of ][]{1992ApJ...395..130P} and a
    $z=0$ elliptical galaxy from \citet{1980ApJS...43..393C}. The
    quasar SED has been truncated at wavelengths longer than
    rest-frame 80~\mum.  The blue solid line is the sum of both
    obscured quasar and host galaxy SEDs at wavelengths shorter than
    20~\mum, fitted using the data points between 3.6 and 24~\mum\,
    (see Section~\ref{sec:broad}).  Source AMS05 has additional 
    data points at 1.2, 1.6 and 2.2~\mum\, from
    \citet{2009MNRAS.393..309S}.   
}}
\end{figure*}

\clearpage

\addtocounter{figure}{-1}
\begin{figure*}
  \includegraphics{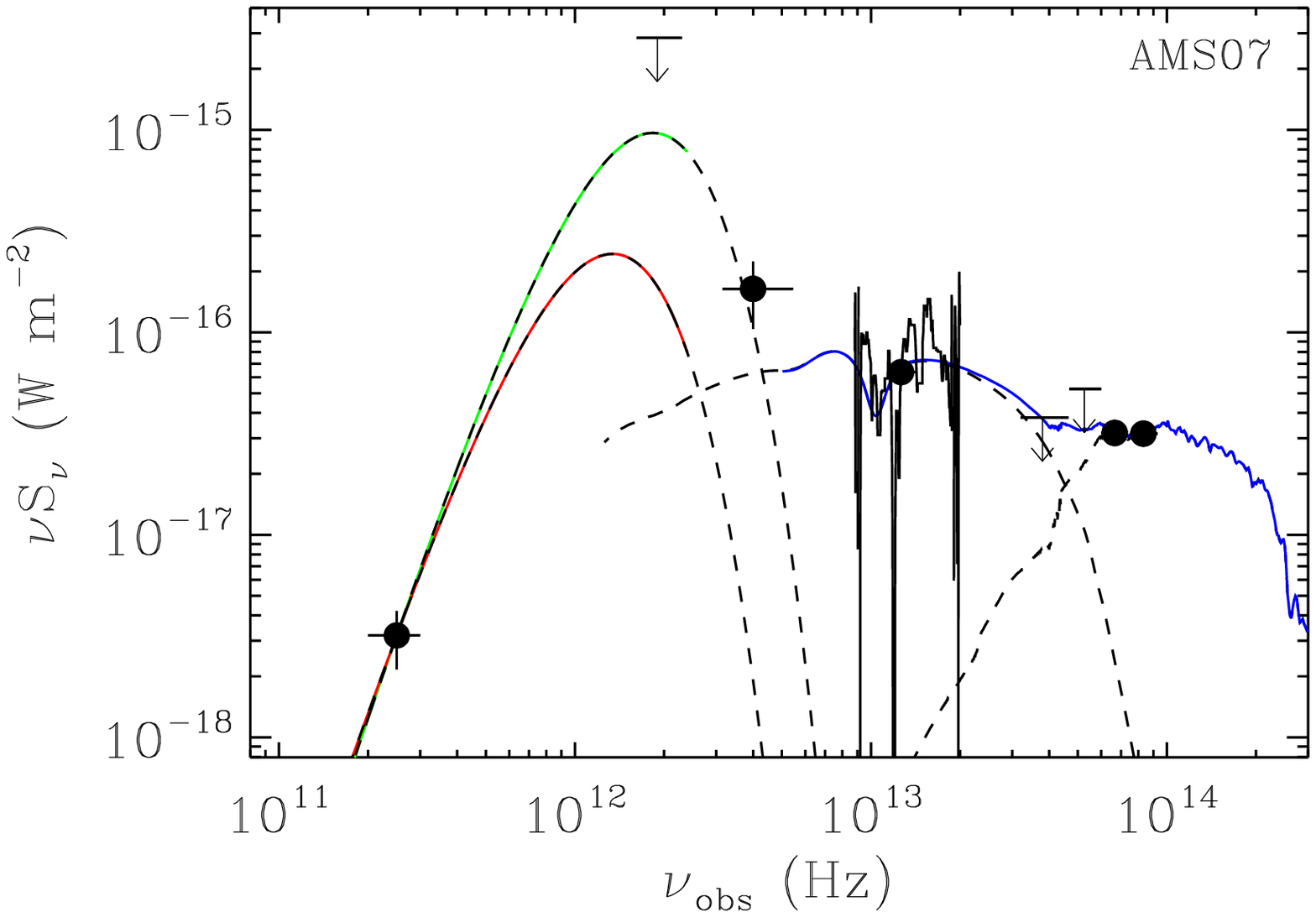} 
  
  \includegraphics{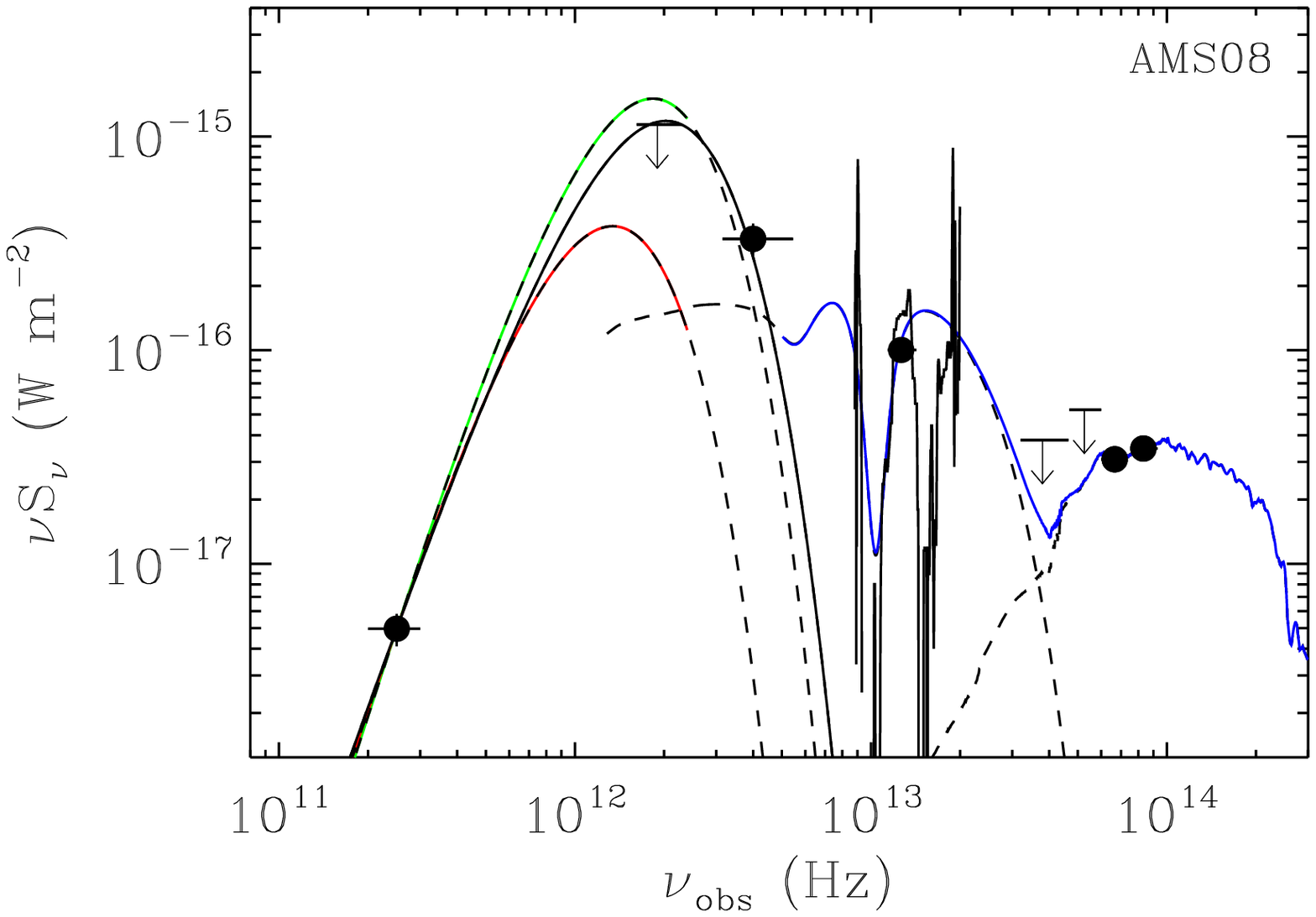} 
  
\includegraphics{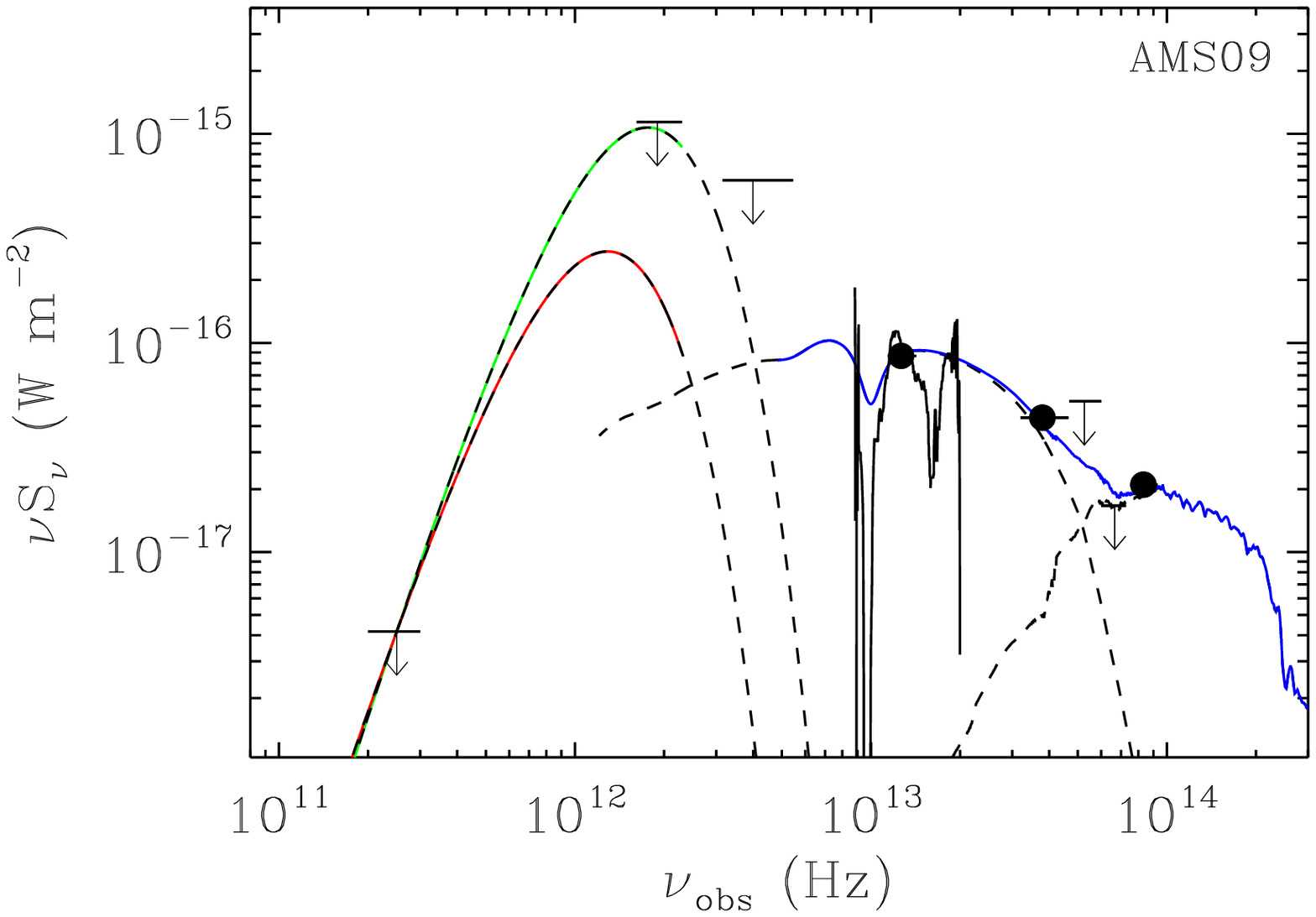} 

\includegraphics{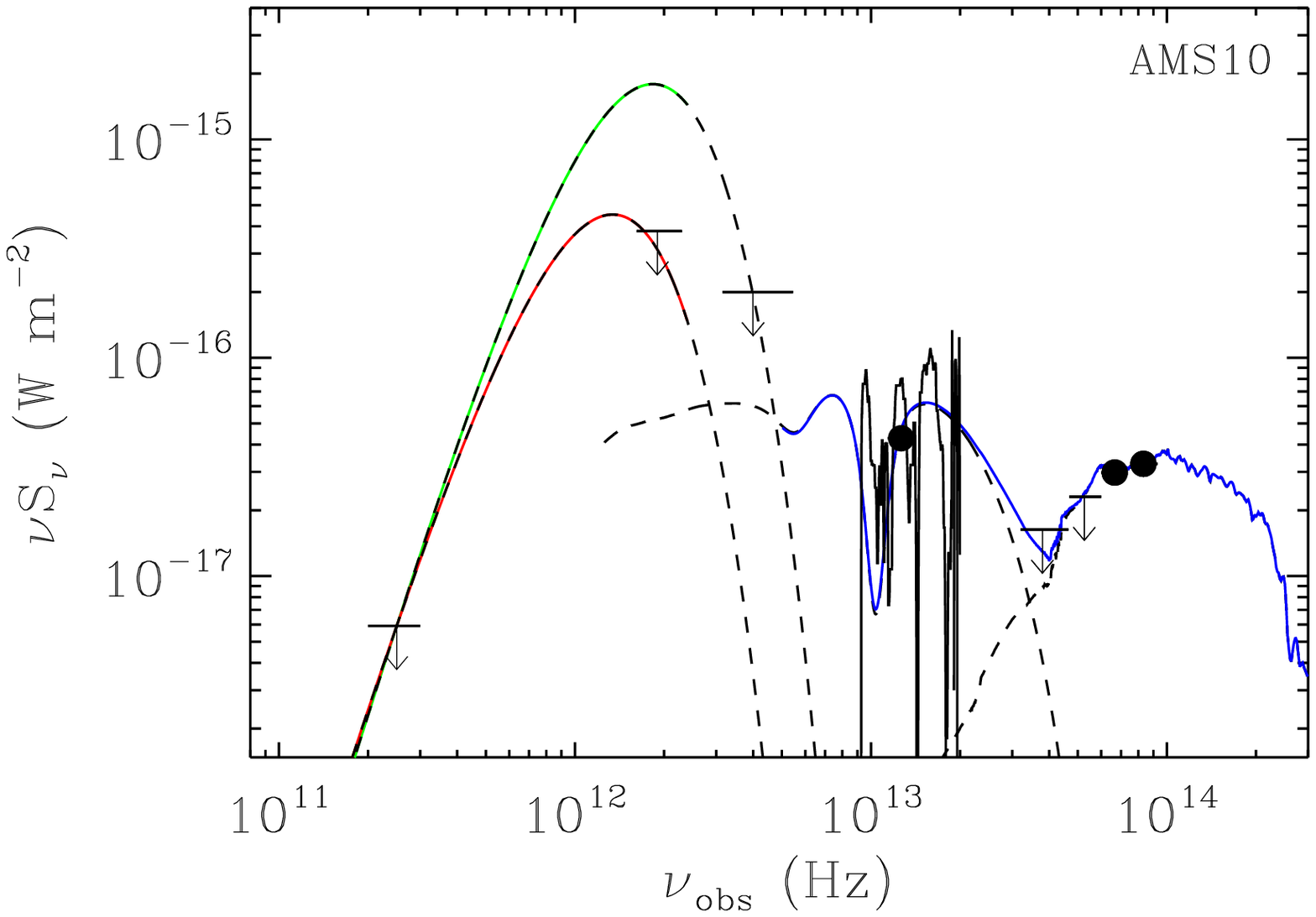} 

\includegraphics{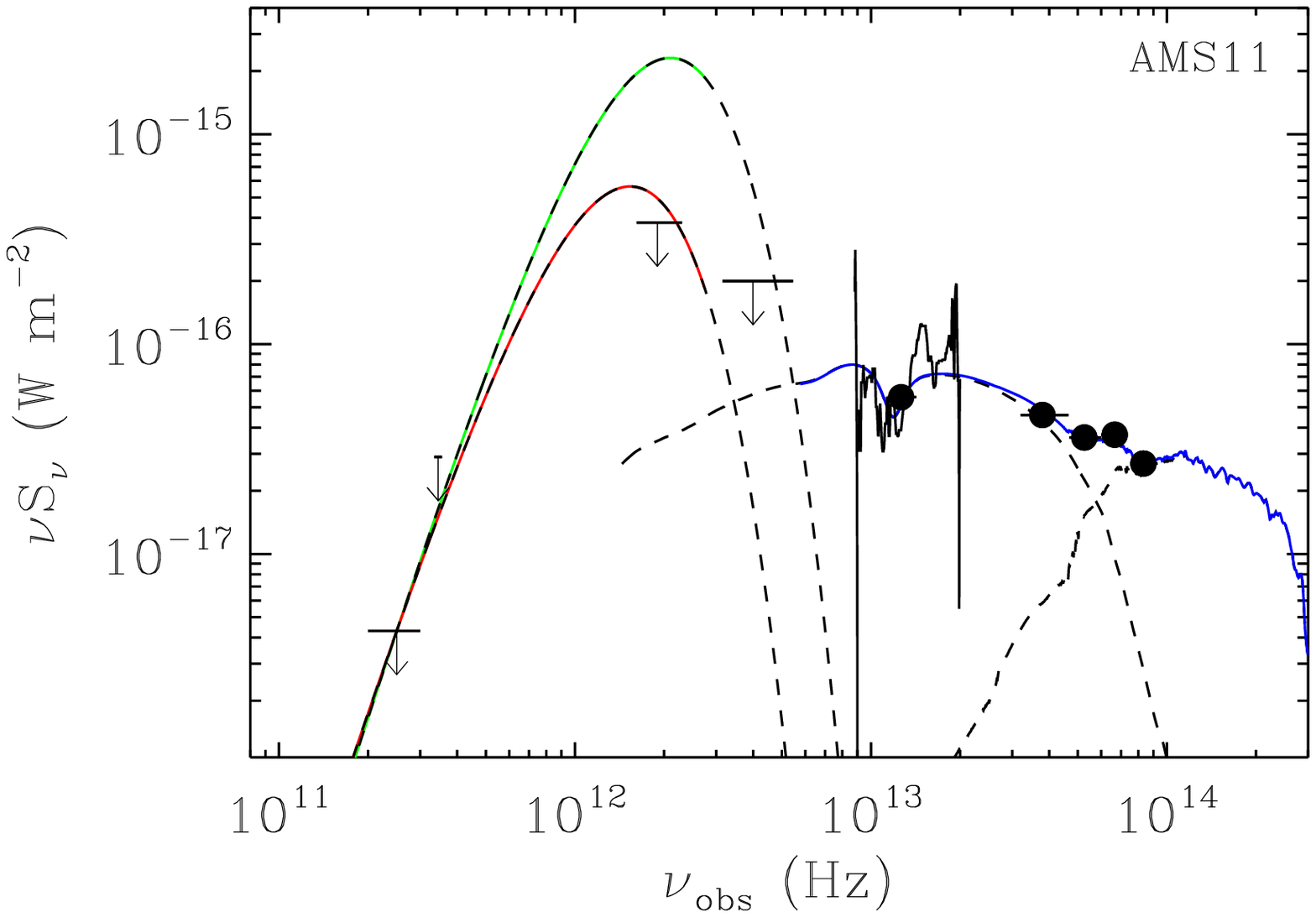} 

\includegraphics{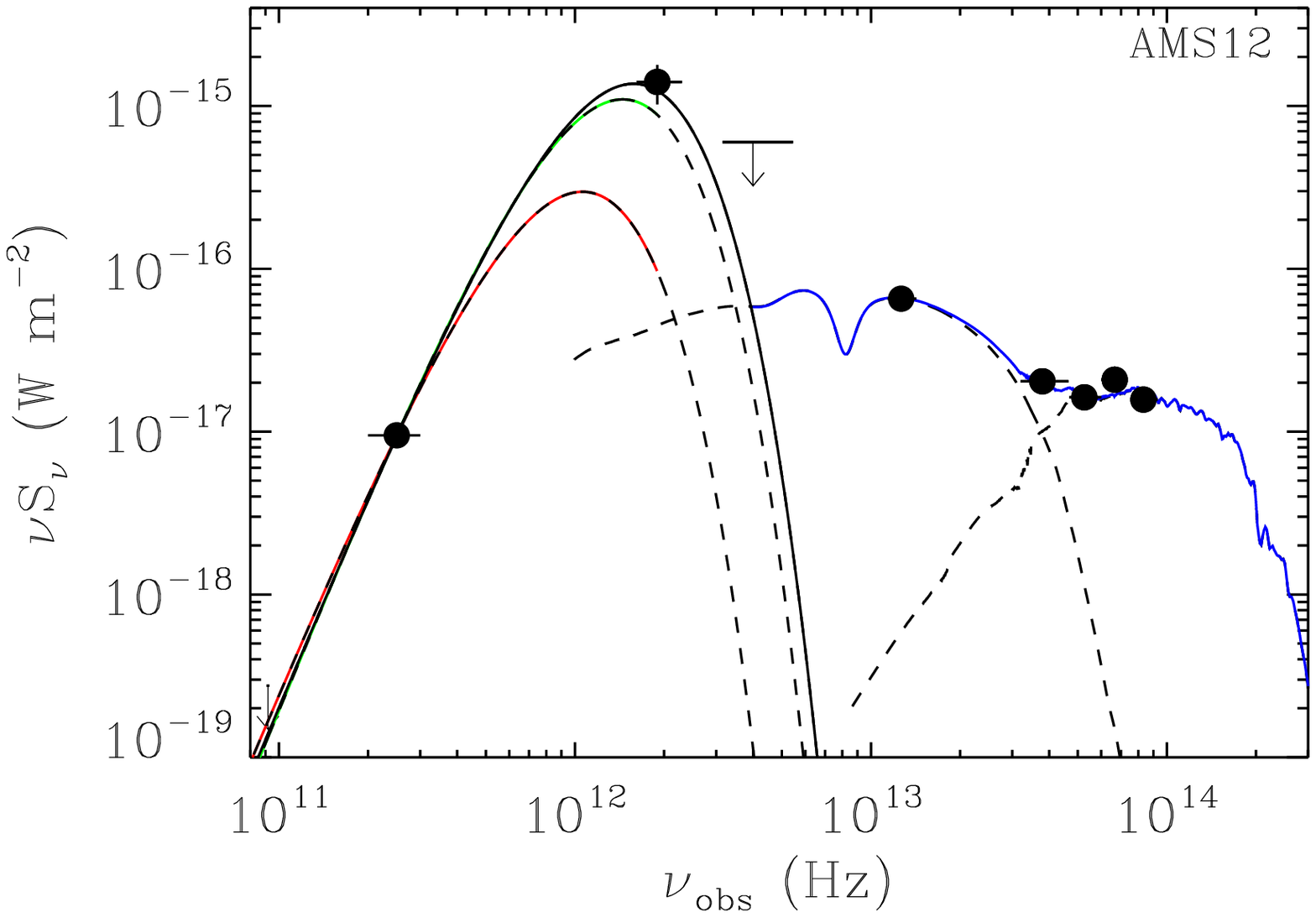} {
\vspace{18.5cm}
  \caption{ Continued. Source AMS11 has an extra point at 850~\mum, from the
    observations of \citet{2004ApJS..154..137F}, while source AMS12 has a limit on the 3~mm flux density from the PdBI observations. Sources AMS08 and AMS12 have  an
    extra gray body overplotted (black solid lines), described by the values in
    Table~\ref{tab:gb_fit}. }}
\end{figure*}

\clearpage

\addtocounter{figure}{-1}
\begin{figure*}
  \includegraphics{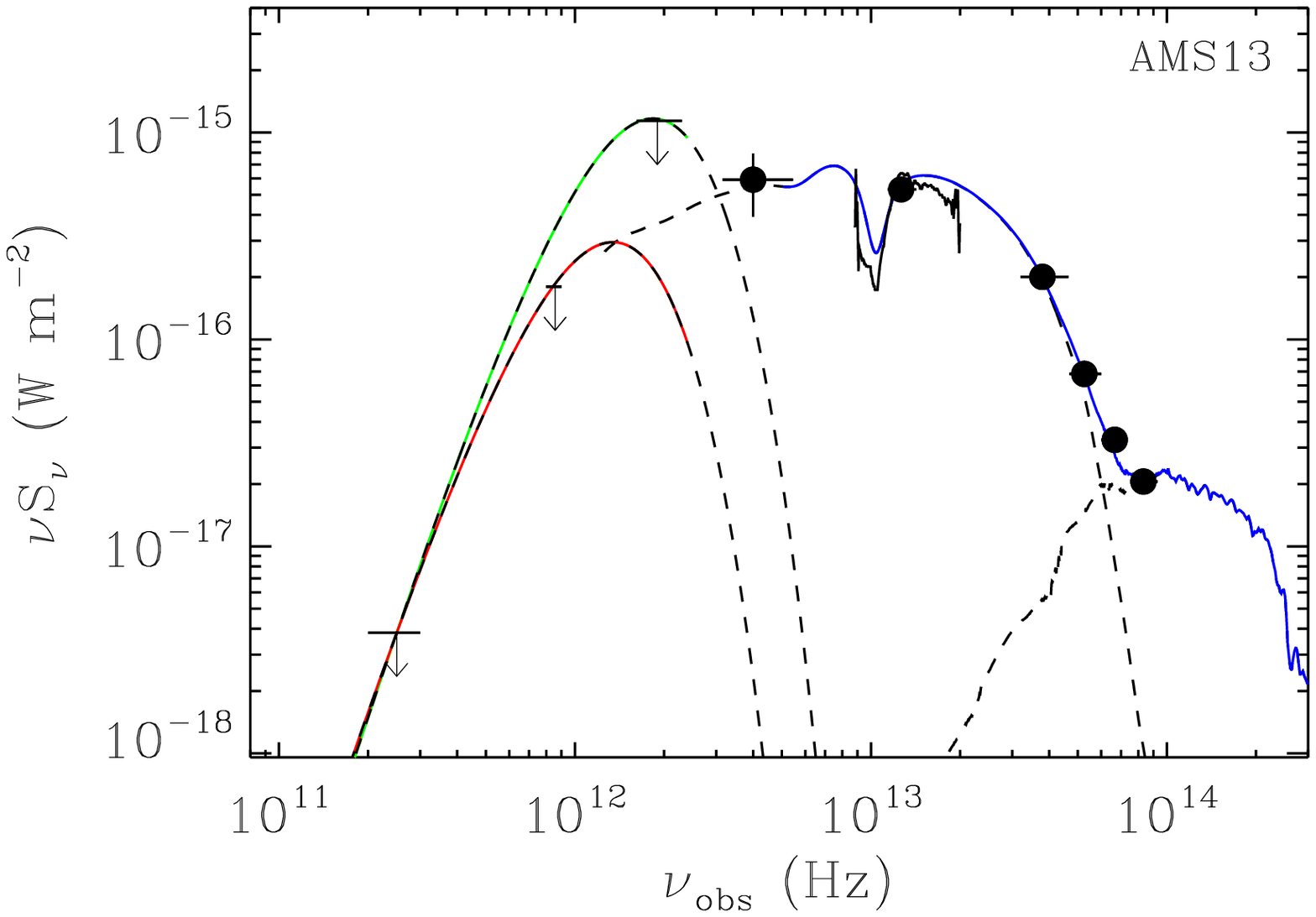} 
  
  \includegraphics{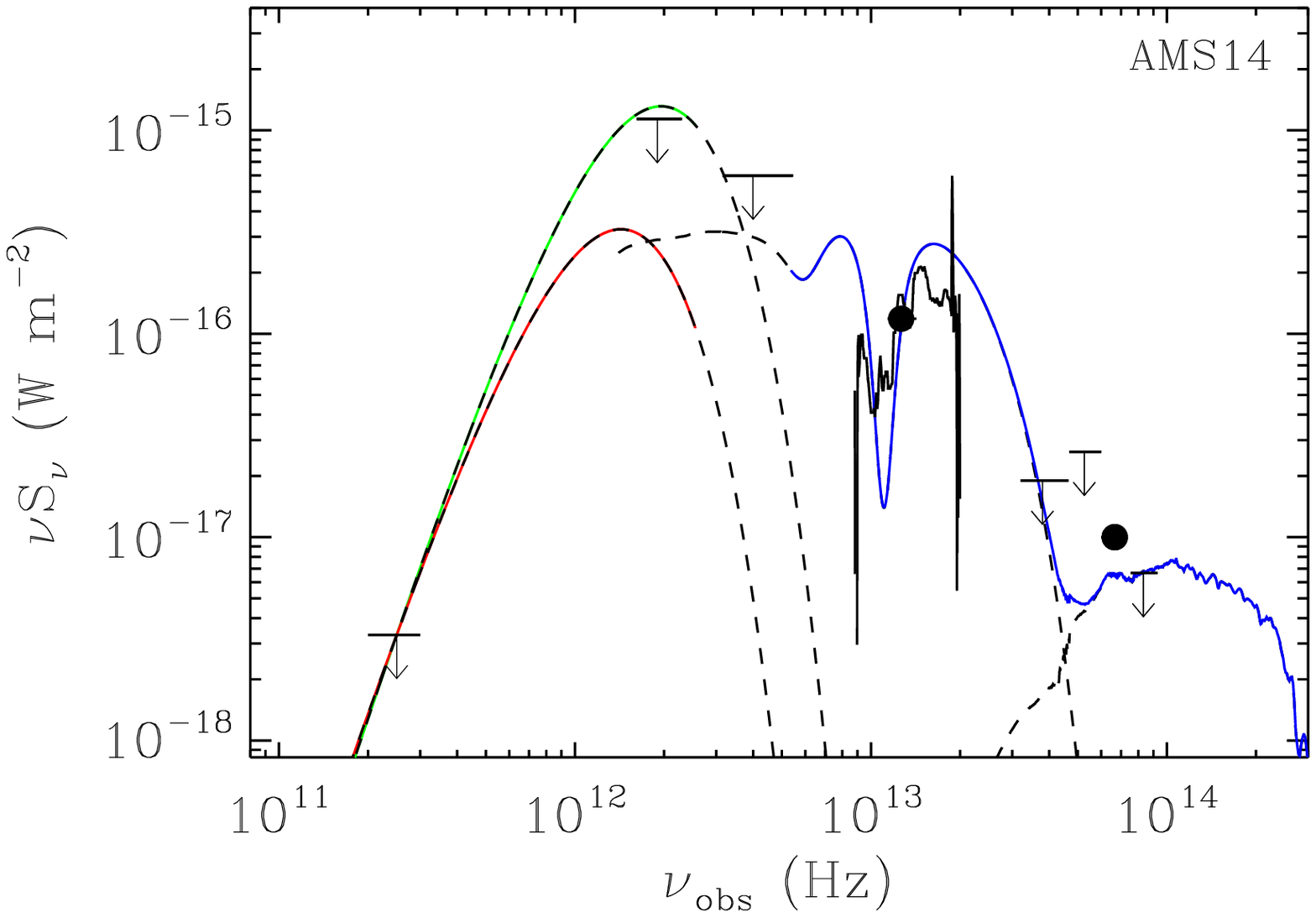} 
  
\includegraphics{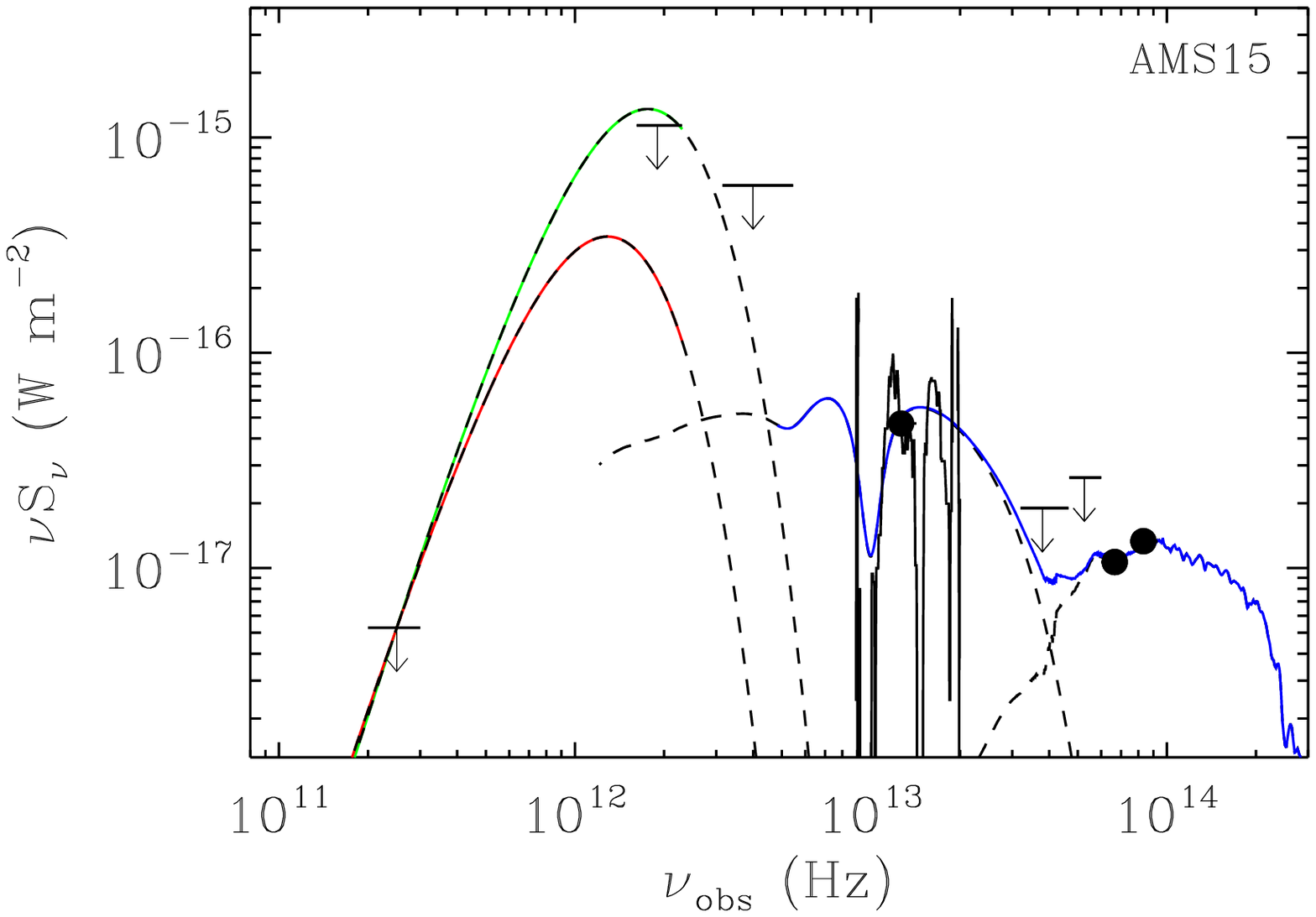} 

\includegraphics{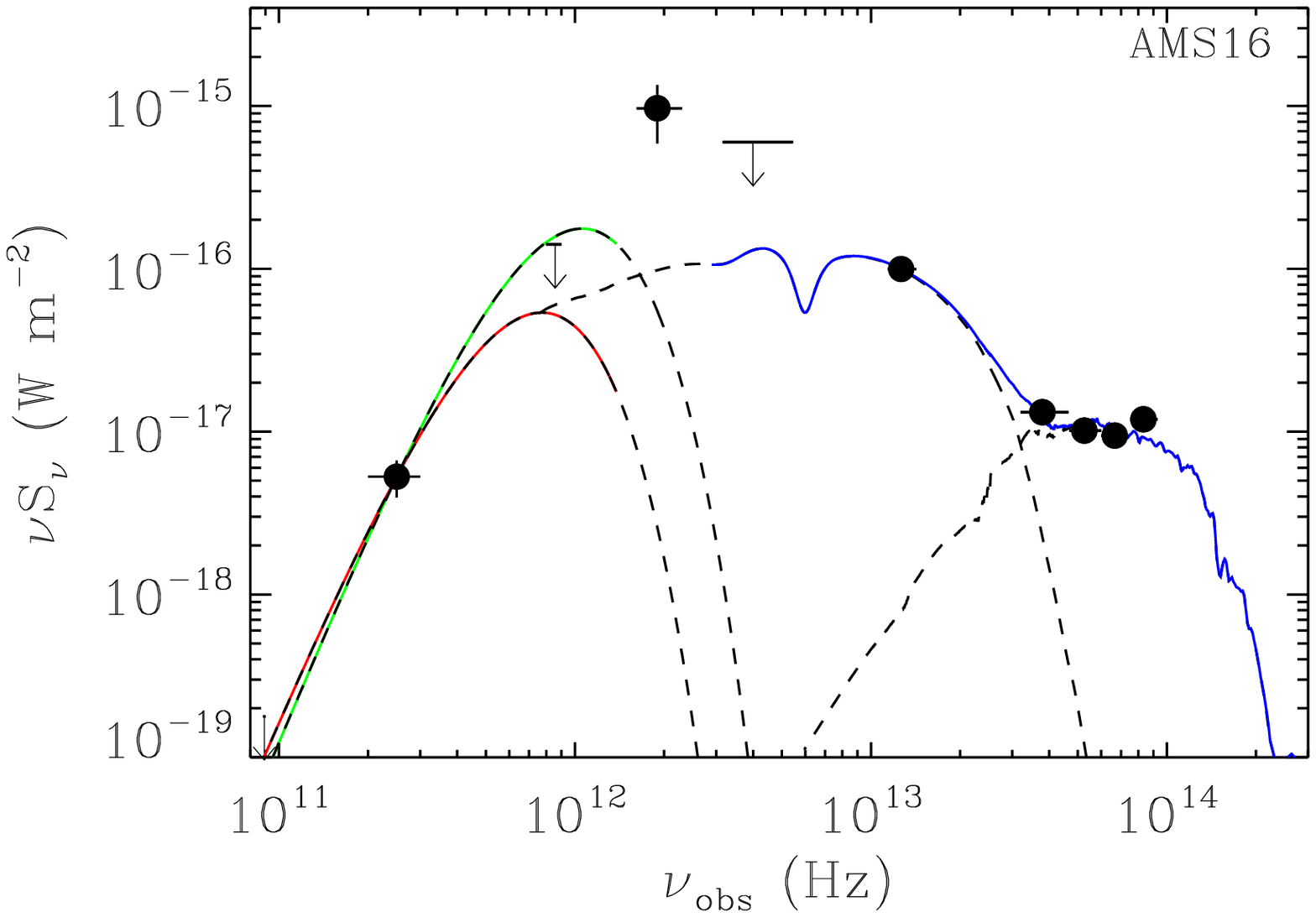} 

\includegraphics{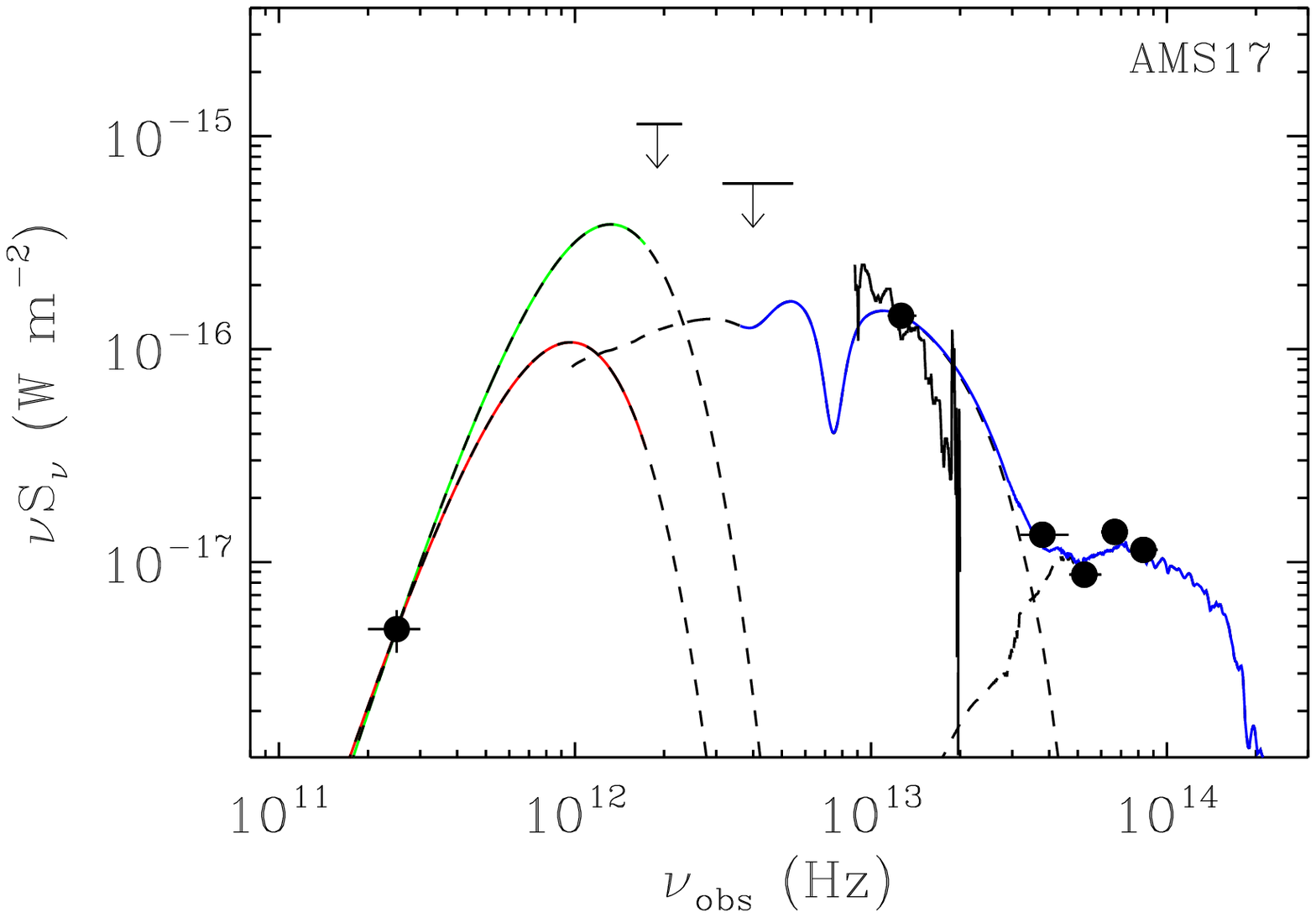} 

\includegraphics{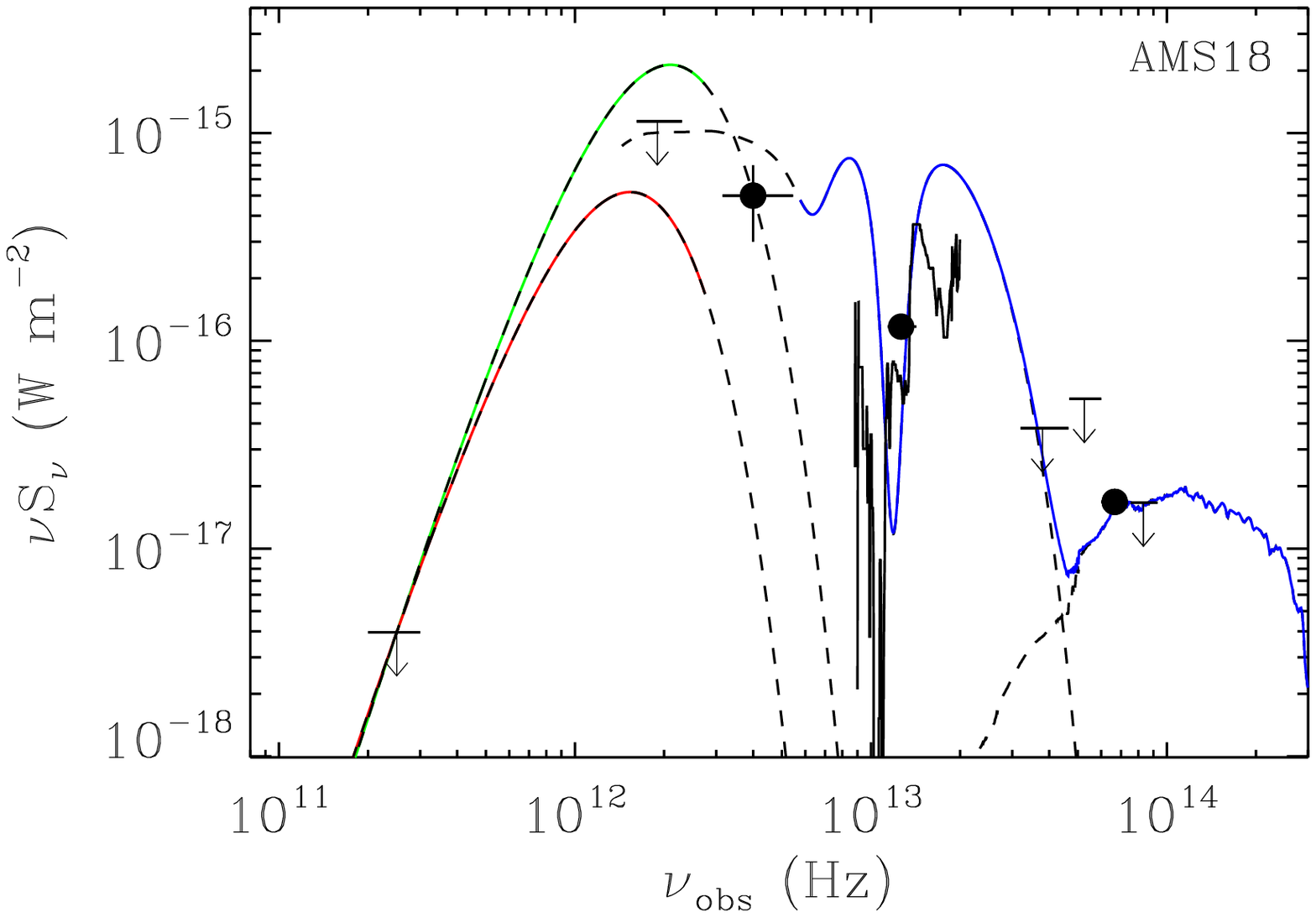} { 
 \vspace{18.5cm}
 \caption{ Continued. Sources AMS13 and AMS16 have additional data at
   350~\mum\, from the SHARC observations. In addition, AMS16 has a limit on the
   3~mm flux density from the PdBI observations.  }}
\end{figure*}
\clearpage

\addtocounter{figure}{-1}
\begin{figure*}
  \includegraphics{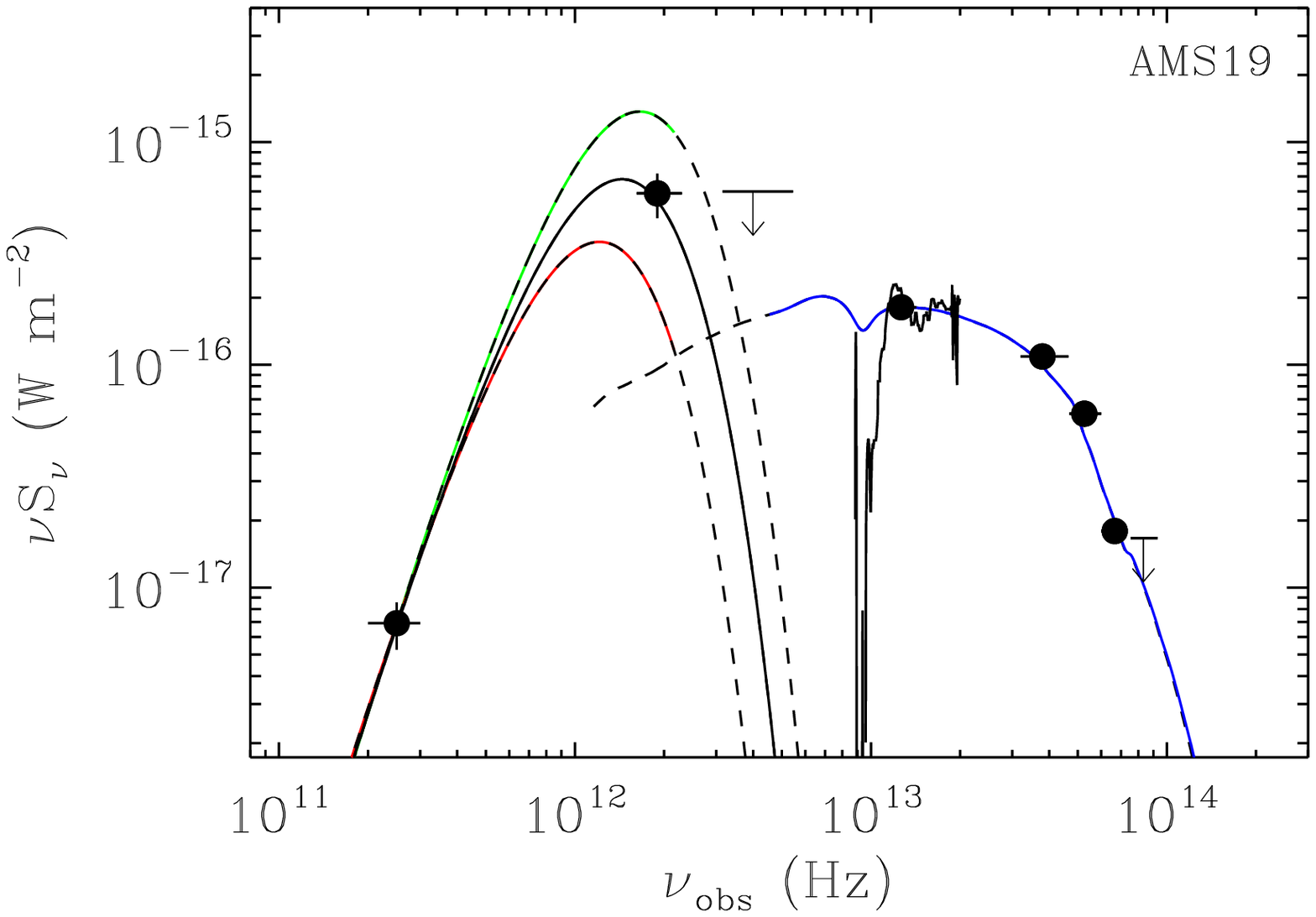} 
  
  \includegraphics{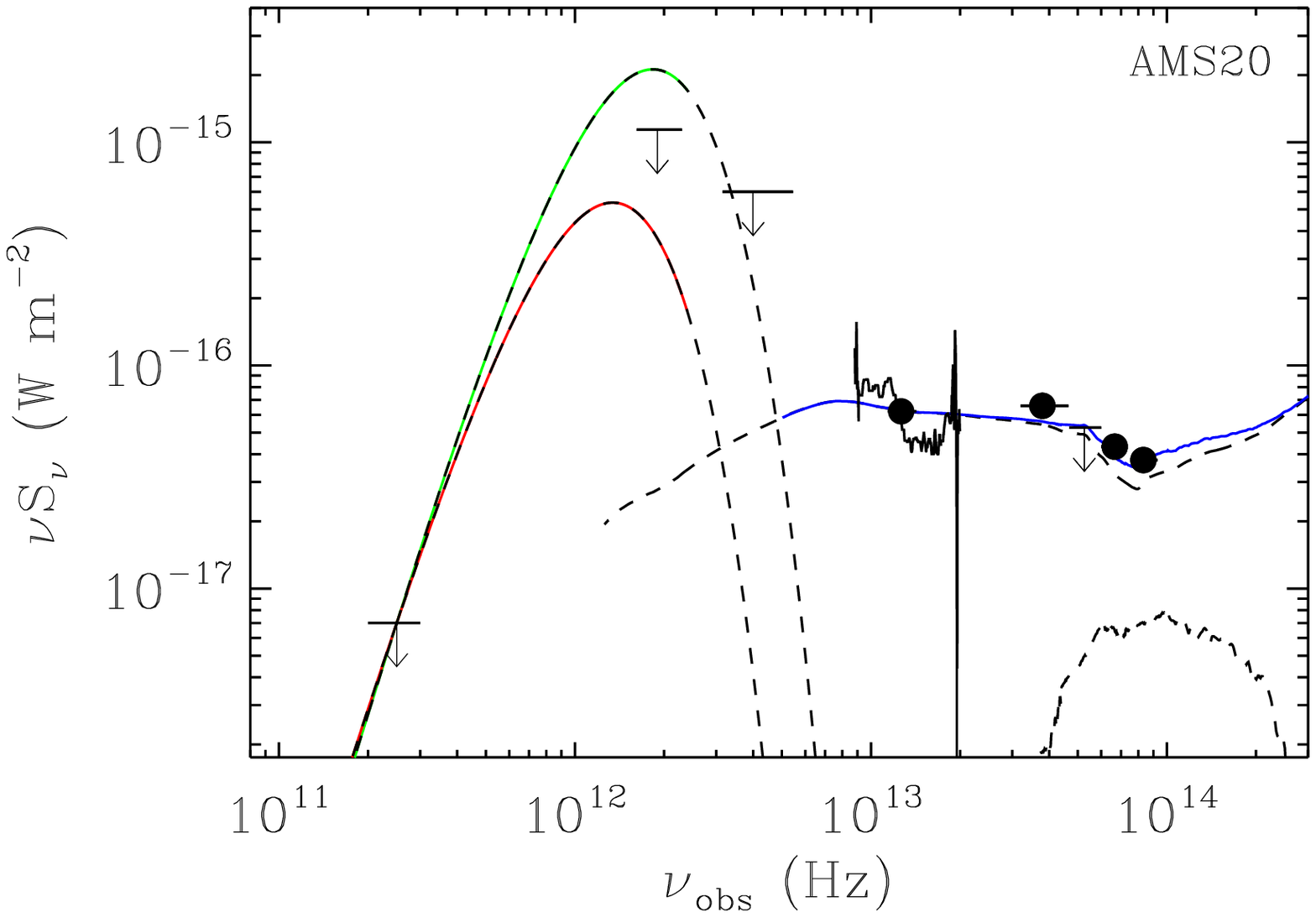} 

\includegraphics{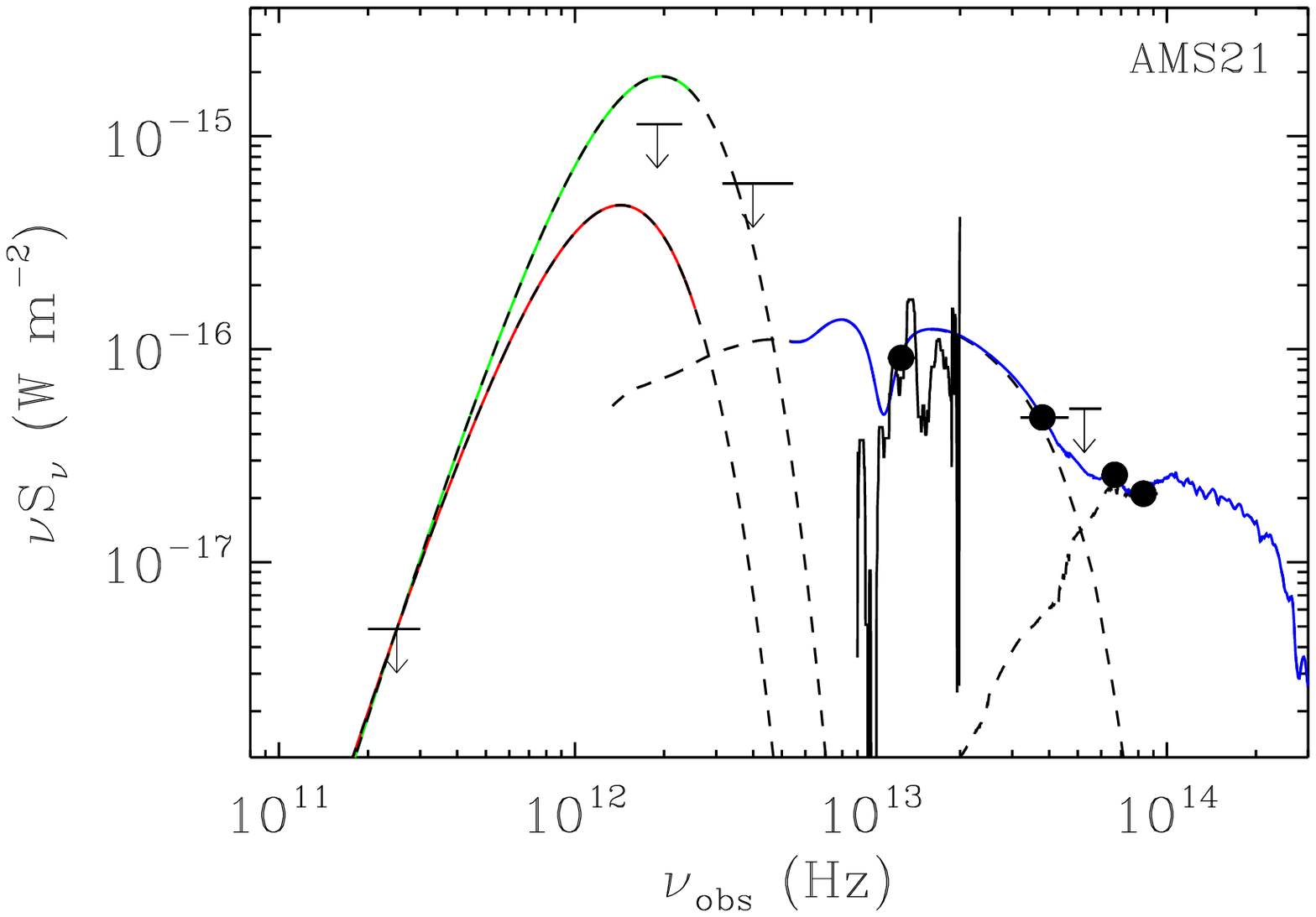} 
\vspace{14.5cm}
{ \caption{ Continued. Source AMS19 has an extra gray body overplotted
    (black solid line), described by the values in
    Table~\ref{tab:gb_fit}. }}
\end{figure*}

\clearpage

\begin{figure} 
\begin{center} 
\plotone{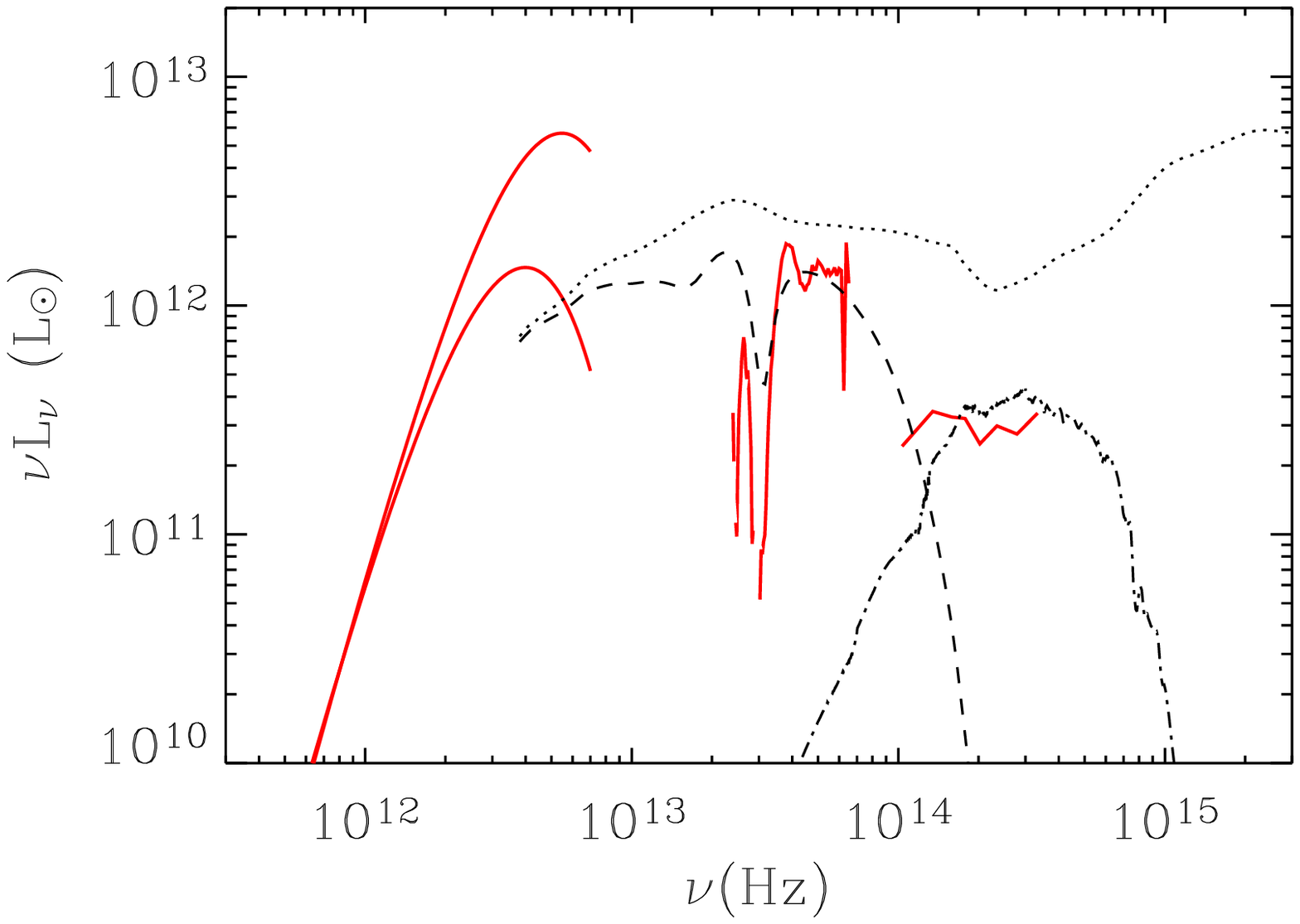}  
\caption{\noindent Mean dust SED for the sources in our sample with
  spectroscopic redshifts (solid red lines, see
  Section~\ref{sec:mean}). The near-infrared component is the mean of
  the observed IRAC photometry ($\nu=1-4\times10^{14}$~Hz, observed at
  3.6-8.0~\mum), the mid-infrared component
  ($\nu=2-6\times10^{13}$~Hz) is constructed using the IRS spectra,
  while the far-infrared component ($\nu\leq1\times10^{13}$~Hz)
  described by either one of the two fiducial gray bodies (the top
  solid red line has $T=$47~K, $\beta=$1.6, the bottom one has
  $T=$35~K, $\beta=$1.5). Overplotted for reference are an unobscured
  quasar SED \citep[dotted line][]{1994ApJS...95....1E} as well as the
  same quasar SED with a screen of MW dust
  \citep[from][]{1992ApJ...395..130P} with the mean \av$=32$ (plotted
  as a dashed line), and a $z=0$ elliptical galaxy \citet[dash-dotted
  line, from ][]{1980ApJS...43..393C}. This last SED has been
  normalised to be the progenitor of a present day $\sim2$$L^{\star}$
  galaxy.  }
\label{fig:mean_sed} 
\end{center} 
\end{figure} 
 
 \begin{figure} 
\begin{center} 
\plottwo{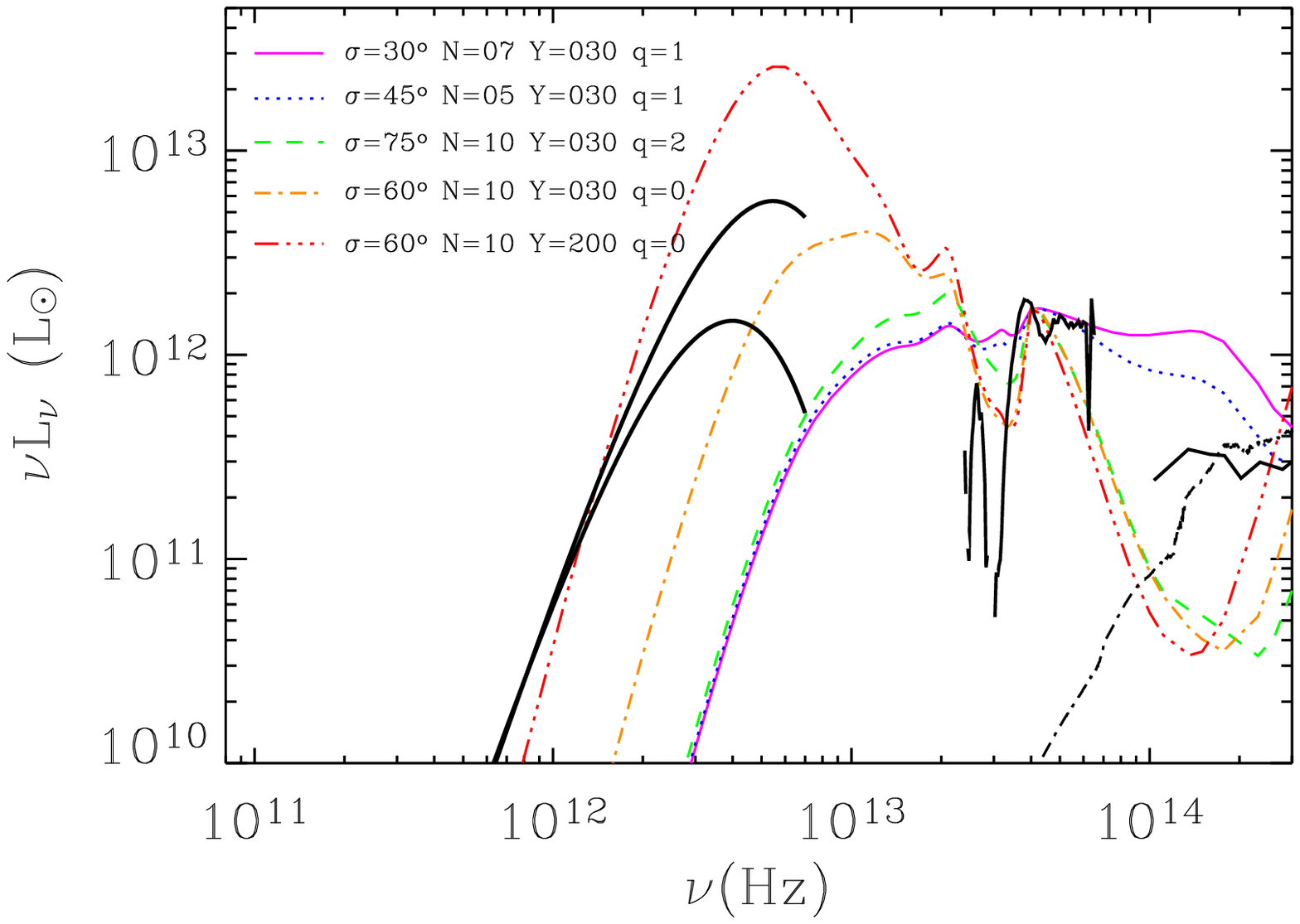} {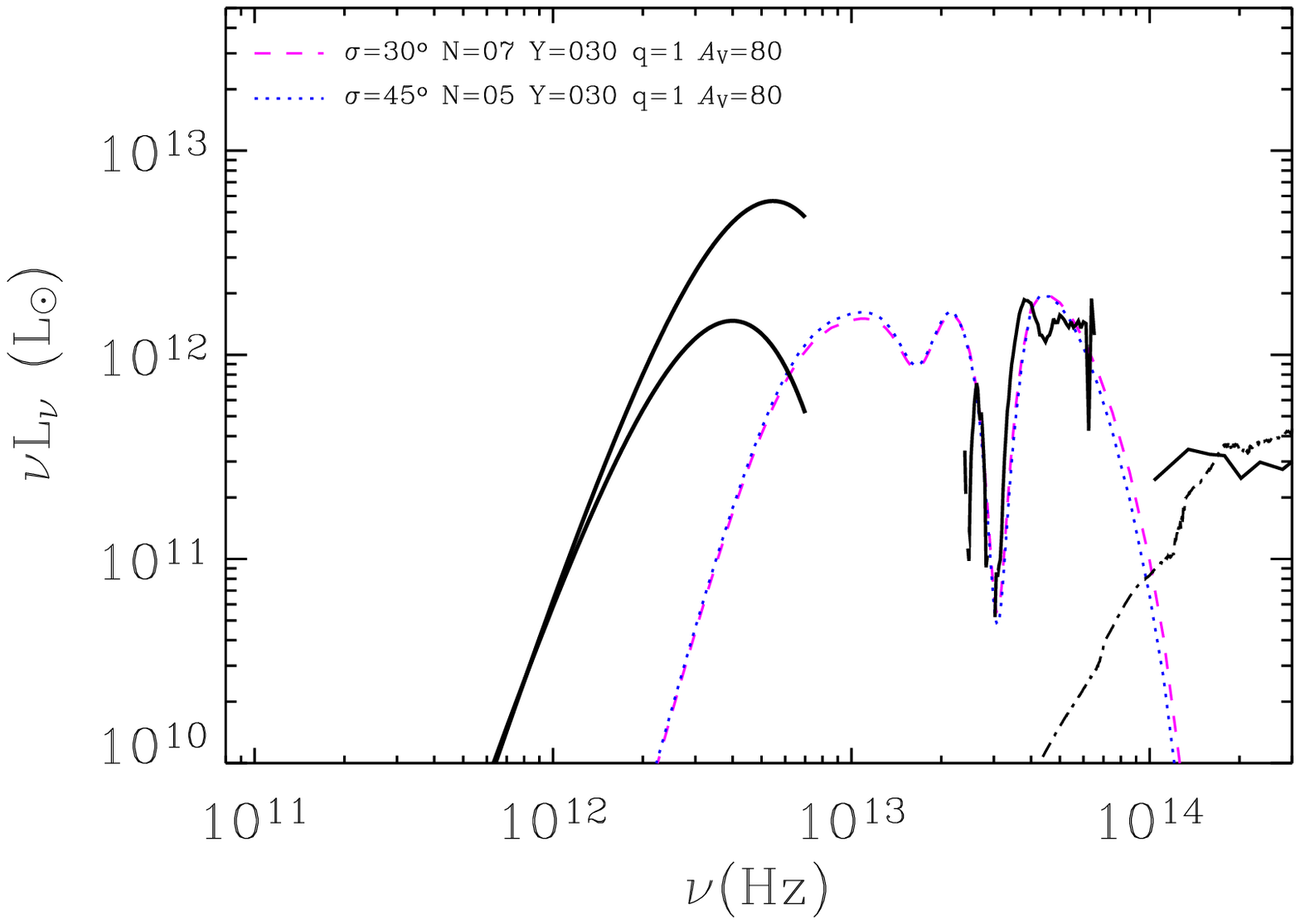}  
\caption{\noindent Comparison of the mean SED (solid black lines) with
  different clumpy models of AGN tori. The progenitor of a
  2$L^{\star}$ galaxy has been overplotted again (dot-dashed
  line). The clumpy models, from
  \citet{2008ApJ...685..147N,2008ApJ...685..160N}, all have
  single-cloud optical depths $\tau_{\rm V}=60$. Left panel: Different
  lines show models with varying parameters for the width of the
  gaussian angular distribution of clumps around the equator,
  $\sigma$, the number of clouds along a ray in the equatorial plane
  $N$, the ratio of inner to outer torus radius $Y$, and the power of
  the radial density distribution, $q$. The models cannot
  simultaneously reproduce the relatively flat continuum between
  $1\times10^{13}$ and $7\times10^{13}$~Hz, the sudden drop at higher
  frequencies, the depth of the silicate feature and the millimetre
  detection. Right panel: Additional extinction from dust along the
  line of sight has been added to the two flatter clumpy models from
  the left panel. The emission from this cool dust is  represented by 
  the two fiducial gray bodies used throughout this work. The typical
  quasar SED of our sample can be reasonably reproduced using the
  clumpy torus models, with foreground extinction \av$=80$ and
  re-emission from a mass of $3\times10^{8}$~\msol\, of cool dust. The emission from this
  cool-dust mass and the foreground extinction are consistent provided the
  dust is within a 2~kpc radius (using Equation~\ref{eq:tau_m} from
  Section~\ref{sec:host}).  }
\label{fig:mean_clumpy} 
\end{center} 
\end{figure} 
 
\begin{figure} 
\begin{center} 
\plotone{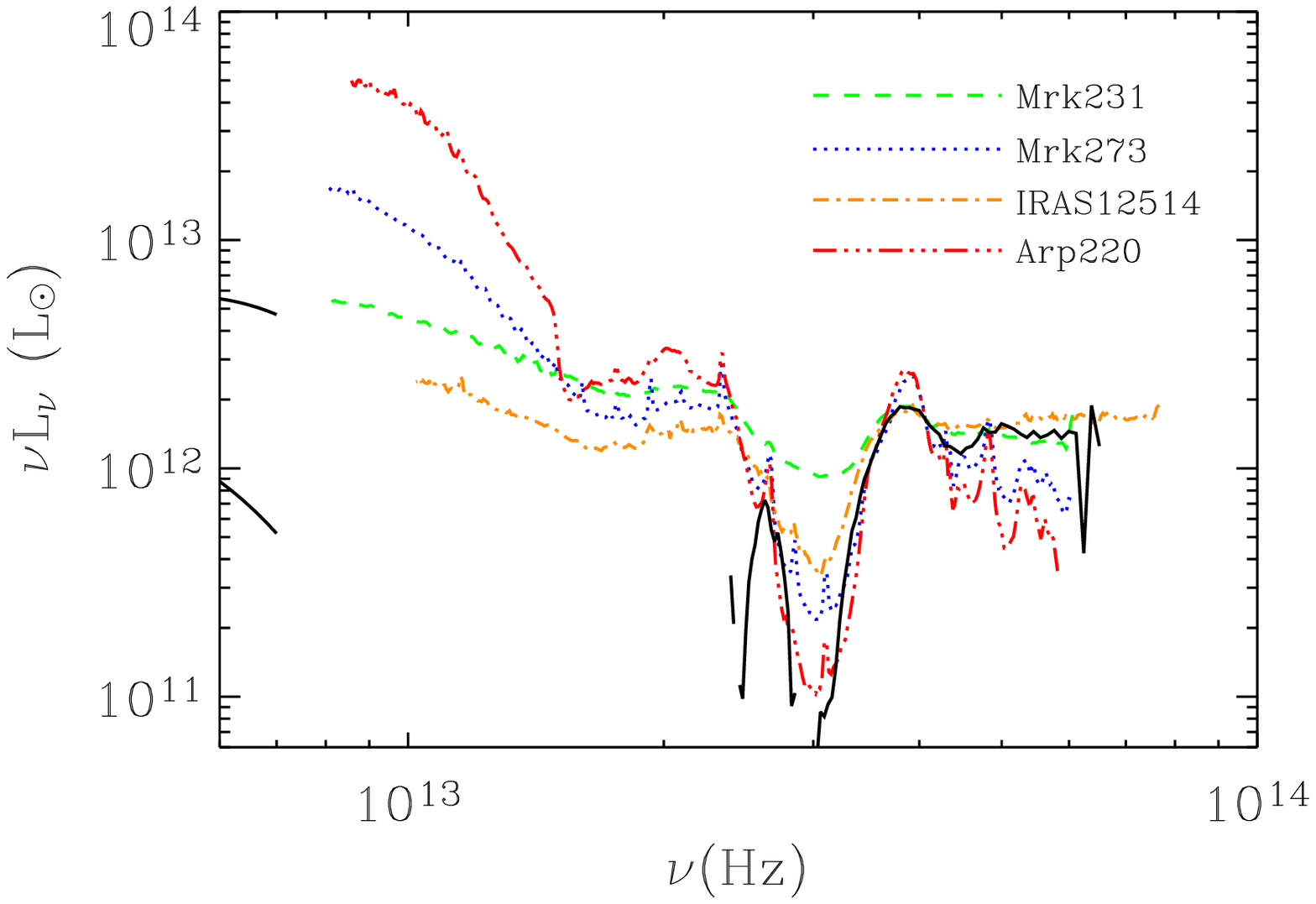}  
\caption{\noindent Comparison of the mid-infrared component of the mean SED (black solid line) with four local ULIRGs: IRAS~12514$+$1027 (hereafter IRAS~12514), Mrk~231, Mrk~273 and
Arp~220. }
\label{fig:mean_comp} 
\end{center} 
\end{figure}

\begin{figure} 
\begin{center} 
\includegraphics[width=8cm, angle=-90]{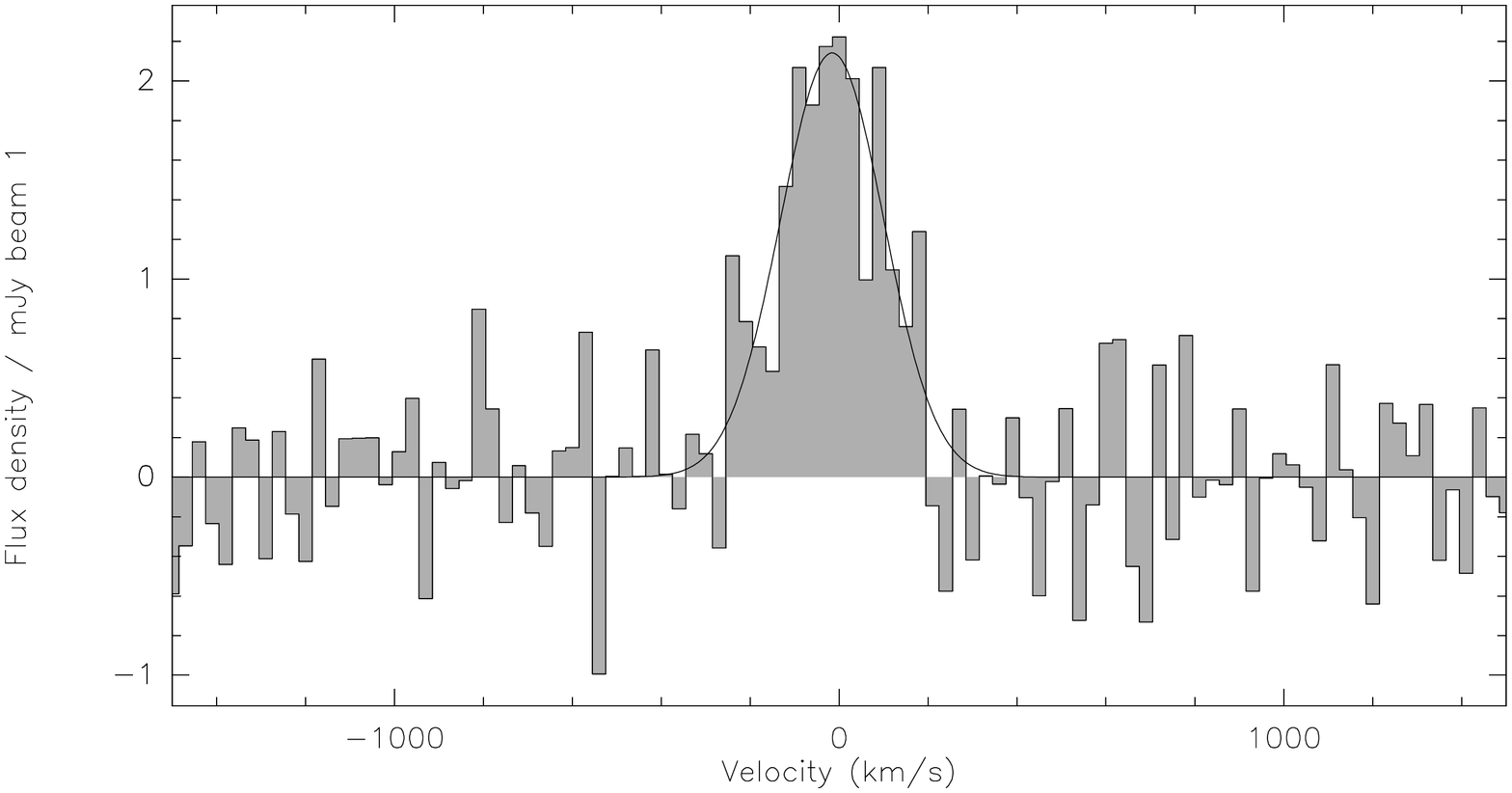}
\caption{\noindent Spectrum of AMS12 obtained with the Plateau de Bure
  Interferometer. The line is well fit by a single gaussian centred on
  -16~\kms\, ($z_{\rm CO}=2.7668$) with a FWHM of 275~\kms. From this
  gaussian fit, the CO(3-2) line has an integrated flux of 630 mJy
  \kms\,corresponding to a frequency-integrated line luminosity of
  $L_{\rm CO (3-2)}=$3.2$\times10^{7}$~\lsol, or a
  brightness-temperature luminosity of $L'_{\rm CO
    (3-2)}=2.4\times10^{10}$ K~\kms~pc$^{2}$.  }
\label{fig:coams12} 
\end{center} 
\end{figure}

\begin{figure} 
\begin{center} 
\includegraphics[width=8cm, angle=-90]{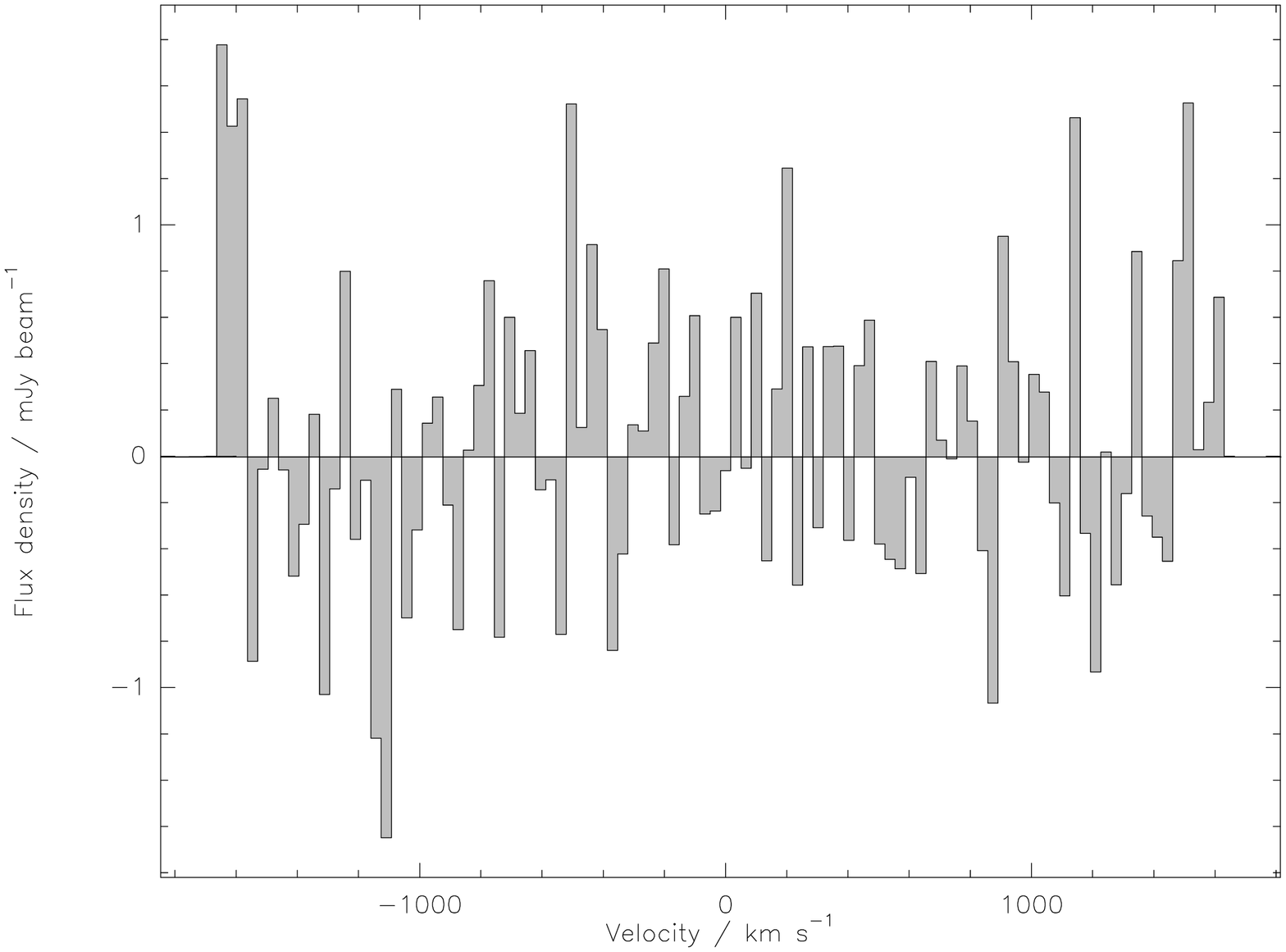}
\caption{\noindent Spectrum of AMS16 obtained with the Plateau de Bure
  Interferometer. The frequency was centred on 89.2002~GHz, the expected
  location of the CO(4-3) rotational transition. The rms noise per 10~MHz bin
  (33 \kms) is 0.65~mJy beam$^{-1}$, assuming a box-car CO line shape with
  full width to zero intensity of 400~\kms\, and the line to be within the
  frequency covered, the 3$\sigma$ limit is $\Delta \nu S_{\rm CO}<$230~mJy
  beam$^{-1}$ \kms, or $L'_{\rm CO (4-3)}<1\times10^{10}$ K~\kms~pc$^{2}$ at
  the redshift of the source. }
\label{fig:coams16} 
\end{center} 
\end{figure}

\begin{figure} 
\begin{center} 
\plotone{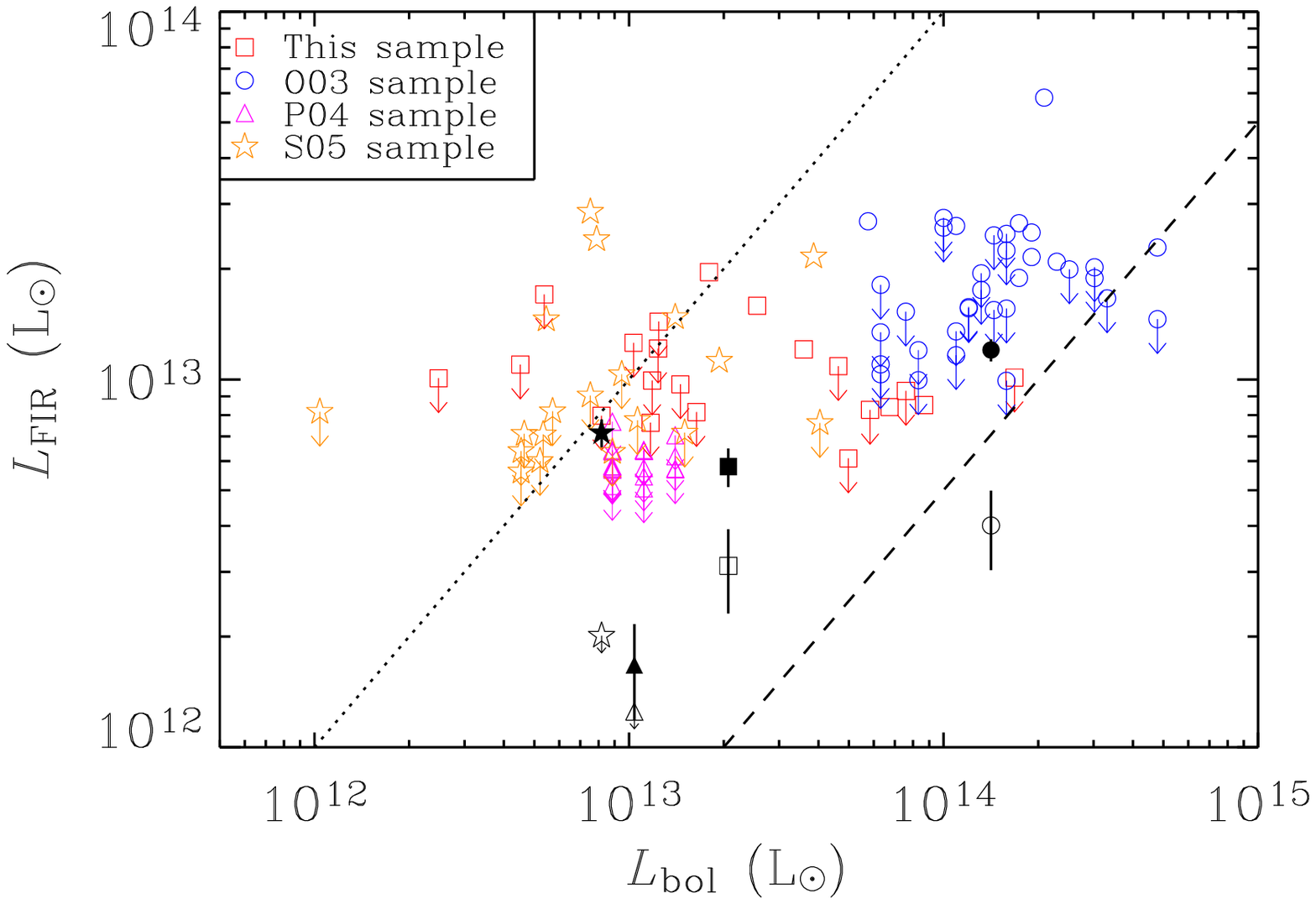}  
\caption{\noindent \lfir\, vs \lbol\, for the four samples of $z\sim2$
  quasars considered. O03 is the sample presented by
  \citet{2003A&A...398..857O}, P04 by \citet{2004ApJ...611L..85P} and
  S05 by \citet{2005MNRAS.360..610S}.  Black filled symbols represent
  the weighted mean of all sources in each sample, black empty symbols
  represent the weighted mean of non-detections in each sample. The
  dashed line represents \lfir$=0.05$\lbol,   the value found in 
  low-redshift quasars considered by \citet{1994ApJS...95....1E}. For reference, the line \lfir$=$\lbol\, has been plotted as a dotted line. For
  all samples, a gray body with $T=47$~K and $\beta=1.6$ has been
  assumed to calculate \lfir.}
\label{fig:lfir_lbol1} 
\end{center} 
\end{figure}

\label{lastpage} 
 
\end{document}